{}

\documentclass[11pt,a4paper]{article}

\newif\ifpublic\publictrue

\ifpublic\else\usepackage{showkeys}\fi
\usepackage[nosort]{cite}
\usepackage[parsep]{collref}
\usepackage[english]{babel}
\usepackage[T1]{fontenc}
\usepackage[utf8]{inputenc}
\usepackage{lmodern}
\usepackage{amsfonts,amsbsy,dsfont,euscript,mathrsfs,fixmath}
\usepackage{amsmath,amssymb}
\usepackage{enumerate}
\usepackage{graphicx}

\def\showkeysrefformat#1{{\normalfont\tiny\ttfamily#1}}
\makeatletter
\def\SK@@ref#1>#2\SK@{{\@inlabelfalse\leavevmode\vbox to\z@{\vss\SK@refcolor\rlap{\vrule\raise .75em \hbox{\showkeysrefformat{#2}}}}}}
\makeatother

\usepackage[a4paper,text={170mm,257mm},centering]{geometry}
\allowdisplaybreaks[3]
\numberwithin{equation}{section}
\def\[{\begin{equation}\begin{aligned}}
\def\]{\end{aligned}\end{equation}}
\newcommand{\nn}{\nonumber}

\usepackage[bookmarks=true,hyperfigures=true,hidelinks,bookmarksnumbered,unicode]{hyperref}

\usepackage[font=small,labelfont=bf,width=0.85\textwidth]{caption}

\expandafter\def\expandafter\bfseries\expandafter{\bfseries\ifmmode\else\boldmath\fi}
\expandafter\def\expandafter\mdseries\expandafter{\mdseries\ifmmode\else\unboldmath\fi}
\expandafter\def\expandafter\normalfont\expandafter{\normalfont\ifmmode\else\unboldmath\fi}

\RequirePackage{verbatim}

\makeatletter
\newwrite\bibinl@out
\newenvironment{bibtex}[1][\jobname]{%
\immediate\openout\bibinl@out #1.bib%
\immediate\write\bibinl@out{\@percentchar generated from `\jobname' starting line \the\inputlineno^^J}%
\def\verbatim@processline{\immediate\write\bibinl@out{\the\verbatim@line}}%
\@bsphack\let\do\@makeother\dospecials\catcode`\^^M\active\verbatim@start%
}
{\immediate\closeout\bibinl@out\@esphack}
\makeatother

\let\barefrac=\frac
\renewcommand{\frac}[2]{\mathinner{\barefrac{#1}{#2}}}

\let\baresqrt=\sqrt
\makeatletter
\renewcommand{\sqrt}{\@ifnextchar[\@sqrt@space@a\@sqrt@space@b}
\def\@sqrt@space@a[#1]#2{\mathinner{\mathchoice{\mkern-3mu}{\mkern-3mu}{}{}\baresqrt[#1]{#2}}}
\def\@sqrt@space@b#1{\mathinner{\mathchoice{\mkern-3mu}{\mkern-3mu}{}{}\baresqrt{#1}}}
\makeatother

\makeatletter
\let\per@dot@old=\.
\def\.{\ifmmode\def\per@dot@sel{\mkern3mu}\else\def\per@dot@sel{\per@dot@old}\fi\per@dot@sel}
\makeatother

\let\barefootnote=\footnote
\renewcommand{\footnote}[1]{\barefootnote{#1\vspace{3pt}}}

\newcommand{\vfrac}[2]{\ifmmode\mathinner{\textstyle^{#1}\!/\!_{#2}}\else$^{#1}\!/\!_{#2}$\fi}

\DeclareMathOperator{\tr}{tr}

\DeclareMathOperator{\im}{im}

\newcommand{\Order}{O}

\newcommand{\Real}{\mathds{R}}
\newcommand{\Complex}{\mathds{C}}
\newcommand{\Integer}{\mathds{Z}}

\newcommand{\Projective}{\mathds{P}}

\makeatletter
\newcommand*\bigcdot{\mathpalette\bigcdot@{.5}}
\newcommand*\bigcdot@[2]{\mathbin{\vcenter{\hbox{\scalebox{#2}{$\m@th#1\bullet$}}}}}
\makeatother

\newcommand{\alg}[1]{\mathfrak{#1}}
\newcommand{\grp}[1]{\mathrm{#1}}
\DeclareMathOperator{\rank}{rank}
\DeclareMathOperator{\Lie}{Lie}
\DeclareMathOperator{\ad}{ad}
\DeclareMathOperator{\Ad}{Ad}

\def\<{\big\langle}
\def\>{\big\rangle}

\DeclareFontEncoding{LS1}{}{}
\DeclareFontSubstitution{LS1}{stix}{m}{n}
\DeclareSymbolFont{stixsymbols}{LS1}{stixscr}{m}{n}
\SetSymbolFont{stixsymbols}{bold}{LS1}{stixscr}{b}{n}
\DeclareMathSymbol{\kay}{\mathalpha}{stixsymbols}{"6B}
\DeclareMathSymbol{\hay}{\mathalpha}{stixsymbols}{"68}
\newcommand{\Rb}{\bar{R}}
\newcommand{\etab}{\bar{\eta}}
\newcommand{\hayb}{\bar{\hay}}
\newcommand{\g}{\alg{g}}
\newcommand{\R}{\scriptstyle{R}}
\newcommand{\Er}{e^{\rho R}}
\newcommand{\Emr}{e^{-\rho R}}

\newcommand{\Erp}{e^{\rho R_+}}
\newcommand{\Erm}{e^{\rho R_-}}
\newcommand{\Erpm}{e^{\rho R_\pm}}
\newcommand{\Oc}{\mathcal{O}}
\newcommand{\alp}{e^\chi}
\newcommand{\Lc}{\mathcal{L}}
\newcommand{\Tc}{\mathcal{T}}
\newcommand{\p}{\partial}
\newcommand{\ti}[1]{_{\mathbf{\underline{#1}}}}
\newcommand{\Rc}{\mathcal{R}}
\newcommand{\vp}{\varphi}
\newcommand{\dd}{\text{d}}
\newcommand{\gh}{\widehat{g}}
\newcommand{\Mg}{\mathcal{M}_g}
\newcommand{\Ng}{\mathcal{N}_g}
\newcommand{\Mgx}{\mathcal{M}_{g(x)}}
\newcommand{\Ngx}{\mathcal{N}_{g(x)}}
\newcommand{\Ngy}{\mathcal{N}_{g(y)}}

\newcommand{\Ergm}{e^{-\rho R_g}}
\newcommand{\Jc}{\mathcal{J}}
\newcommand{\Id}{1}
\DeclareMathOperator{\res}{res}
\newcommand{\cK}{K}
\newcommand{\tra}{{\mkern2mu {}^t\mkern-1mu}}
\newcommand{\trab}{{\mkern2mu {}^{(t)}\mkern-1mu}}
\newcommand{\trap}{{\mkern2mu {}^t\mkern-3mu}}
\newcommand{\llangle}{\langle\mkern-2.5mu\langle}
\newcommand{\rrangle}{\rangle\mkern-2.5mu\rangle}
\newcommand{\dotp}{\mkern3mu \dot{+} \mkern3mu}
\DeclareMathAlphabet{\mathdsl}{U}{bbm}{m}{sl}
\newcommand{\gdsl}{\mathdsl{g}}
\newcommand{\udsl}{\mathdsl{u}}
\newcommand{\lh}{{\bar{v}}}
\newcommand{\er}{e^{\rho R}}

\newcommand{\Uc}{\mathcal{U}}
\newcommand{\Vc}{\mathcal{V}}

\providecommand{\href}[2]{#2}

\makeatletter
\def\mr@ignsp#1 {\ifx\:#1\@empty\else #1\expandafter\mr@ignsp\fi}
\newcommand{\multiref}[1]{\begingroup%
\xdef\mr@no@sparg{\expandafter\mr@ignsp#1 \: }%
\def\mr@comma{}\def\mr@dash{-}%
\@for\mr@refs:=\mr@no@sparg\do{%
\ifx\mr@refs\mr@dash\def\mr@comma{}--\else%
\mr@comma\def\mr@comma{,}\ref{\mr@refs}\fi}%
\endgroup}
\renewcommand{\eqref}[1]{(\multiref{#1})}
\makeatother

\makeatletter
\newcommand{\namedref}[2]{\hyperref[#2]{#1~\ref*{#2}}}
\newcommand{\secref}{\@ifstar{\namedref{Section}}{\namedref{sec.}}}
\newcommand{\ssecref}{\@ifstar{\namedref{Subsection}}{\namedref{subsec.}}}
\newcommand{\appref}{\@ifstar{\namedref{Appendix}}{\namedref{app.}}}
\newcommand{\tabref}{\@ifstar{\namedref{Table}}{\namedref{tab.}}}
\newcommand{\figref}{\@ifstar{\namedref{Figure}}{\namedref{fig.}}}
\newcommand{\footref}{\@ifstar{\namedref{Footnote}}{\namedref{footnote}}}
\makeatother

\usepackage{amsthm}
\theoremstyle{definition}

\let\oldbib=\thebibliography
\def\thebibliography{\phantomsection\addcontentsline{toc}{section}{\refname}\oldbib}

\let\oldtoc=\tableofcontents
\def\tableofcontents{\phantomsection\addcontentsline{toc}{section}{\contentsname}\oldtoc}

\providecommand{\hypersetup}[1]{}
\providecommand{\texorpdfstring}[2]{#1}
\hypersetup{plainpages=false}
\hypersetup{pdfpagemode=UseNone}
\hypersetup{bookmarksnumbered=true}
\hypersetup{pdfstartview=FitH}
\hypersetup{colorlinks=false}
\hypersetup{citebordercolor={1 1 1}}
\hypersetup{urlbordercolor={1 1 1}}
\hypersetup{linkbordercolor={1 1 1}}

\makeatletter
\let\@keywords\@empty
\let\@subject\@empty
\providecommand{\keywords}[1]{\gdef\@keywords{#1}}
\providecommand{\subject}[1]{\gdef\@subject{#1}}
\def\thetitle{\@title}
\def\theauthor{\@author}
\def\thesubject{\@subject}
\def\thedate{\@date}
\def\thekeywords{\@keywords}
\makeatother
\AtBeginDocument{
\hypersetup{pdftitle={\thetitle}}
\hypersetup{pdfauthor={\theauthor}}
\hypersetup{pdfsubject={\thesubject}}
\hypersetup{pdfkeywords={\thekeywords}}}

\newif\ifshownote
\shownotetrue

\ifpublic\shownotefalse\fi

\ifshownote

\ifpdf\else\RequirePackage[active]{srcltx}\fi
\RequirePackage{xcolor}
\RequirePackage{ulem}

\newcommand{\remark}[2][]{{\normalfont\sffamily\hspace{1ex}
\def\emph{\textsl}\def\textbullet{$\bullet$}
\def\tmparga{#1}
\def\tmpargb{BH}\ifx\tmparga\tmpargb\color[rgb]{0,0.7,0.3}\fi
\def\tmpargb{SL}\ifx\tmparga\tmpargb\color[rgb]{0,0.3,0.7}\fi
\def\tmpargb{}\ifx\tmparga\tmpargb\normalfont\color{red}\fi
\def\tmpargb{}\ifx\tmparga\tmpargb\else \textbf{#1:}\fi
#2\hspace{1ex}}}

\else

\newcommand{\remark}[2][]{\ignorespaces}

\fi

\title{Yang-Baxter deformations of the \texorpdfstring{\\}{} Principal Chiral Model plus Wess-Zumino term}
\author{B. Hoare\texorpdfstring{\textsuperscript{1}}{} and S. Lacroix\texorpdfstring{\textsuperscript{2}}{}}

\begin{document}

\pdfbookmark[1]{Title Page}{title}
\thispagestyle{empty}

\begingroup\raggedleft
[ZMP-HH/20-17]
\par\endgroup

\vspace*{2cm}
\begin{center}
\begingroup\Large\bfseries\thetitle\par\endgroup
\vspace{1cm}

\begingroup\theauthor\par\endgroup
\vspace{1cm}

\textit{\textsuperscript{1}Institut f\"ur Theoretische Physik, Eidgen\"ossische Technische Hochschule Z\"urich,\\
Wolfgang-Pauli-Strasse 27, 8093 Z\"urich, Switzerland}
\vspace{1mm}

\begingroup\ttfamily\small
bhoare@ethz.ch\par
\endgroup
\vspace{5mm}

\textit{\textsuperscript{2}
II. Institut f\"ur Theoretische Physik, Universit\"at Hamburg, \\ Luruper Chaussee 149, 22761 Hamburg, Germany
\vspace{1mm}
\\ Zentrum f\"ur Mathematische Physik, Universit\"at Hamburg, \\ Bundesstrasse 55, 20146 Hamburg, Germany}
\vspace{1mm}

\begingroup\ttfamily\small
sylvain.lacroix@desy.de\par
\endgroup
\vspace{5mm}

\vfill

\textbf{Abstract}\vspace{5mm}

\begin{minipage}{13cm}\small
A large class of integrable deformations of the Principal Chiral Model, known as the Yang-Baxter deformations, are governed by skew-symmetric R-matrices solving the (modified) classical Yang-Baxter equation.
We carry out a systematic investigation of these deformations in the presence of the Wess-Zumino term for simple Lie groups, working in a framework that treats both inhomogeneous and homogeneous deformations on the same footing.
After analysing the cohomological conditions under which such a deformation is admissible, we consider an action for the general Yang-Baxter deformation of the Principal Chiral Model plus Wess-Zumino term and prove its classical integrability. We also show how the model is found from a number of alternative formulations: affine Gaudin models, $\mathcal{E}$-models, 4-dimensional Chern-Simons theory and, for homogeneous deformations, non-abelian T-duality.
\end{minipage}

\vspace*{4cm}

\end{center}

\newpage

\setcounter{tocdepth}{2}
\tableofcontents

\section{Introduction}
\label{sec:intro}

Integrable $\sigma$-models are an important class of 2-dimensional integrable fields theories.
The prototypical example is the Principal Chiral Model (PCM), a $\sigma$-model whose target space is a simple Lie group $\grp{G}$.
Yang-Baxter (YB) deformations~\cite{Klimcik:2002zj,Klimcik:2008eq} of the PCM are continuous deformations that preserve its integrability.
The deformed models were first constructed in~\cite{Klimcik:2002zj,Klimcik:2008eq} and are characterised by a skew-symmetric R-matrix on the Lie algebra of $\grp{G}$.
Such deformations can be further categorised as either inhomogeneous, when the R-matrix solves the modified classical Yang-Baxter equation (mcYBE) or homogeneous, when it solves the classical Yang-Baxter equation (cYBE).

The PCM is part of a family of integrable $\sigma$-models with target space $\grp{G}$.
The action of these models is given by that of the PCM plus the standard topological Wess-Zumino (WZ) term~\cite{Novikov:1982ei,Witten:1983tw,Witten:1983ar}. Each model in this family is labelled by the level $k$ of the WZ term and has a single coupling constant. For a critical value of this coupling, proportional to $k$, this model is a 2-dimensional conformal field theory.
More precisely, it is the Wess-Zumino-Witten (WZW) model~\cite{Novikov:1982ei,Witten:1983tw,Witten:1983ar} at level $k$.
The existence of this family of integrable $\sigma$-models, henceforth known as the PCM plus WZ term, motivates us to investigate their integrable deformations; in particular, their YB deformations.

Progress has been made in the construction of such deformations for two classes of R-matrices:
\begin{itemize}
\item The $\sigma$-model on the squashed 3-sphere~\cite{Cherednik:1981df} is an example of a YB deformation of the $\grp{SU}(2)$ PCM governed by the standard Drinfel'd-Jimbo R-matrix.
It was shown in~\cite{Kawaguchi:2011mz,Kawaguchi:2013gma} that the model remains integrable on additionally adding a WZ term.
In~\cite{Delduc:2014uaa} this deformation, based on the standard Drinfel'd-Jimbo R-matrix, was generalised to arbitrary $\grp{G}$.
\item For homogeneous R-matrices, \textit{i.e.} solving the cYBE, the deformed model has been formulated by exploiting the relation to non-abelian T-duality~\cite{Borsato:2018idb}.
This requires that the WZ term can be written in a form that can be dualised.
This is not always the case and indicates that some YB deformations of the PCM plus WZ term may not be admissible.
\end{itemize}
These two constructions are different in approach and both hide aspects of the underlying algebraic structure.
Furthermore, even together they do not cover all possible skew-symmetric R-matrices.

In this article we address these questions by carrying out a systematic analysis of YB deformations, \textit{i.e.} integrable deformations governed by solutions of the (m)cYBE, in the presence of a WZ term.
We work in a unifying framework that treats both inhomogeneous and homogeneous deformations on an equal footing.
Our goals are (i) to determine conditions for such a deformation to be admissible, (ii) construct an action for the YB deformation of the PCM plus WZ term and (iii) prove its classical integrability.
Our construction is motivated by recent work~\cite{Klimcik:2019kkf,Klimcik:2020fhs} of Klim\v{c}\'ik, which revisits the case of the standard Drinfel'd-Jimbo R-matrix and, working in the context of $\mathcal{E}$-models, explores the underlying algebraic structure.
This leads to an alternative, more compact, formulation of its action, which provides the basis for our generalisation to all those skew-symmetric R-matrices for which the deformation is admissible.

\medskip

The integrability of a 2-dimensional $\sigma$-model relies on the existence of a Lax connection encoding its dynamics, \textit{i.e.} a connection depending on an auxiliary complex spectral parameter $z$ whose flatness is equivalent to the equations of motion.
In all the models discussed thus far, this Lax connection takes a particular form relying on the existence of a flat and conserved current valued in the Lie algebra $\g$ of $\grp{G}$.
This current is the Noether current associated with a global $\grp{G}$-symmetry acting on the target space by right translations.
Motivated by this common structure, our starting point is the general ansatz
\begin{equation}\label{Eq:AnsatzIntro}
\mathsf{S} = \int d^2x \; \kappa (g^{-1}\partial_+ g, \mathcal{O}_g g^{-1}\partial_- g) + \frac{\kay}{6} \int d^3 x \; \epsilon^{abc} \kappa(g^{-1}\partial_a g, [g^{-1}\partial_b g , g^{-1} \partial_c g]),
\end{equation}
where $g$ is a $\grp{G}$-valued field, $\kappa$ is the normalised Killing form on $\g$ and $\p_\pm$ are light-cone derivatives.
$\mathcal{O}$ is a constant linear operator on $\alg{g}$ characterising the model, $\mathcal{O}_g$ is defined as $\Ad_g^{-1} \mathcal{O} \Ad_g$ and the second term is the familiar 3-dimensional WZ term with level $k = 4\pi \kay$.
The action \eqref{Eq:AnsatzIntro} is invariant under right multiplication, $g\mapsto gg_0$, and hence its equations of motion are equivalent to the conservation of a Noether current.
In \secref{sec:general} we investigate the conditions on the operator $\mathcal{O}$ under which this current is also flat on-shell, ensuring the existence of a Lax connection.

Starting from the YB deformation of the PCM, \textit{i.e.} $\kay = 0$ and $\mathcal{O} = \frac{\hay}{1-\eta R}$ where $R$ is a skew-symmetric R-matrix on $\g$, we develop perturbation theory in $\kay/\hay$.
At the first sub-leading order we find that an additional condition on $R$ is needed to ensure the integrability of the model, which we formulate in cohomological terms as follows.
It is a standard result that a solution of the (m)CYBE defines a second Lie bracket $[X,Y]_R=[RX,Y]+[X,RY]$ on $\g$ and thus a Lie algebra $\g_R$.
In the Lie algebra cohomology of $\g_R$, the 3-cochains $\Omega(X,Y,Z)=\kappa(X,[Y,Z])$ and $\Omega_R(X,Y,Z) = \kappa( R_\pm X, [R_\pm Y, R_\pm Z] )$ are closed
(here $R_\pm = R \pm c$, where $c\neq0$ and $c=0$ for solutions of the mcYBE and cYBE respectively).
The first-order integrability condition then admits a solution if and only if $\Omega_R + \alpha\,\Omega$ is exact for some choice of $\alpha$.
While not all R-matrices satisfy this condition, there are large classes that do, including the standard Drinfel'd-Jimbo R-matrix and those homogeneous R-matrices for which the deformed model can be found by non-abelian T-duality.
For all the R-matrices we consider satisfying this condition, it turns out that $\Omega_R$ itself vanishes.
This condition admits an interesting algebraic reformulation.
Recall that, if $R$ a solution of the (m)cYBE, then $\alg{h}_\pm = \im R_\pm$ are subalgebras of $\g$, and hence $\Omega_R = 0$ if and only if $\alg{h}_\pm$ is solvable.
Therefore, for those $R$ with $\alg{h}_\pm$ solvable there exists an integrable YB deformation of the PCM plus WZ term at first order in $\kay/\hay$.

In \secref{sec:solvable} we show that this integrable deformation extends to all values of $\kay$.
More precisely, we show that if $\alg{h}_\pm$ is solvable then the action
\begin{equation}\label{Eq:SolvSIntro}
\mathsf{S} = \frac{\kay}{2} \int d^2x \; \kappa \Big(g^{-1}\partial_+ g, \frac{\alp+e^{\rho R_g}}{\alp-e^{\rho R_g}} g^{-1}\partial_- g \Big) + \frac{\kay}{6} \int d^3 x \; \epsilon^{abc} \kappa(g^{-1}\partial_a g, [g^{-1}\partial_b g , g^{-1} \partial_c g]),
\end{equation}
defines an integrable $\sigma$-model.
In particular, assuming that $R$ solves the (m)cYBE and $\alg{h}_\pm$ is solvable, the orthogonal operator $e^{\rho R}$ satisfies a simple algebraic identity, proven in \appref{sec:identity}, that ensures the existence of a Lax connection.
In \appref{app:gendef} we investigate the reverse logic and show that this identity and the skew-symmetry of $R$ are also necessary conditions for integrability under certain assumptions.
The action \eqref{Eq:SolvSIntro} was first proposed in~\cite{Klimcik:2019kkf} as a reformulation of the model initially constructed in~\cite{Delduc:2014uaa} when $R$ is the standard Drinfel'd-Jimbo R-matrix.
Therefore, \secref{sec:solvable} can be thought of as generalising this result to all skew-symmetric R-matrices with solvable $\alg{h}_\pm$.

The proof of classical integrability is completed in \secref{sec:hamiltonian}.
The Hamiltonian analysis of the YB deformed PCM with and without WZ term was performed in~\cite{Delduc:2014uaa} and~\cite{Delduc:2013fga} respectively for the standard Drinfel'd-Jimbo R-matrix.
In these cases, the Poisson bracket of the Lax matrix is a Maillet bracket~\cite{Maillet:1985fn,Maillet:1985ek} taking a particular form that is encoded by a rational function of the spectral parameter known as the twist function~\cite{Maillet:1985ec,Reyman:1988sf,Sevostyanov:1995hd,Vicedo:2010qd} (see also~\cite{Lacroix:2018njs}).
This ensures that the conserved charges extracted from the monodromy of the Lax matrix are in involution.
In \secref{sec:hamiltonian}, we perform the Hamiltonian analysis of YB deformation of the PCM plus WZ term for any R-matrix with solvable $\alg{h}_\pm$.
In particular, we show that the Lax matrix also satisfies a Maillet bracket with twist function, with the proof given in \appref{sec:Maillet}.
Furthermore, this also allows us to interpret this model as a realisation of affine Gaudin model~\cite{Vicedo:2017cge}.

In \secref{sec:altform}, we show that the action \eqref{Eq:SolvSIntro} can also be obtained using various alternative formulations that have been proposed to study integrable $\sigma$-models.
These include $\mathcal{E}$-models, 4-dimensional Chern-Simons theory and, for homogeneous R-matrices, non-abelian T-duality.
For the first two, we investigate the structure of the underlying Drinfel'd double, observing that these deformations can equivalently be understood as governed by an asymmetric solution of the (m)cYBE.
We conclude in \secref{sec:comments} with a discussion of the results and possible extensions.

\section{General construction}
\label{sec:general}

The YB deformation of the PCM for simple Lie group $\grp{G}$ exists for any choice of R-matrix that solves the (m)cYBE on the Lie algebra $\alg{g} = \Lie(\grp{G})$ over $\Real$.
Our aim in this section is to investigate when it is possible to similarly deform the PCM plus WZ term. Motivated by this, we start from the general ansatz
\begin{equation}\label{ansatz}
\mathsf{S} = \int d^2x \; \kappa (g^{-1}\partial_+ g, \mathcal{O}_g g^{-1}\partial_- g) + \frac{\kay}{6} \int d^3 x \; \epsilon^{abc} \kappa(g^{-1}\partial_a g, [g^{-1}\partial_b g , g^{-1} \partial_c g]),
\end{equation}
where $g$ is a field valued in $\grp{G}$, the light-cone derivatives are defined as $\partial_\pm = \partial_t \pm \partial_x$ and
$\mathcal{O} : \g \to \g$ is a constant and invertible linear operator with $\mathcal{O}_g = \Ad_g^{-1} \mathcal{O} \Ad_g^{\vphantom{-1}}$.
$\kappa$ is proportional to the Killing form on $\alg{g}$:
\begin{equation}\label{kappa}
\kappa(X,Y) = - \frac{1}{2h^{\vee}} \tr[\ad_X \ad_Y] , \qquad \forall ~ X,Y \in \g,
\end{equation}
where $h^\vee$ is the dual Coxeter number.
The choice of sign gives a non-degenerate bilinear form on $\alg{g}$ that is positive-definite for compact $\grp{G}$.
Taking $\mathcal{O} = \frac{\kay}{2}$ the action \eqref{ansatz} becomes that of the WZW model at level $k = 4\pi\kay$.

\paragraph{Equations of motion and Lax connection.} The action \eqref{ansatz} is invariant under right multiplication, $g \mapsto g g_0$.
Its equations of motion are equivalent to the conservation equation
\begin{equation}\label{eom}
\partial_+ \cK_- + \partial_- \cK_+ = 0
\end{equation}
of the corresponding Noether current
\begin{equation}\label{currents}
\cK_+ = \frac{1}{\xi}\Big(\tra\mathcal{O}_g - \frac{\kay}{2}\Big) j_+, \qquad
\cK_- = \frac{1}{\xi}\Big(\mathcal{O}_g + \frac{\kay}{2}\Big) j_-,
\end{equation}
where $j_\pm = g^{-1}\partial_\pm g$ are the light-cone components of the Maurer-Cartan one-form pulled back to the two-dimensional worldsheet and we have introduced an overall constant $\xi$ parametrising the freedom in normalising $\cK_\pm$.
Note that $t$ denotes the transpose with respect to the Killing form, \textit{i.e.}
\begin{equation}
\kappa(X,\mathcal{O}Y) = \kappa(\tra \mathcal{O} X,Y), \qquad \forall ~ X,Y \in \g.
\end{equation}
When the conserved current $\cK_\pm$ is flat it immediately follows that there exists a Lax connection
\begin{equation}\label{Eq:Lax}
\mathcal{L}_\pm(z) = \frac{\cK_\pm}{1\mp z}.
\end{equation}
for the model \eqref{ansatz}.

Let us now investigate under which condition on the operator $\mathcal{O}$ this is the case. We first rewrite eq.~\eqref{currents} as
\begin{equation}\label{intq}
j_\pm = \mathcal{Q}^\pm_g \cK_\pm,
\end{equation}
where $Q^\pm_g = \Ad_g^{-1} \mathcal{Q}^\pm \Ad_g^{\vphantom{-1}}$ and
\begin{equation}\label{qpm}
\mathcal{Q}^- = \xi \Big(\mathcal{O} + \frac{\kay}{2}\Big)^{-1}, \qquad
\mathcal{Q}^+ = \xi \Big(\tra \mathcal{O} - \frac{\kay}{2}\Big)^{-1}.
\end{equation}
The flatness of the Maurer-Cartan one-form implies
\begin{equation}
\partial_+ j_- - \partial_- j_+ + [j_+,j_-] = 0,
\end{equation}
and upon substituting in for $\cK_\pm$ using eq.~\eqref{intq} we find
\begin{equation}
\mathcal{Q}_g^- (\partial_+ \cK_- + [\mathcal{Q}_g^+ \cK_+, \cK_-]) -
\mathcal{Q}_g^+ (\partial_- \cK_+ - [\cK_+, \mathcal{Q}_g^- \cK_-]) -
[\mathcal{Q}_g^+ \cK_+ ,\mathcal{Q}_g^-\cK_-] = 0.
\end{equation}
This can be rewritten as
\begin{equation}\label{final1}
\frac{\mathcal{Q}_g^+ + \mathcal{Q}_g^-}{2} F_{+-} (\cK) - \frac{\mathcal{Q}_g^+ - \mathcal{Q}_g^-}{2} (\partial_+\cK_- + \partial_- \cK_+) = \Ad_g^{-1} \mathcal{Z}(\Ad_g^{\vphantom{-1}} \cK_+, \Ad_g^{\vphantom{-1}}\cK_-),
\end{equation}
where we have defined the field strength for $\cK_\pm$
\begin{equation}
F_{+-}(\cK) = \partial_+ \cK_- - \partial_- \cK_+ + [\cK_+,\cK_-],
\end{equation}
and
\begin{equation}\label{zeq}
\mathcal{Z}(X,Y) = [\mathcal{Q}^+ X, \mathcal{Q}^- Y] - \mathcal{Q}^+ [X,\mathcal{Q}^- Y] - \mathcal{Q}^- [\mathcal{Q}^+ X,Y] + \frac{\mathcal{Q}^+ + \mathcal{Q}^-}{2}[X,Y].
\end{equation}
Assuming that $\mathcal{Q}^+ + \mathcal{Q}^-$ is invertible, which will generically be the case for the models we consider, eq.~\eqref{final1} tells us that, on the equations of motion \eqref{eom}, the flatness of the Maurer-Cartan one-form implies that the current $\cK_\pm$ is flat if and only if $\mathcal{Z}(\Ad_g^{\vphantom{-1}} \cK_+, \Ad_g^{\vphantom{-1}}\cK_-) = 0$.
This gives us a condition on the operators $\mathcal{Q}^\pm$, and thus on the operator $\mathcal{O}$, that ensures the existence of a Lax connection for the model \eqref{ansatz}:
\begin{equation}\label{Eq:IntCond}
\mathcal{Z}(X,Y) = 0, \qquad \forall ~ X,Y \in \alg{g}.
\end{equation}

\subsection{Perturbation theory around \texorpdfstring{$\kay=0$}{k=0}}\label{subsec:pertk}

Considering a setup in which we recover the YB deformation of the PCM~\cite{Klimcik:2002zj,Klimcik:2008eq} when the coefficient of the WZ term is set to zero, we now use perturbation theory to investigate when it is possible to construct the corresponding deformation of the PCM plus WZ term.
In particular, we assume that the Lax connection remains of the form \eqref{Eq:Lax} with $K_\pm$ defined in \eqref{currents}, and hence that the integrability of the model is determined by the condition~\eqref{Eq:IntCond}.

\paragraph{The YB deformation of the PCM.} We start by reviewing the case without WZ term, \textit{i.e.} when $\kay = 0$.
The action of the YB deformed PCM was first given in~\cite{Klimcik:2002zj} in terms of a skew-symmetric R-matrix, that is a constant linear operator $R:\alg{g}\to \alg{g}$ that satisfies $\trap R = -R$ and solves the (m)cYBE
\begin{equation}\label{cybe}
[RX,RY] - R[X,Y]_{\R} + c^2 [X,Y] = 0, \qquad \forall ~ X,Y\in \g,
\end{equation}
where we have introduced the R-bracket
\begin{equation}\label{rbracket}
[X,Y]_{\R} = [X,RY] + [RX,Y], \qquad X,Y \in \g.
\end{equation}
Since the R-matrix preserves the real Lie algebra $\g$ the parameter $c$ is such that $c^2 \in \Real$.
Therefore, there are three cases of interest:
\begin{itemize}
\item $c\in \Real^*$: $R$ is a split solution of the mcYBE;
\item $i c \in \Real^*$: $R$ is a non-split solution of the mcYBE;
\item $c=0$: $R$ is a solution of the cYBE.
\end{itemize}
Rescaling $R$, we can choose $c=1$, $c=i$ and $c=0$ as representatives of each case without loss of generality. We call the $c\neq 0$ case inhomogeneous and the $c=0$ case homogeneous. Unless otherwise stated, we will treat both the inhomogeneous and homogeneous cases simultaneously.

The YB deformation of the PCM is given by the action \eqref{ansatz} with $\kay = 0$ and
\begin{equation}\label{Eq:OYB}
\mathcal{O} = \frac{\hay}{1 - \eta R},
\end{equation}
where $\eta \in \Real$ controls the strength of the deformation and $\hay$ is the PCM coupling for $\eta = 0$.
Thus it follows from eq.~\eqref{qpm} that for $\kay = 0$
\begin{equation}
\mathcal{Q}^\pm = \frac{\xi}{\hay} (1 \pm \eta R).
\end{equation}
Substituting into eq.~\eqref{zeq} and using the (m)cYBE \eqref{cybe} we find that requiring $\mathcal{Z}(X,Y)$ to vanish implies that
\begin{equation}
\Big(1-c^2 \eta^2 - \frac{\hay}{\xi} \Big) [X,Y] = 0,
\end{equation}
and hence we are able to fix the constant $\xi$ as
\begin{equation}\label{Eq:XiYB}
\xi = \frac{\hay}{1-c^2\eta^2},
\end{equation}
to ensure the existence of a Lax connection.

\paragraph{First order in $\kay/\hay$.} Since the level $k = 4\pi\kay$ is integer-valued for certain Lie groups $\grp{G}$ we use the expansion parameter $\kay/\hay$, parametrising the leading corrections to $\mathcal{O}$ and $\xi$ as
\begin{equation}\label{oex}
\mathcal{O}^{-1} = \frac{1}{\hay}\Big((1 - \eta R) + \frac{\kay}{\hay} \widehat{\mathcal{O}} + \Order\Big(\frac{\kay^2}{\hay^2}\Big)\Big), \qquad \xi = \hay\Big(\frac{1}{1-c^2\eta^2} + \Order\Big(\frac{\kay^2}{\hay^2}\Big)\Big),
\end{equation}
where $\widehat{\mathcal{O}} : \g \to \g$ is again a constant linear operator.
Note that any $\Order(\kay/\hay)$ term in $\xi$ can be absorbed into a redefinition of $\widehat{\mathcal{O}}$ and hence without loss of generality we assume there are no such corrections.
The resulting expansions of the operators $\mathcal{Q}^\pm$ \eqref{qpm} are
\begin{equation}\begin{split}\label{qex}
\mathcal{Q}^- & = \frac{1}{1-c^2\eta^2} \Big((1 - \eta R) + \frac{\kay}{\hay}(\widehat{\mathcal{O}} - \frac{1}{2}(1 - \eta R)^2) \Big) + \Order\Big(\frac{\kay^2}{\hay^2}\Big),\qquad
\\
\mathcal{Q}^+ & = \frac{1}{1-c^2\eta^2} \Big((1 + \eta R) + \frac{\kay}{\hay}(\tra \widehat{\mathcal{O}} + \frac{1}{2}(1 + \eta R)^2) \Big) + \Order\Big(\frac{\kay^2}{\hay^2}\Big).
\end{split}\end{equation}
Again substituting into eq.~\eqref{zeq} and using the (m)cYBE \eqref{cybe} we find that requiring $\mathcal{Z}(X,Y)$ to vanish at order $\Order(\kay/\hay)$ implies that
\begin{equation}\begin{split}\label{orderkay}
& \eta\Big([RX,\widehat{\mathcal{O}}Y]- R[X,\widehat{\mathcal{O}}Y] -\widehat{\mathcal{O}}[RX,Y]
- [\tra \widehat{\mathcal{O}}X,RY] + \tra \widehat{\mathcal{O}}[X,RY]+ R[X,\tra \widehat{\mathcal{O}}Y] \Big) \\ & \hspace{30ex} = \frac{\eta^3}{2}R_+R_-[X,Y]_{\R} + \frac{1+c^2\eta^2}{2}(\widehat{\mathcal{O}} + \tra \widehat{\mathcal{O}}) [X,Y],
\end{split}\end{equation}
where
\begin{equation}
R_\pm = R \pm c ,
\end{equation}
in terms of which the (m)cYBE~\eqref{cybe} takes the simple form
\begin{equation}\label{cybe2}
[R_\pm X,R_\pm Y] = R_\pm[X,Y]_R, \qquad \forall ~ X,Y \in \alg{g}.
\end{equation}

At this point it is useful to split $\widehat{\mathcal{O}}$ into its symmetric and skew-symmetric parts: $\widehat{\mathcal{O}} = \tfrac12(\mathcal{B} + \mathcal{A})$ ($\tra \mathcal{B} = \mathcal{B}$, $\tra \mathcal{A} = -\mathcal{A}$), and consider the sum and difference of eq.~\eqref{orderkay} with itself with $X$ and $Y$ interchanged:
\begin{subequations}\label{orderkaysplit}
\begin{align}
& \eta\Big([RX,\mathcal{B}Y]- R[X,\mathcal{B}Y] - \mathcal{B}[RX,Y]
- [\mathcal{B}X,RY] + \mathcal{B}[X,RY]+ R[\mathcal{B}X,Y] \Big) = 0, \label{orderkaysplita}
\\
\begin{split} \label{orderkaysplitb}
& \eta\Big([RX,\mathcal{A}Y]- R[X,\mathcal{A}Y] - \mathcal{A}[RX,Y]
+ [\mathcal{A}X,RY] - \mathcal{A}[X,RY]- R[\mathcal{A}X,Y] \Big) \\ & \hspace{40ex} = \frac{\eta^3}{2} R_+R_-[X,Y]_{\R} + \frac{1+c^2\eta^2}{2}\mathcal{B} [X,Y].
\end{split}
\end{align}
\end{subequations}
This set of $(\dim \grp{G})^3$ linear equations for $(\dim \grp{G})^2$ free coefficients in $\widehat{\mathcal{O}}$ is overdetermined.
As we will see, there are certain R-matrices for which there is a solution, and others where none exists.
Most of the interesting structure is contained in the second of these equations; however, before we proceed to analyse this equation let us note that the first has at least one simple solution, which is to take $\mathcal{B}$ proportional to the identity.

\paragraph{Lie algebra cohomology of \texorpdfstring{$\alg{g}_{\R}$}{gR}.} When an R-matrix solves the (m)cYBE, it is a standard result that the R-bracket \eqref{rbracket} satisfies the Jacobi identity and hence defines a Lie algebra, denoted $\alg{g}_{\R}$.
We can therefore introduce the associated Lie algebra cohomology for the trivial representation of $\alg{g}_{\R}$.
In particular, $n$-cochains are alternating linear functions $\psi: \bigwedge^n \alg{g}_{\R} \to \Real$ whose differential is the $(n+1)$-cochain given by
\begin{equation}\begin{split}
& (d_{\R}\,\psi)(X_1,\dots,X_{n+1}) \\ & \hspace{10ex} = \sum_{i<j}(-1)^{i+j-1} \psi([X_i,X_j]_{\R},X_1,\dots, X_{i-1},X_{i+1},\dots, X_{j-1},X_{j+1},\dots, X_{n+1}),
\end{split}\end{equation}
such that $d_{\R}^2 = 0$.

Contracting \eqref{orderkaysplitb} with $Z\in \g$ using the normalised Killing form~\eqref{kappa} we can equivalently rewrite it as
\begin{equation}\label{orderkaydiff}
(d_{\R}\,\psi_\mathcal{A})(X,Y,Z) = - \frac{\eta^2}{2} \Omega_R(X,Y,Z) + \frac{1+c^2\eta^2}{2\eta} \kappa(\mathcal{B}[X,Y],Z),
\end{equation}
where
\begin{equation}\label{2cochain}
\psi_\mathcal{A}(X,Y) = \kappa(X,\mathcal{A}Y)
\end{equation}
is a 2-cochain and
\begin{equation}\label{Eq:OmegaR}
\Omega_R (X,Y,Z) = -\kappa\bigl(R_+R_-[X,Y]_R,Z \bigr) = \kappa\bigl([R_\pm X,R_\pm Y],R_\pm Z\bigr)
\end{equation}
is a 3-cochain. To obtain the second equality in eq.~\eqref{Eq:OmegaR} we use $\trap R_\pm=-R_\mp$ and the (m)cYBE written in the form \eqref{cybe2}.

Since $d_R\,\psi_\mathcal{A}$ and $\Omega_R$ are both 3-cochains, the consistency of \eqref{orderkaydiff} requires that $\kappa(\mathcal{B}[X,Y],Z)$ is also a 3-cochain.
In particular, it should be the case that $\kappa(\mathcal{B}[X,Y],Z) = \kappa(\mathcal{B}[Y,Z],X)$, or, equivalently, $\mathcal{B}[X,Y] = [\mathcal{B}X,Y]$.
The latter is the statement that $\mathcal{B}$ is an intertwining map between the adjoint representation of $\alg{g}$ and itself.
Therefore, given that $\alg{g}$ is assumed to be simple, Schur's lemma immediately tells us that $\mathcal{B}$ is proportional to the identity, which also solves eq.~\eqref{orderkaysplita}.

Defining
\begin{equation}
\Omega(X,Y,Z) = \kappa([X,Y],Z).
\end{equation}
we arrive at the following equation for the skew-symmetric operator $\mathcal{A}$
\begin{equation}\label{diffeq}
d_R\,\psi_\mathcal{A} = - \frac{\eta^2}{2} (\Omega_R + \alpha \Omega) ,
\end{equation}
where $\alpha$ is a free parameter.
One can straightforwardly check that $\Omega_R$ and $\Omega$ are $d_R$-closed by construction: $d_R\,\Omega_R = d_R\,\Omega = 0$.
Consequently, a solution to eq.~\eqref{diffeq} only exists if $\Omega_R + \alpha \Omega$ is $d_R$-exact for some $\alpha$.
\unskip\footnote{Note that when $R$ solves the mcYBE ($c \neq 0$) we have $d_R\psi_R = -2c^2 \Omega$ and hence $\Omega$ is $d_R$-exact.
The condition that $\Omega_R + \alpha \Omega$ is $d_R$-exact is then equivalent to $\Omega_R$ being $d_R$-exact.
On the other hand, the situation is more involved for solutions to the cYBE.
In particular, there are examples for which $\Omega$ is $d_R$-exact and examples where it is not.} As we will see shortly, there are R-matrices for which this is not the case, and hence, under the assumption that the Lax connection remains of the form \eqref{Eq:Lax}, we cannot construct the YB deformation of the PCM plus WZ term perturbatively in $\kay/\hay$.

\medskip

We will investigate these cohomology questions by studying various examples of R-matrices in the rest of this section. Before that, let us point out that there exists a natural class of R-matrices for which the condition \eqref{diffeq} admits a solution. These are those that satisfy
\begin{equation}\label{omr}
\Omega_R = 0.
\end{equation}
Indeed, in this case, $\Omega_R + \alpha \Omega$ is trivially $d_R$-exact if we set $\alpha = 0$. Recall that, since the (m)cYBE \eqref{cybe} can be written in the form~\eqref{cybe2}, $\alg{h}_{\pm} = \im R_\pm$ form subalgebras of $\alg{g}$, while $\alg{p}^\star_\pm = \ker R_\pm$ are ideals of $\alg{g}_R$.
\unskip\footnote{This implies that $\alg{g}_R$ is isomorphic as a Lie algebra to an extension of $\alg{h}_\pm$ by $\alg{p}^\star_\pm$, \textit{i.e.} $\alg{h}_\pm \cong \alg{g}_R/\alg{p}_\pm^\star$.}
The condition $\Omega_R = 0$ is equivalent to the subalgebras $\alg{h}_\pm$ being solvable, by the Cartan criterion for solvability.
Note that for $c=0$, \textit{i.e.} $R$ satisfies the cYBE, we have $R_\pm =R$.
When specifically discussing this case we denote the image and kernel as $\alg{h}$ and $\alg{p}^\star$.
In \secref{sec:solvable}, we will demonstrate that the form of the action for the YB deformation of the PCM plus WZ term proposed in~\cite{Klimcik:2019kkf,Klimcik:2020fhs} for standard Drinfel'd-Jimbo R-matrices is integrable for all R-matrices with solvable $\alg{h}_\pm$, thus promoting the $\Order(\kay/\hay)$ results of this section to all orders.
The Hamiltonian analysis of this model will be carried out in \secref{sec:hamiltonian}.

\subsection{Examples of \texorpdfstring{$\alg{sl}(2)$}{sl(2)} and \texorpdfstring{$\alg{sl}(3)$}{sl(3)}}

To understand better when it is possible to construct the YB deformation of the PCM plus WZ term let us consider two examples: $\alg{sl}(2)$ and $\alg{sl}(3)$.
We will work with the complexified Lie algebras, \textit{i.e.} we drop the distinction between split and non-split solutions of the mcYBE, with the following defining relations:
\begin{equation}\begin{gathered}\hspace{0pt}
[h_i,e_j] = a_{ij} e_j, \qquad [h_i,f_j] = - a_{ij} f_j, \qquad [e_i,f_j] = \delta_{ij} h_i.
\\ \ad_{e_i}^{1-a_{ij}} e_j = 0, \qquad \ad_{f_i}^{1-a_{ij}} f_j = 0,
\\ e_{i_1\dots i_n} = \ad_{e_{i_1}} \ad_{e_{i_2}} \dots e_{i_n}, \qquad f_{i_1\dots i_n} = \ad_{f_{i_1}} \ad_{f_{i_2}} \dots f_{i_n},
\end{gathered}\end{equation}
where $i,j,\ldots= 1,\dots,\rank \alg{g}$, $a_{ij}$ is the Cartan matrix, $h_i$ are the Cartan generators and $e_i$ and $f_i$ are the positive and negative simple roots generators.

The skew-symmetric R-matrices below are written in terms of their kernel $r \in \alg{g} \wedge \alg{g}$, defined through
\begin{equation}
RX = \kappa\ti{2}(r,(1\otimes X)),
\end{equation}
where we use the normalised Killing form \eqref{kappa} in the second entry of the tensor product.
We work up to automorphisms, with $\beta$, $\beta_i$ and $\beta_{ij}$ denoting parameters that cannot be eliminated via such transformations.
Furthermore, for solutions of the mcYBE we fix the normalisation of the R-matrix so that it solves the mcYBE \eqref{cybe} with $c=1$.

For $\alg{sl}(2)$ we have Cartan matrix $a=2$.
Skew-symmetric solutions of the (m)cYBE for $\alg{sl}(2)$ are straightforward to classify~\cite{Belavin:1982,Belavin:1984,Stolin:1991a}, and up to $\alg{sl}(2)$ automorphisms there are two.
The jordanian solution of the cYBE is $r = h_1 \wedge e_1$.
The corresponding subalgebra $\alg{h} = \operatorname{span}\{h_1,e_1\}$ is solvable and Frobenius.
Similarly, for the Drinfel'd-Jimbo solution to the mcYBE, $r = e_1 \wedge f_1$, the subalgebras $\alg{h}_+ = \operatorname{span}\{h_1,f_1\}$ and $\alg{h}_- = \operatorname{span}\{h_1,e_1\}$ are also solvable.
Therefore, for both solutions we have $\Omega_R = 0$ and hence eq.~\eqref{diffeq} can be solved straightforwardly.
\unskip\footnote{The simplest solution is $\mathcal{A} = \alpha = 0$; however, for both the jordanian and Drinfel'd-Jimbo solutions $\Omega$ is $d_R$-exact and hence a solution exists for any $\alpha$.}

For $\alg{sl}(3)$ the Cartan matrix is
\begin{equation}
a = \begin{pmatrix} 2 & -1 \\ -1 & 2 \end{pmatrix}.
\end{equation}
From the classification of skew-symmetric solutions to the cYBE in~\cite{Stolin:1991a} we see that the subalgebra $\alg{h}$ is solvable for all such R-matrices with the exception of the rank-6 R-matrix
\begin{equation}\label{eq:r5}
r= (h_1+2h_2) \wedge e_{12} + 3 e_1 \wedge e_2 + (h_1-h_2)\wedge f_1,
\end{equation}
for which $\alg{h} = \operatorname{span}\{h_1,h_2,e_1,e_2,e_{12},f_1\}$ is parabolic and Frobenius, but not solvable, \textit{e.g.} we have $\kappa([e_1,f_1],h_1) \neq 0$.
For this R-matrix
\begin{equation}\begin{gathered}
\Omega_R(h_1,f_2,f_{21}+2e_1) + \alpha \Omega(h_1,f_2,f_{21}+2e_1) \neq 0,
\end{gathered}\end{equation}
for any $\alpha$, while for a general 2-cochain \eqref{2cochain} one can check that
\begin{equation}
d_R\psi_\mathcal{A}(h_1,f_2,f_{21}+2 e_1) = 0.
\end{equation}
Therefore, there is no choice of $\alpha$ such that $\Omega_R + \alpha \Omega$ is $d_R$-exact and we cannot construct the YB deformation of the PCM plus WZ term perturbatively in $\kay/\hay$ for the R-matrix \eqref{eq:r5}.

For the R-matrix \eqref{eq:r5} one can check that $\Omega$ is $d_R$-exact and hence $\Omega_R$ is not.
As a curiosity, let us also note that for all the rank-2 solutions of the cYBE for $\alg{sl}(3)$
\begin{equation}\begin{aligned}
& \text{quasi-Frobenius:} &\ \ & r = \beta h_1 \wedge h_2 , \quad r = (h_1+2h_2) \wedge e_1 , \quad r = e_1 \wedge e_{12} , \quad r = (e_1+e_2) \wedge e_{12},
\\
& \text{Frobenius:} &\ \ & r = (h_2 + \beta (h_1+2h_2)) \wedge e_1 , \qquad r=(h_2 + 2h_1+3e_1)\wedge e_{12} ,
\\ & && r = (h_1+h_2) \wedge (e_1+e_2) , \hspace{36pt} r= (h_2+2h_1)\wedge(e_1+e_{12}) ,
\end{aligned}\end{equation}
$\Omega$ is not $d_R$-exact.
For those rank-4 R-matrices with Frobenius $\alg{h}$
\begin{equation}
r = (\beta_1 h_1 - \beta_2 h_2)\wedge e_{12} + (\beta_1 - \beta_2) e_1 \wedge e_2 , \qquad \beta_1 > \beta_2 , \quad \beta_1\beta_2 \neq 0, \quad \beta_1 \neq 2 \beta_2, \quad \beta_2 \neq 2 \beta_1 ,
\end{equation}
$\Omega$ is $d_R$-exact, but when $\alg{h}$ is quasi-Frobenius
\begin{equation}\begin{aligned}
&\begin{aligned}r &= h_2\wedge e_2 + e_1\wedge e_{12} , \qquad & r &= h_2 \wedge e_{12} + e_1 \wedge e_2 ,
\\
r &= (h_1 +2 h_2) \wedge e_{12} + 3 (e_1 + e_{12})\wedge e_2 , \qquad & r &= (h_1 +2 h_2) \wedge e_{12} + 3 e_1\wedge e_2 ,
\end{aligned}
\\
& r = (h_1 - h_2) \wedge e_1 + (h_1 + 2h_2)\wedge e_{12} + \beta e_1 \wedge e_{12},
\end{aligned}\end{equation}
this is no longer the case.
Nevertheless, for all rank-2 and rank-4 R-matrices $\alg{h}$ is solvable and hence eq.~\eqref{diffeq} is solved with $\mathcal{A} = \alpha = 0$, with a solution existing for any $\alpha$ when $\Omega$ is $d_R$-exact.

Let us note as a curiosity that for the cases above for which $\Omega$ is $d_R$-exact it turns out that we can write $\Omega = d_R \psi_{R'}$ where $R'$ is related to $R$ by an $\alg{sl}(3)$ automorphism.
To be precise, we have the pairings
\begin{equation}\begin{aligned}
r &= (\beta_1 h_1 - \beta_2 h_2)\wedge e_{12} + (\beta_1 - \beta_2) e_1 \wedge e_2 , \\
r' &= \frac{1}{(\beta_1 - \beta_2)^2} \big((\beta_1 h_2 -\beta_2 h_1) \wedge f_{21} + (\beta_1 - \beta_2) f_1 \wedge f_2\big) ,
\end{aligned}
\end{equation}
and
\begin{equation}\begin{aligned}
r&= (h_1+2h_2) \wedge e_{12} + 3 e_1 \wedge e_2 + (h_1-h_2)\wedge f_1 , \\
r'&= \frac{1}{9}\big( (h_1-h_2) \wedge e_{1} + 3 e_{12} \wedge f_2 + (h_1+2h_2)\wedge f_{21}\big).
\end{aligned}\end{equation}
The same is also true for the jordanian $\alg{sl}(2)$ R-matrix with
\begin{equation}
r = h_1 \wedge e_1 , \qquad r' = \frac{1}{4} h_1\wedge f_1 .
\end{equation}
Given that $R$ and $R'$ are both solutions of the cYBE, it follows from $\Omega = d_R \psi_{R'}$ that $R-c^2R'$ solves the mcYBE \eqref{cybe}.

There are two skew-symmetric solutions of the mcYBE for $\alg{sl}(3)$~\cite{Belavin:1982,Belavin:1984}.
The first is the Drinfel'd-Jimbo solution
\begin{equation}\label{eq:dj3}
r = e_1\wedge f_1 + e_2 \wedge f_2 + e_{12}\wedge f_{21} + \beta h_1 \wedge h_2,
\end{equation}
for which $\alg{h}_+ = \operatorname{span}\{h_1,h_2,f_1,f_2,f_{21}\}$ and $\alg{h}_- = \operatorname{span}\{h_1,h_2,e_1,e_2,e_{12}\}$ when $\beta^2 \neq -\frac{1}{3}$.
At these special points $R^2 = 1$ and the subalgebras $\alg{h}_+ = \operatorname{span}\{h_1+\beta(h_1+2h_2),f_1,f_2,f_{21}\}$ and $\alg{h}_- = \operatorname{span}\{h_2 + \beta(h_2+2h_1),e_1,e_2,e_{12}\}$ correspond to the eigenspaces of $R$ with eigenvalues $+1$ and $-1$ respectively.
The subalgebras $\alg{h}_\pm$ are solvable for all $\beta$.
Therefore, $\Omega_R = 0$ and eq.~\eqref{diffeq} admits a solution.

The second solution of the mcYBE is
\begin{equation}\label{eq:bd}
r = e_1 \wedge f_1 + e_2 \wedge f_2 + e_{12} \wedge f_{21} + \frac{1}{3} h_1 \wedge h_2 + e_1 \wedge f_2,
\end{equation}
for which $\alg{h}_+ = \operatorname{span}\{h_1,h_2,e_1,f_1,f_2,f_{21}\}$ and $\alg{h}_- = \operatorname{span}\{h_1,h_2,e_1,e_2,e_{12},f_2\}$.
These are not solvable since both $\kappa([e_1,f_1],h_1)$ and $\kappa([e_2,f_2],h_2)$ are non-vanishing.
For this R-matrix
\begin{equation}
\Omega_R(h_1-h_2,e_1-2e_2,f_2-2f_1) + \alpha \Omega(h_1-h_2,e_1-2e_2,f_2-2f_1) \neq 0,
\end{equation}
for any $\alpha$, while for a general 2-cochain \eqref{2cochain} one can check that
\begin{equation}
d_R\psi_\mathcal{A}(h_1-h_2,e_1-2e_2,f_2-2f_1) = 0.
\end{equation}
Therefore, also in this case, there is no choice of $\alpha$ such that $\Omega_R + \alpha \Omega$ is $d_R$-exact and we cannot construct the YB deformation of the PCM plus WZ term perturbatively in $\kay/\hay$ for the R-matrix \eqref{eq:bd}.

For completeness, we recall that for any solution of the mcYBE we have that $\Omega$ is $d_R$-exact with $\Omega = -\frac{1}{2c^2}d_R \psi_R$.
Therefore, for the Drinfel'd-Jimbo R-matrix \eqref{eq:dj3} a solution to \eqref{diffeq} exists for any $\alpha$, while for the second R-matrix \eqref{eq:bd} it follows that $\Omega_R$ is not $d_R$-exact.

\subsection{Examples for general simple algebras}

We conclude with some general comments on solutions of the (m)cYBE for simple Lie algebras and the condition \eqref{diffeq}.
First, it is well-known that any R-matrix whose image is an abelian subalgebra of $\alg{g}$ solves the cYBE.
Since $\alg{h}$ is abelian, it trivially follows that it is solvable and eq.~\eqref{diffeq} can be solved.
Another important class of solutions of the cYBE are those of extended jordanian type~\cite{Ogievetsky:1992ph,Kulish:1998be,Kulish:1999ua,Tolstoy}
\begin{equation}\begin{gathered}
r = h_{\theta_0} \wedge e_{\theta_0} + \sum_{l=1}^N e_{\theta_l} \wedge e_{\theta_{-l}} .
\end{gathered}\end{equation}
Here $h_{\theta_{0}}$ is an element of the Cartan subalgebra and $\alg{h} = \operatorname{span}\{h_{\theta_0},e_{\theta_0},e_{\theta_{\pm l}}\}$ has non-vanishing commutation relations
\begin{equation}\begin{gathered}\hspace{0pt}
[h_{\theta_0}, e_{\theta_0}] = e_{\theta_0}, \qquad
[h_{\theta_0}, e_{\theta_l}] = (1-t_{\theta_l}) e_{\theta_l}, \qquad
[h_{\theta_0}, e_{\theta_{-l}}] = t_{\theta_l} e_{\theta_{-l}}, \qquad
[e_{\theta_l}, e_{\theta_{-l}}] = e_{\theta_0},
\end{gathered}\end{equation}
where $t_{\theta_l}$ are complex numbers.
Given that its derived series terminates, $\alg{h}$ is solvable and contained within a Borel subalgebra of $\alg{g}$, and hence, for these R-matrices, eq.~\eqref{diffeq} admits a solution.
Furthermore, for any solution of the cYBE such that $\alg{h}$ is contained within a Borel subalgebra, we have that $\alg{h}$ is solvable and eq.~\eqref{diffeq} can be solved.

\medskip

Turning now to the mcYBE, for every simple Lie algebra $\alg{g}$ there is a Drinfel'd-Jimbo solution~\cite{Drinfeld:1985rx,Jimbo:1985zk}
\begin{equation}\label{rmatdj}
r = \sum_m \hat{e}_m \wedge \hat{f}_m + \sum_{i,j} \beta_{ij} h_i \wedge h_j,
\end{equation}
where $\hat{e}_m$ and $\hat{f}_m$ are the positive and negative roots generators of $\alg{g}$, normalised such that $\kappa(\hat{e}_m,\hat{f}_n) = -1$, and $h_i$ are elements of the Cartan subalgebra of $\alg{g}$.
When $\beta_{ij} = 0$ we refer to the solution \eqref{rmatdj} as the standard Drinfel'd-Jimbo R-matrix.
The corresponding YB deformation of the PCM plus WZ term can then be constructed using the property $R^3 = R$~\cite{Delduc:2014uaa}.
\unskip\footnote{Note that in~\cite{Delduc:2014uaa} the compact real form was considered.
In order to preserve the real form the R-matrix \eqref{rmatdj} should be multiplied by $i$.
Therefore, in this case the analogous property is $R^3 = -R$.}
This is consistent with the perturbative analysis since the subalgebras $\alg{h}_+ = \operatorname{span}\{h_i,f_m\}$ and $\alg{h}_- = \operatorname{span}\{h_i,e_m\}$ are both solvable.
For $\beta_{ij} \neq 0$, \textit{i.e.} including a Reshetikhin twist~\cite{Reshetikhin:1990ep}, the subalgebras $\alg{h}_\pm$ are unchanged, except at certain special points where particular Cartan elements may no longer be in $\im R_\pm$.
Nevertheless, $\alg{h}_\pm$ remain solvable for all $\beta_{ij}$.
This agrees with the expectation that these parameters can equivalently be introduced through TsT transformations in directions associated to the Cartan elements~\cite{Matsumoto:2014nra,Osten:2016dvf}, which are symmetries of \eqref{rmatdj}, \textit{i.e.} $(1\otimes \ad_{h_i} + \ad_{h_i} \otimes 1) r = 0$ or $[\ad_{h_i},R] = 0$.
For other solutions of the mcYBE~\cite{Belavin:1982,Belavin:1984}, of which \eqref{eq:bd} is an example, $\alg{h}_\pm$ will typically not be solvable.

\medskip

More generally, we can consider a setup in which we have two R-matrices, one of which is subordinate to the other.
Let $R_1$ and $R_2$ be solutions of the (m)cYBE and cYBE respectively such that $R_2$ is subordinate to $R_1$, \textit{i.e.} $[\ad_{\im R_2},R_1] = 0$.
It is then a standard result that $R = R_1 + \beta R_2$ is a solution of the (m)cYBE~\cite{Belavin:1982,Belavin:1984}.
If $\alg{h}_{1\pm} = \im R_{1\pm}$ and $\alg{h}_2 = \im R_2$ are solvable it follows that $\alg{h}_\pm = \im R_\pm \subset \alg{h}_{1\pm} + \alg{h}_{2}$ is also solvable.
To see this we observe that the subordinate property can be written as $[R_{1\pm}X,R_2Y] = R_{1\pm}[X,R_2Y]$, $X,Y\in\alg{g}$, which implies that $\alg{h}_{1\pm} + \alg{h}_{2}$ is an algebra with commutation relations
\begin{equation}\hspace{0pt}
[\alg{h}_{1\pm},\alg{h}_{1\pm}]\subset \alg{h}_{1\pm}, \qquad
[\alg{h}_{2},\alg{h}_{2}]\subset \alg{h}_{2}, \qquad
[\alg{h}_{1\pm},\alg{h}_{2}]\subset \alg{h}_{1\pm}.
\end{equation}
It then follows that both $\alg{h}_{1\pm} + \alg{h}_{2}$ and its subalgebra $\alg{h}_\pm$ are solvable.
Indeed, by the solvability of $\alg{h}_2$, at some point in the derived series of $\alg{h}_{1\pm} + \alg{h}_{2}$ we will find a subalgebra of $\alg{h}_{1\pm}$, and hence the derived series will terminate by the solvability of $\alg{h}_{1\pm}$.
This construction covers both the Reshetikhin twist of the standard Drinfel'd-Jimbo R-matrix \eqref{rmatdj} and the almost abelian R-matrices of~\cite{Borsato:2016ose,vanTongeren:2016eeb}.
Finally, let us note that almost abelian R-matrices are examples of a larger class of homogeneous R-matrices that have solvable $\alg{h}$ known as unimodular R-matrices, \textit{i.e.} such that $\tr (R\ad_X) = 0$ for all $X\in\g$~\cite{Chu,Lichnerowicz}.
\unskip\footnote{We thank L.~Wulff for pointing this result out to us.}

\section{Solvable \texorpdfstring{$\alg{h}_\pm$}{h±}: action and Lax connection}
\label{sec:solvable}

Let us now focus on YB deformations whose underlying R-matrix is such that $\alg{h}_\pm$ is solvable, \textit{i.e.} $\Omega_R$ vanishes.
In this case the results of the previous section show that one can add a WZ term to the deformed action while preserving its integrability at first order in the expansion parameter $\kay/\hay$.
In this section we aim to show that this is also true for all values of $\kay$.
Inspired by the results of Klim\v{c}\'{i}k~\cite{Klimcik:2019kkf,Klimcik:2020fhs}, we will explicitly construct an operator $\mathcal{O}$ depending on $R$ and $\kay$ that satisfies the integrability condition \eqref{Eq:IntCond} and reduces to the operator \eqref{Eq:OYB} when $\kay=0$.
Let us emphasise that, unless otherwise stated, we treat both $c\neq0$ and $c=0$ simultaneously, using l'H\^opital's rule to evaluate expressions at $c=0$ when necessary.

\subsection{Algebraic consequences of the solvability of \texorpdfstring{$\alg{h}_\pm$}{h±}.}

Let us first study various algebraic consequences of the solvability of $\alg{h}_\pm$.

\paragraph{Properties of the R-matrix.}
The solvability of $\alg{h}_\pm$ is equivalent to $\Omega_R$ vanishing.
From the definition of $\Omega_R$ \eqref{Eq:OmegaR} and the non-degeneracy of the bilinear form $\kappa$ it follows that $\alg{h}_\pm$ is solvable if and only if
\begin{equation}\label{Eq:IdentityR1}
R_+R_- [X,Y]_R = 0, \qquad \forall ~ X,Y \in \g.
\end{equation}
Furthermore, using the ad-invariance of $\kappa$ and the skew-symmetry of $R$ we can rewrite eq.~\eqref{Eq:OmegaR} as
\begin{equation}
\Omega_R(X,Y,Z) = -
\kappa\big( X, [RY,R_+R_-Z] - R[Y,R_+R_-Z] \big), \qquad \forall ~ X,Y,Z \in \g.
\end{equation}
Therefore, the solvability of $\alg{h}_\pm$ is also equivalent to
\begin{equation}\label{Eq:IdentityR2}
[RX,R_+R_-Y] = R[X,R_+R_-Y], \qquad \forall ~ X,Y \in \g.
\end{equation}
Applying the (m)cYBE twice for a general R-matrix one can show that
\begin{equation}
[R_+R_-X,R_+R_-Y] = R_+R_- \big( [RX,Y]_R + [X,RY]_R \big), \qquad \forall ~ X,Y \in \g,
\end{equation}
and hence $\im R_+R_-$ is a subalgebra of $\g$.
The identity \eqref{Eq:IdentityR2} can then be reinterpreted as the statement that the subalgebra $\im R_+R_-$ is a symmetry of the R-matrix when $\alg{h}_\pm$ is solvable.
Moreover, in this case, eq.~\eqref{Eq:IdentityR1} implies that $\im R_+R_-$ is abelian
\begin{equation}\label{Eq:IdentityR3}
[R_+R_-X,R_+R_-Y] = 0, \qquad \forall ~ X,Y \in \g.
\end{equation}

\paragraph{The integrability identity for $\Er$.}
Let us consider the operators $\Er$ and $\Erpm = e^{\rho (R\pm c)}$, where $R$ is a skew-symmetric R-matrix with solvable $\alg{h}_\pm$ and $\rho$ is a real parameter.
From eqs.~\eqref{Eq:IdentityR1}, \eqref{Eq:IdentityR2} and \eqref{Eq:IdentityR3}, one can prove that the operator $e^{\rho R}$ satisfies the identity
\begin{equation}\begin{split}\label{Eq:IdExp}
\big[\Er X, \Er Y \big] - \Er \big[\Er X, Y\big] & - \Er \big[X , \Er Y\big] \\ & - [X,Y] + (\Erp+\Erm)[X,Y]=0, \qquad \forall ~ X,Y \in \g.
\end{split}\end{equation}
This identity will be the most important tool in the remainder of this section.
As its proof is somewhat technical we present it in \appref{sec:identity}.
Let us note that the same identity has been used in~\cite{Klimcik:2020fhs} for $R$ equal to the standard Drinfel'd-Jimbo R-matrix on a compact Lie algebra, \textit{i.e} assuming $R^3 = -R$.
As explained in \secref{sec:general}, in this case $\alg{h}_{\pm}$ is indeed solvable.

It will also be useful to introduce the operator
\begin{equation}\label{eq:defrhat}
\hat{R} = \frac{c}{\sinh c\rho} (e^{\rho R} - \cosh c\rho) .
\end{equation}
It follows from \eqref{Eq:IdExp} that if $R$ is a skew-symmetric R-matrix with solvable $\alg{h}_\pm$, then $\hat{R}$ is also solution of the (m)cYBE \eqref{cybe}, although it is typically not skew-symmetric.
Rather, it satisfies the symmetry property
\begin{equation}\label{eq:transposer}
\frac{\sinh c\rho}{c}\,\big(\trap \hat{R} \hat R + c^2\big) + \cosh c\rho\, \big(\trap \hat{R} + \hat{R}\big) = 0 .
\end{equation}

Let us now show that, conversely, if a skew-symmetric operator $R$ satisfies the identity~\eqref{Eq:IdExp} for all $\rho \in \Real$, then it follows that $R$ solves the (m)cYBE and $\alg{h}_\pm$ is solvable.
To do so, we expand $\hat R$ for small $\rho$
\begin{equation}
\hat R = R + \frac{\rho}{2}R_+R_- + \Order(\rho^2),
\end{equation}
and substitute into the (m)cYBE~\eqref{cybe}.
At leading order we find the (m)cYBE for $R$, while at order $\Order(\rho)$ we have
\begin{equation}\begin{split}\label{convstart}
& [R_+R_-X, RY] - R[R_+R_-X,Y]
\\ & \quad + [RX,R_+R_-Y] -R[X,R_+R_-Y] - R_+R_-[X,Y]_R = 0, \qquad \forall\, X,Y\in\g.
\end{split}\end{equation}
If $\alg{h}_\pm$ is solvable then, as expected, this equality holds by eqs.~\eqref{Eq:IdentityR2} and \eqref{Eq:IdentityR1}.
To prove the converse, namely that eq.~\eqref{convstart} implies that $\alg{h}_\pm$ is solvable, we contract with $Z\in\g$ using the bilinear form $\kappa$.
Using the ad-invariance of $\kappa$ and the skew-symmetry of $R$, we find
\begin{equation}
\kappa\big( X, R_+R_-[Y,Z]_R \big) + \kappa\big( Y, R_+R_-[Z,X]_R \big) - \kappa\big( Z, R_+R_-[X,Y]_R \big) = 0.
\end{equation}
Recalling the expression for the 3-cochain $\Omega_R(X,Y,Z)$ in eq.~\eqref{Eq:OmegaR}, this is equivalent to
\begin{equation}
\Omega_R(X,Y,Z) = 0,
\end{equation}
which indeed implies that $\alg{h}_\pm$ is solvable.

\subsection{The YB deformation of the PCM plus WZ term for solvable \texorpdfstring{$\alg{h}_\pm$}{h±}}
\label{ssec:ybdef}

\paragraph{Action.}
For an R-matrix with solvable $\alg{h}_\pm$ we define the YB deformation of the PCM plus WZ term by the action \eqref{ansatz} with the operator $\Oc$ given by
\begin{equation}\label{Eq:OExp}
\Oc = \frac{\kay}{2} \frac{\alp+\Er}{\alp-\Er},
\end{equation}
where $\chi$ is a free parameter.
The action then explicitly reads
\begin{equation}\label{Eq:SolvS}
\mathsf{S} = \frac{\kay}{2} \int d^2x \; \kappa \Big(g^{-1}\partial_+ g, \frac{\alp+e^{\rho R_g}}{\alp-e^{\rho R_g}} g^{-1}\partial_- g \Big) + \frac{\kay}{6} \int d^3 x \; \epsilon^{abc} \kappa(g^{-1}\partial_a g, [g^{-1}\partial_b g , g^{-1} \partial_c g]).
\end{equation}
Here we propose this as the action of the YB deformation of the PCM plus WZ term for all R-matrices with solvable $\alg{h}_\pm$.
This form, in the special case that $R$ is given by the standard Drinfel'd-Jimbo R-matrix on a compact Lie algebra, first appeared in~\cite[eq.~(1.9)]{Klimcik:2019kkf} (with the parameters $\alpha$, $\rho_L$ and $\rho_R$ there equal to $\alp$, $\rho$ and 0 respectively).
As explained in~\cite{Klimcik:2019kkf}, expanding the operator $\Oc$ as a polynomial in $R$ using that $R^3=-R$, the action originally constructed in~\cite{Delduc:2014uaa} is recovered.
\unskip\footnote{\label{Foot:ParamDJ}More precisely, introducing
\begin{equation*}
K_\textsubscript{DMV} = \kay \frac{\sinh\chi}{\cosh\chi-\cos\rho},
\qquad \eta_\textsubscript{DMV}^2 = \frac{1-\cos\rho}{\cosh\chi-1},
\qquad k_\textsubscript{DMV} = \frac{\cosh\chi-\cos\rho}{\sinh\chi},
\qquad A_\textsubscript{DMV} = \frac{\sin\rho}{\sinh\chi},
\end{equation*}
and using $R^3=-R$, the operator $\mathcal{O}$~\eqref{Eq:OExp} can be written as
$\mathcal{O} = \frac12 K_\textsubscript{DMV} (1+\eta_\textsubscript{DMV}^2 + A_\textsubscript{DMV}R + \eta_\textsubscript{DMV}^2 R^2)$.
Substituting into~\eqref{ansatz} we recover the action introduced in~\cite{Delduc:2014uaa}.
It is also useful to recall that the parameters of~\cite{Delduc:2014uaa} are related through
\begin{equation*}
A_\textsubscript{DMV}^2 = \eta_\textsubscript{DMV}^2 \Big( 1 - \frac{k_\textsubscript{DMV}^2}{1+\eta^2_\textsubscript{DMV}} \Big).
\end{equation*}}

\paragraph{Integrability.}
To show that the model defined by the action \eqref{Eq:SolvS} is integrable we first need to show that the operator $\Oc$ defined in eq.~\eqref{Eq:OExp} satisfies the integrability condition \eqref{Eq:IntCond} derived in \secref{sec:general}.
Using the definitions of the operators $\mathcal{Q}^\pm$ in terms of $\Oc$ in eq.~\eqref{qpm} and the skew-symmetry of $R$, we have that
\begin{equation}\label{Eq:QSolv}
\mathcal{Q}^- = \frac{\xi}{\kay}(1-e^{-\chi}e^{\rho R}), \qquad
\mathcal{Q}^+ = - \frac{\xi}{\kay}(1-\alp\,e^{\rho R}).
\end{equation}
Computing $\mathcal{Z}(X,Y)$, defined in eq.~\eqref{zeq}, we find
\begin{equation}\begin{split}
\frac{\kay^2}{\xi} \mathcal{Z}[X,Y] & = \big( \kay\sinh\chi - 2\xi\cosh\chi \big) \Er[X,Y]
\\ &\hspace{11pt} -\xi \Big( \big[\Er X, \Er Y \big] - \Er \big[\Er X, Y\big] - \Er \big[X , \Er Y\big] - [X,Y] \Big),
\end{split}\end{equation}
and then using the identity \eqref{Eq:IdExp} gives
\begin{equation}
\frac{\kay^2}{\xi} \mathcal{Z}[X,Y] = \big( \kay\sinh\chi - 2\xi(\cosh\chi-\cosh{c\rho}) \big) \Er[X,Y].
\end{equation}
It is now clear that the integrability condition \eqref{Eq:IntCond} is satisfied if and only if we choose the free parameter $\xi$ to be
\begin{equation}\label{Eq:XiSolv}
\xi = \frac{\kay}{2} \frac{\sinh\chi}{\cosh\chi-\cosh{c\rho}}.
\end{equation}

Therefore, we have constructed an integrable deformation of the PCM plus WZ term based on any R-matrix for which $\alg{h}_\pm$ is solvable.
In principle, we have only used that $R$ is skew-symmetric and $e^{\rho R}$ satisfies the identity~\eqref{Eq:IdExp}, and hence the action \eqref{Eq:SolvS} admits a Lax connection for any operator $R$ with these two properties.
However, as discussed above, if \eqref{Eq:IdExp} is satisfied for all $\rho \in \Real$ then it follows that $R$ is a solution of the (m)cYBE with solvable $\alg{h}_\pm$, and hence we return to our original setup.
Nevertheless, there may be other isolated skew-symmetric solutions of~\eqref{Eq:IdExp} for fixed $\rho$.
These would not necessarily correspond to deformations of the PCM plus WZ term, but may give rise to new examples of integrable $\sigma$-models.
In \appref{app:gendef} we investigate the reverse logic and show that the identity~\eqref{Eq:IdExp} and the skew-symmetry of $R$ are also necessary conditions for integrability under certain assumptions, including that the Lax connection takes of the form \eqref{Eq:Lax}.

Thus far we have demonstrated the existence of an infinite number of conserved quantities, which can be extracted from the Lax connection
\begin{equation}
\Lc_\pm(z) = \pm \frac{\kay}{\xi(1\mp z)} \frac{1}{e^{\pm \chi}\,e^{\rho R_g}-1} g^{-1}\p_\pm g.
\end{equation}
To complete the proof of integrability, we also have to show that these conserved charges are in involution by studying the Poisson bracket of the Lax matrix $\Lc(z)=\frac{1}{2}\big(\Lc_+(z)-\Lc_-(z)\big)$.
We will perform this analysis in the Hamiltonian formulation in \secref{sec:hamiltonian}.

\paragraph{Limits.}
To recover the standard YB deformation of the PCM from the action \eqref{Eq:SolvS} in the limit $\kay\to 0$ we parametrise $\chi$ and $\rho$
in terms of two new parameters, $\hay$ and $\eta$,
\begin{equation}\label{eq:chirho}
\chi = \frac{\kay}{\hay}, \qquad \rho = \frac{\eta\kay}{\hay}.
\end{equation}
We now take $\kay \to 0$ while keeping $\hay$ and $\eta$ fixed.
A direct computation shows that the operator $\Oc$ defined in eq.~\eqref{Eq:OExp} behaves as
\begin{equation}\label{Eq:ExpandO}
\Oc = \hay\Big(\frac{1}{1-\eta R} + \Order\Big(\frac{\kay^2}{\hay^2}\Big)\Big),
\end{equation}
and in the limit $\kay \to 0$ we indeed recover the operator \eqref{Eq:OYB} that characterises the YB deformation of the PCM~\cite{Klimcik:2002zj,Klimcik:2008eq}
\begin{equation}\label{eq:ybstandard}
\mathsf{S} = \hay \int d^2x \; \kappa \Big(g^{-1}\partial_+ g, \frac{1}{1-\eta R_g} g^{-1}\partial_- g \Big) .
\end{equation}
Furthermore, the constant $\xi$, chosen above to be \eqref{Eq:XiSolv}, satisfies
\begin{equation}\label{Eq:ExpandXi}
\xi = \hay\Big(\frac{1}{1-c^2\eta^2} + \Order\Big(\frac{\kay^2}{\hay^2}\Big)\Big),
\end{equation}
coinciding with eq.~\eqref{Eq:XiYB} in the limit $\kay\to 0$.
In this limit the identity \eqref{Eq:IdExp} simply reduces to the (m)cYBE \eqref{cybe} for $R$.
This is consistent since the YB deformation of the PCM~\eqref{eq:ybstandard} is integrable for any skew-symmetric R-matrix, with no additional constraints such as the solvability of $\alg{h}_\pm$.

Let us compare these results with the general analysis of \ssecref{subsec:pertk}, in which we investigate how to construct an integrable YB deformation of the PCM plus WZ term to first order in the expansion parameter $\kay/\hay$.
In particular, in \ssecref{subsec:pertk} we introduced the operator $\widehat{\Oc}$ to parametrise the leading corrections to the operator $\Oc$ and the constant $\xi$ \eqref{oex}.
The construction of the integrable deformation at order $\Order(\kay/\hay)$ was then possible if an operator $\widehat{\Oc}$ can be found that satisfies the condition \eqref{orderkay}, which depends on the choice of R-matrix.
Comparing the expansions \eqref{Eq:ExpandO} and \eqref{Eq:ExpandXi} with eq.~\eqref{oex}, we see that for $\Oc$ as defined in eq.~\eqref{Eq:OExp} we have $\widehat{\Oc} = 0$.
Recalling that we are considering an R-matrix for which $\alg{h}_\pm$ is solvable, \textit{i.e.} $R_+R_-[X,Y]_R=0$ (see eq.~\eqref{Eq:IdentityR1}), $\widehat{\Oc} = 0$ is indeed a solution to the condition \eqref{orderkay}.
\unskip\footnote{Recalling that $\widehat{\Oc} = \frac12(\mathcal{B} + \mathcal{A})$, this corresponds to the solution $\alpha = \mathcal{A} = 0$ of \eqref{diffeq} for solvable $\alg{h}_\pm$, \textit{i.e.} $\Omega_R = 0$.}
Therefore, when $\alg{h}_\pm$ is solvable, there exists an extension of this solution valid for all $\kay$.

We end this discussion of limits by briefly mentioning the undeformed limit, which corresponds to taking $\rho \to 0$.
In this limit, the action \eqref{Eq:SolvS} becomes the action of the PCM plus WZ term
\begin{equation}\label{eq:pcmpluswz}
\mathsf{S} = \frac{\kay}{2} \coth{\frac{\chi}{2}} \int d^2x \; \kappa (g^{-1}\partial_+ g, g^{-1}\partial_- g ) + \frac{\kay}{6} \int d^3 x \; \epsilon^{abc} \kappa(g^{-1}\partial_a g, [g^{-1}\partial_b g , g^{-1} \partial_c g]).
\end{equation}
Let us note that in this parametrisation, the conformal point corresponding to the WZW model is recovered in the limit $\chi\to+\infty$.
Taking this limit in the deformed action \eqref{Eq:SolvS}, the dependence on the R-matrix drops out and hence, in this sense, the YB deformation of the WZW model is trivial.

\paragraph{Asymmetric R-matrix.}
Let us conclude this section by rewriting the action \eqref{Eq:SolvS} in terms of the asymmetric R-matrix \eqref{eq:defrhat}, which generalises that introduced in~\cite{Delduc:2019whp} in the case of the standard Drinfel'd-Jimbo R-matrix.
Introducing the parameters $\gamma$ and $\hat{\rho}$, defined in terms of $\chi$ and $\rho$ as
\begin{equation}\label{eq:gampm}
\gamma = e^\chi, \qquad \hat\rho = \frac{\sinh c\rho}{c},
\end{equation}
we have that
\begin{equation}\label{Eq:SolvSasym}
\mathsf{S} = \frac{\kay}{2} \int d^2x \; \kappa \Big(g^{-1}\partial_+ g, \frac{\gamma + \sqrt{1+c^2\hat\rho^2} + \hat\rho \hat{R}_g}{\gamma - \sqrt{1+c^2\hat\rho^2} - \hat \rho\hat{R}_g} g^{-1}\partial_- g \Big) + \frac{\kay}{6} \int d^3 x \; \epsilon^{abc} \kappa(g^{-1}\partial_a g, [g^{-1}\partial_b g , g^{-1} \partial_c g]).
\end{equation}
The symmetry property \eqref{eq:transposer} implies that the operators $\mathcal{Q}^\pm$, defined in eq.~\eqref{qpm}, take the form
\begin{equation}\begin{aligned}\label{eq:ooqqgam}
\mathcal{Q}^-& = \frac{\xi}{\kay}\Big(1 - \gamma^{-1}\big( \sqrt{1+c^2\hat\rho^2} + \hat\rho \hat{R}\big)\Big), \qquad
&\mathcal{Q}^+& = - \frac{\xi}{\kay}\Big(1 -\gamma\big( \sqrt{1+c^2\hat\rho^2} + \hat\rho \hat{R}\big)\Big).
\end{aligned}\end{equation}
Noting that $\mathcal{Q}^\pm$ are affine functions of $\hat{R}$, when we substitute into \eqref{zeq} we find that the integrability condition \eqref{Eq:IntCond} reduces to the (m)cYBE for $\hat{R}$ if
\begin{equation}\label{fixxi}
\xi = \frac{\kay}{2}\frac{\gamma - \gamma^{-1}}{\gamma + \gamma^{-1} -2 \sqrt{1+c^2\hat{\rho}^2}} ,
\end{equation}
which agrees with \eqref{Eq:XiSolv} using the definitions of $\gamma$ and $\hat\rho$ in eq.~\eqref{eq:gampm}.

Let us briefly review the various limits in this formulation.
First, the limit without WZ term is given by setting
\begin{equation}\label{eq:limwowz}
\gamma = 1 + \frac{\kay}{\hay} + \Order\Big(\frac{\kay^2}{\hay^2}\Big) ,\qquad
\hat\rho = \frac{\eta \kay}{\hay} + \Order\Big(\frac{\kay^2}{\hay^2}\Big) ,
\end{equation}
and sending $\kay \to 0$.
Taking this limit in the action~\eqref{Eq:SolvSasym} and the symmetry property \eqref{eq:transposer} we recover the standard YB deformation of the PCM~\eqref{eq:ybstandard} defined in terms of a skew-symmetric R-matrix.
The undeformed limit is given by taking $\hat\rho \to 0$.
In this limit the action~\eqref{Eq:SolvSasym} becomes the action of the PCM plus WZ term~\eqref{eq:pcmpluswz} with $\frac{\gamma+1}{\gamma-1} = \coth \frac{\chi}{2}$.
Finally the WZW model is recovered in the limit $\gamma \to + \infty$.

Here we have simply re-established the existence of a Lax connection for the model~\eqref{Eq:SolvSasym} if $\hat{R}$ solves the (m)cYBE and satisfies the symmetry property \eqref{eq:transposer}.
We know that if we have a skew-symmetric R-matrix with $\alg{h}_\pm$ solvable then we can construct such an $\hat{R}$ using eq.~\eqref{eq:defrhat} and vice versa if we have an $\hat{R}$ with these properties for all $\rho$.
Just as in the previous formulation, there may be isolated solutions of the (m)cYBE with the symmetry property \eqref{eq:transposer} for fixed $\rho$ that do not fall into this class.
Nevertheless, it is interesting to observe that an alternative way to interpret the YB deformation of the PCM plus WZ term is in terms of an asymmetric R-matrix.

\section{Hamiltonian formulation}
\label{sec:hamiltonian}

In this section we investigate the YB deformation of the PCM plus WZ term~\eqref{Eq:SolvS} in the Hamiltonian formulation, generalising the analysis of~\cite{Delduc:2014uaa} for the case of the standard Drinfel'd-Jimbo R-matrix and~\cite{Delduc:2013fga} for the case without WZ term.
We first perform the Hamiltonian analysis of the general model \eqref{ansatz}, describing its phase space and Hamiltonian, and then use this to study the integrable structure of the YB deformed model~\eqref{Eq:SolvS}.
We conclude by outlining the relation to the formalism of affine Gaudin models.

\subsection{Hamiltonian analysis of the general model}\label{subsec:HamGen}

We start by performing the Hamiltonian analysis of the general model \eqref{ansatz}.
As the first part of this analysis is standard, we restrict ourselves to giving an overview of the key steps.
For more details see, for instance,~\cite[subsec.~3.1]{Delduc:2019bcl}.

\paragraph{Phase space.}
This model describes the dynamics of a $\grp{G}$-valued Lagrangian field $g(x,t)$.
In the Hamiltonian language, its phase space corresponds to canonical fields on the cotangent bundle $T^\star_{\vphantom{g}} \grp{G}$, depending on the space coordinate $x$ (the time coordinate $t$ being induced by the Hamiltonian).
By left translation the cotangent space $T^\star_g\grp{G}$ at a point $g\in \grp{G}$ can be canonically mapped to the cotangent space $T^\star_{\Id}\grp{G}$ at the identity $\Id \in \grp{G}$, which is the dual $\g^\star$ of the Lie algebra $\g$.
Moreover, since $\g$ is equipped with the non-degenerate bilinear form $\kappa$, $\g^\star$ is canonically isomorphic to $\g$ itself.
Thus, the cotangent bundle $T^\star_{\vphantom{g}} \grp{G}$ can be identified with the direct product $\grp{G}\times\g$.
In particular, the Hamiltonian model that we are considering can be described by two fields, $g(x) \in \grp{G}$ and $X(x) \in \g$, with the latter encoding the conjugate momenta of the scalar fields parametrising the former.

As a cotangent bundle, $T^\star_{\vphantom{g}} \grp{G}$ is equipped with a canonical symplectic form, which translates to a Poisson bracket on the canonical fields in $T^\star_{\vphantom{g}} \grp{G}$.
When these canonical fields are parametrised by $g(x)$ and $X(x)$ this Poisson bracket reads
\begin{subequations}\label{Eq:PBTstarG}
\begin{align}
\left\lbrace g\ti{1}(x), g\ti{2}(y) \right\rbrace & = 0, \\
\left\lbrace X\ti{1}(x), g\ti{2}(y) \right\rbrace & = g\ti{2}(x) C\ti{12} \delta_{xy},\label{Eq:PBXg} \\
\left\lbrace X\ti{1}(x), X\ti{2}(y) \right\rbrace & = \left[ C\ti{12}, X\ti{1}(x) \right] \delta_{xy},\label{Eq:PBXX}
\end{align}
\end{subequations}
where we use standard tensorial notation, $\delta_{xy}=\delta(x-y)$ is the Dirac distribution and $C\ti{12}$ denotes the quadratic split Casimir of $\g$, defined as the unique element of $\g\otimes\g$ that satisfies
\begin{equation}\label{eq:splitcas}
\kappa\ti{2}(C\ti{12},X\ti{2}) = X\ti{1} , \qquad \forall ~ X\in \alg{g}.
\end{equation}

\paragraph{Maurer-Cartan spatial current.}
Let us consider the spatial Maurer-Cartan current
\begin{equation}\label{Eq:j}
j(x) = g(x)^{-1}\p_x g(x),
\end{equation}
which is valued in $\g$.
Its Poisson brackets can be computed from the canonical brackets \eqref{Eq:PBTstarG} giving
\begin{subequations}\label{Eq:PBj}
\begin{align}
\left\lbrace g\ti{1}(x), j\ti{2}(y) \right\rbrace &= 0,\label{Eq:PBgj}\\
\left\lbrace j\ti{1}(x), j\ti{2}(y) \right\rbrace &= 0, \\
\left\lbrace X\ti{1}(x), j\ti{2}(y) \right\rbrace &= \big[ C\ti{12}, j\ti{1}(x) \big] \delta_{xy} - C\ti{12} \delta'_{xy},
\end{align}
\end{subequations}
where $\delta'_{xy}=\p_x\delta(x-y)$ denotes the derivative of the Dirac distribution.

\paragraph{WZ term.}
Let us consider the WZ term in the action \eqref{ansatz}.
It is defined as the integral of a closed 3-form on a 3-dimensional extension of the 2-dimensional space-time of the model.
Thus, it can be written, at least locally, as an integral over the space-time coordinates $(x,t)$, which takes the form
\begin{equation}\label{Eq:WZ}
\frac{1}{6} \int d^3 x \; \epsilon^{abc} \kappa(g^{-1}\partial_a g, [g^{-1}\partial_b g , g^{-1} \partial_c g]) = \int d^2 x \; \kappa (g^{-1}\p_t g, W),
\end{equation}
for some $\g$-valued current $W$, constructed from the scalar fields parametrising $g$ and their spatial derivatives.
In the Hamiltonian formalism, the current $W$ can be shown to satisfy the following Poisson brackets~\cite{Delduc:2019bcl}
\begin{equation}\label{Eq:PbW1}
\big\lbrace g\ti{1}(x), W\ti{2}(y) \big\rbrace = 0, \qquad \big\lbrace j\ti{1}(x), W\ti{2}(y) \big\rbrace = 0,
\end{equation}
and
\begin{equation}\label{Eq:PbW2}
\big\lbrace X\ti{1}(x), W\ti{2}(y) \big\rbrace + \big\lbrace W\ti{1}(x), X\ti{2}(y) \big\rbrace = \big[ C\ti{12}, W\ti{1}(x)-j\ti{1}(x) \big] \delta_{xy}.
\end{equation}
Moreover, it satisfies the orthogonality relation
\begin{equation}\label{Eq:Wj}
\kappa(W,j) = 0.
\end{equation}

\paragraph{Eliminating temporal derivatives.}
In order to perform the Hamiltonian analysis of the model, we eliminate the temporal derivatives of the Lagrangian field $g(x,t)$ in favour of the conjugate momenta encoded in the field $X$.
Recalling that $\partial_\pm = \partial_t \pm \partial_x$ and using the expression~\eqref{Eq:WZ} for the WZ term, we can rewrite the action~\eqref{ansatz} as
\begin{equation}\begin{split}\label{Eq:Stx}
\mathsf{S} &= \int d^2x \; \Big( \frac{1}{2} \kappa\big( g^{-1}\p_t g, (\tra \Oc_g + \Oc_g) g^{-1}\p_t g \big) + \kappa\big( g^{-1}\p_t g, (\tra \Oc_g - \Oc_g) g^{-1}\p_x g + \kay\,W \big)
\\
&\hspace{250pt} - \frac{1}{2} \kappa\big( g^{-1}\p_x g, (\tra \Oc_g + \Oc_g) g^{-1}\p_x g \big) \Big).
\end{split}\end{equation}
Computing the conjugate momenta of the scalar fields, we find that $X$ is given by
\begin{equation}\label{Eq:X}
X = (\tra \Oc_g + \Oc_g) g^{-1}\p_t g + (\tra \Oc_g - \Oc_g) g^{-1}\p_x g + \kay\,W.
\end{equation}
Inverting this relation to express the temporal Maurer-Cartan current $g^{-1}\partial_t g$ in terms of $X$ we find
\begin{equation}\label{Eq:Leg}
g^{-1}\p_t g = \Tc_g \,(X-\kay\,W) + \Tc_g \,(\Oc_g - \tra \Oc_g)j,
\end{equation}
where $j$ is the spatial Maurer-Cartan current~\eqref{Eq:j} and $\Tc_g = \Ad_g^{-1}\Tc\Ad_g^{\vphantom{-1}}$ with the linear operator $\Tc:\g\rightarrow\g$ defined as
\begin{equation}\label{Eq:T}
\Tc = \frac{1}{\Oc + \tra \Oc}.
\end{equation}
The relation \eqref{Eq:X} can be written in terms of the light-cone currents $g^{-1}\p_\pm g$ as
\begin{equation}\label{Eq:Xlc}
X = \tra \Oc_g\, g^{-1}\p_+g + \Oc_g\, g^{-1}\p_-g + \kay\,W.
\end{equation}

\paragraph{Hamiltonian.}
Let us now compute the Hamiltonian of the model.
It is defined by the Legendre transform
\begin{equation}
\mathsf{H} = \int dx \; \Big( \kappa(X,g^{-1}\p_t g) - \mathsf{L} \Big),
\end{equation}
where $\mathsf{L}$ is the Lagrangian density for the action \eqref{Eq:Stx}.
The phase space expression of the Hamiltonian, \textit{i.e.} using eq.~\eqref{Eq:Leg} to eliminate $g^{-1}\partial_t g$, is
\begin{equation}
\mathsf{H} = \frac{1}{2} \int dx \; \Big( \kappa\Big(X-\kay\,W+(\Oc_g - \tra \Oc_g)j, \Tc_g \big(X-\kay\,W+(\Oc_g - \tra \Oc_g)j\big) \Big) + \kappa\Big( j, (\Oc_g + \tra \Oc_g) j\Big) \Big).
\end{equation}
This Hamiltonian also has a simple form when expressed in terms of the light-cone currents $g^{-1}\p_\pm g$.
Indeed, starting from the action \eqref{ansatz} and using eq.~\eqref{Eq:Xlc} we find
\begin{equation}\label{Eq:HamLC}
\mathsf{H} = \frac{1}{4} \int dx \; \Big( \kappa\big(g^{-1}\p_+ g,(\Oc_g + \tra \Oc_g)g^{-1}\p_+ g\big) + \kappa\big(g^{-1}\p_- g,(\Oc_g + \tra \Oc_g)g^{-1}\p_- g\big) \Big).
\end{equation}

\paragraph{Light-cone currents in phase space variables.}
It will also be useful to express the light-cone currents $g^{-1}\p_\pm g$ in terms of phase space variables.
Using eq.~\eqref{Eq:Leg} to eliminate $g^{-1}\p_t g$, we find
\begin{equation}\begin{split}\label{eq:lc1}
g^{-1}\p_+ g &= \Tc_g\,(X-\kay\,W) + 2 \Tc_g\,\Oc_g\, j, \\
g^{-1}\p_- g &= \Tc_g\,(X-\kay\,W) - 2 \Tc_g\,\tra \Oc_g\, j.
\end{split}\end{equation}
Introducing the $\g$-valued current
\begin{equation}\label{Eq:Z}
Z = X - \kay\,W - \kay\,j,
\end{equation}
and recalling the definition~\eqref{qpm} of the operators $\mathcal{Q}^\pm$, we can then express the light-cone currents as
\begin{equation}\label{Eq:JlcHam}
g^{-1}\p_\pm g = \Tc_g\,Z \pm 2\xi \,\Tc_g\,\mathcal{Q}_g^{\mp\,-1}\,j.
\end{equation}

\subsection{Integrable structure of the YB deformed PCM plus WZ term}
\label{ssec:HamInt}

Let us now study the integrable structure of the YB deformed PCM plus WZ term based on an R-matrix with solvable $\alg{h}_\pm$.
The action of this model was constructed in \secref{sec:solvable} and is given in eq.~\eqref{Eq:SolvS}.
We will show that the Lax matrix satisfies a Maillet bracket with twist function.
Therefore, the conserved charges extracted from the monodromy of this Lax matrix are in involution, thus completing the proof of integrability.

\paragraph{Lax matrix.}
Let us consider the Lax matrix of the model, \textit{i.e.} the spatial component of its Lax connection, $\Lc(z)=\frac{1}{2}(\Lc_+(z)-\Lc_-(z))$.
From eqs.~\eqref{Eq:Lax} and \eqref{intq}, we have
\begin{equation}\label{Eq:LaxLC}
\Lc(z) = \frac{1}{2} \frac{\mathcal{Q}^{+\,-1}_g}{1 - z} g^{-1}\p_+ g - \frac{1}{2} \frac{\mathcal{Q}^{-\,-1}_g}{1 + z} g^{-1}\p_- g.
\end{equation}
Using eq.~\eqref{Eq:JlcHam} to eliminate the light-cone currents in favour of phase space variables, we find
\begin{equation}\label{Eq:LaxTQ}
\Lc(z) = \frac{1}{2} \bigg( \frac{\mathcal{Q}^{+\,-1}_g \Tc_g}{1 - z} - \frac{\mathcal{Q}^{-\,-1}_g \Tc_g}{1 + z} \bigg) Z + \xi \bigg( \frac{\mathcal{Q}^{+\,-1}_g\Tc_g\mathcal{Q}^{-\,-1}_g}{1 - z} + \frac{\mathcal{Q}^{-\,-1}_g\Tc_g\mathcal{Q}^{+\,-1}_g}{1 + z} \bigg) j.
\end{equation}
For the model of interest~\eqref{Eq:SolvS} the operators $\Oc$ and $\mathcal{Q}^\pm$ are given in eqs.~\eqref{Eq:OExp} and \eqref{Eq:QSolv} respectively, which lead to the following expression for $\Tc$, defined in terms of $\Oc$ in eq.~\eqref{Eq:T}:
\begin{equation}
\Tc = \frac{(e^\chi-e^{\rho R})(e^\chi e^{\rho R}-1)}{\kay(e^{2\chi}-1)e^{\rho R}}.
\end{equation}
Substituting in for $\Tc$ and $\mathcal{Q}^\pm$ in eq.~\eqref{Eq:LaxTQ} then gives
\begin{equation}\label{Eq:LaxHam}
\Lc(z) = \big( \alpha(z)\,e^{-\rho R_g} + \beta(z) \big) Z + 2\kay \,\alpha(z)\,e^{-\rho R_g}\, j,
\end{equation}
where we have defined
\begin{equation}
\alpha(z) = \frac{1}{2\xi\sinh\chi\, (1-z^2)}, \qquad \beta(z) = \frac{z \sinh \chi - \cosh\chi}{2\xi\sinh\chi\, (1-z^2)}.
\end{equation}

\paragraph{Maillet bracket with twist function.}
To compute the Poisson bracket of the Lax matrix \eqref{Eq:LaxHam} with itself it is useful to know the Poisson brackets satisfied by $Z$~\eqref{Eq:Z}.
From the brackets \eqref{Eq:PBXX}, \eqref{Eq:PBj} and \eqref{Eq:PbW2}, we find that $Z$ is a Kac-Moody current
\begin{equation}\label{Eq:PbZ1}
\left\lbrace Z\ti{1}(x), Z\ti{2}(y) \right\rbrace = \big[ C\ti{12}, Z\ti{1}(x) \big] \delta_{xy} + 2\kay\, C\ti{12} \delta'_{xy}.
\end{equation}
Moreover, from the brackets \eqref{Eq:PBXg}, \eqref{Eq:PBj} and \eqref{Eq:PbW1}, we also have
\begin{equation}\label{Eq:PbZ2}
\left\lbrace Z\ti{1}(x), g\ti{2}(y) \right\rbrace = g\ti{2}(x) C\ti{12} \delta_{xy}, \qquad \left\lbrace Z\ti{1}(x), j\ti{2}(y) \right\rbrace = \big[ C\ti{12}, j\ti{1}(x) \big] \delta_{xy} - C\ti{12} \delta'_{xy}.
\end{equation}
Starting from these brackets, it is possible to compute the Poisson bracket of the Lax matrix \eqref{Eq:LaxHam} with itself.
As shown in \appref{sec:Maillet}, we find that the Lax matrix satisfies a non-ultralocal Maillet bracket~\cite{Maillet:1985fn,Maillet:1985ek}
\begin{equation}\begin{split}\label{Eq:Maillet}
\big\lbrace \Lc\ti{1}(z,x), \Lc\ti{2}(w,y) \big\rbrace
&= \big[ \Rc\ti{12}(z,w), \Lc\ti{1}(z,x) \big] \delta_{xy} - \big[ \Rc\ti{21}(w,z), \Lc\ti{2}(w,x) \big] \delta_{xy}\\
& \hspace{150pt} - \left( \Rc\ti{12}(z,w) + \Rc\ti{21}(w,z) \right) \delta'_{xy},
\end{split}\end{equation}
with $\Rc\ti{12}(z,w)$, the $\Rc$-matrix characterising this bracket, taking the form~\cite{Maillet:1985ec, Reyman:1988sf, Sevostyanov:1995hd, Vicedo:2010qd,Lacroix:2018njs}
\begin{equation}\label{Eq:RMat}
\Rc\ti{12}(z,w) = \frac{C\ti{12}}{w-z} \vp(w)^{-1},
\end{equation}
where the twist function $\vp(z)$ is the following rational function of the spectral parameter
\begin{equation}\label{Eq:Twist}
\vp(z) = \frac{2\xi (1-z^2)}{\left(z-\dfrac{\kay}{2\xi}\right)^2 - \dfrac{\sinh^2{\rho c}}{\sinh^2{\chi}}}.
\end{equation}
For the case of the standard Drinfel'd-Jimbo R-matrix with $c=i$ it is straightforward to check that, after rewriting in terms of the parameters introduced in \footref{Foot:ParamDJ}, we recover the twist function computed in~\cite{Delduc:2014uaa}. Considering the limit without WZ term, \textit{i.e.} taking $\kay\to 0$ while keeping $\hay$ and $\eta$ fixed in the parametrisation \eqref{eq:chirho}, we recover the twist function of the standard YB deformation of the PCM without WZ term~\cite{Delduc:2013fga}.

\paragraph{Inhomogeneous case -- Kac-Moody currents and poles of the twist function.}
In this paragraph we take $R$ to be an inhomogeneous R-matrix, \textit{i.e.} $c\neq 0$.
The twist function~\eqref{Eq:Twist} has simple poles at
\begin{equation}
z_\pm = \frac{\kay}{2\xi} \pm \frac{\sinh{\rho c}}{\sinh{\chi}}
\end{equation}
with the corresponding residues given by
\begin{equation}\label{Eq:LevelsI}
\ell_\pm = \res_{z=z_\pm} \vp(z)\, dz = \kay \left( \pm \coth{c\rho} - 1 \right).
\end{equation}
It is a standard result~\cite{Vicedo:2015pna} that simple poles of the twist function are associated with Kac-Moody currents, which can be constructed from the Lax matrix~\eqref{Eq:LaxHam} as follows
\begin{equation}\label{Eq:KM}
\Jc^\pm = \ell_\pm\,\Lc(z_\pm) = \pm\frac{1}{2\sinh{c\rho}} \Big( (\Ergm - e^{\mp c\rho}) Z + 2\kay\, \Ergm j \Big),
\end{equation}
generalising the currents defined in~\cite{Delduc:2014uaa} for the standard Drinfel'd-Jimbo R-matrix.
The definition~\eqref{Eq:KM} of the currents $\Jc^\pm$ is equivalent to the following partial fraction decomposition of $\vp(z)\Lc(z)$:
\begin{equation}
\vp(z)\Lc(z) = \frac{\Jc^+}{z-z_+} + \frac{\Jc^-}{z-z_-}.
\end{equation}
From the Maillet bracket \eqref{Eq:Maillet} it follows that $\Jc^\pm$ are Poisson commuting Kac-Moody currents with levels $\ell_\pm$:
\begin{subequations}\begin{align}
\big\lbrace \Jc^\pm\ti{1}(x),\Jc^\pm\ti{2}(y) \big\rbrace &= \big[C\ti{12},\Jc^\pm\ti{1}(x) \big]\delta_{xy} - \ell_\pm \,C\ti{12} \,\delta'_{xy}, \\
\big\lbrace \Jc^\pm\ti{1}(x),\Jc^\mp\ti{2}(y) \big\rbrace &= 0.
\end{align}\end{subequations}
These brackets can be checked explicitly following a similar approach to the computation of the Poisson bracket of the Lax matrix with itself described in \appref{sec:Maillet}.

Let us briefly discuss the reality conditions obeyed by these Kac-Moody currents and their levels.
If we have a split R-matrix, \textit{i.e.} $c=1$, then the poles $z_\pm$ and the levels $\ell_\pm$ are real, and the Kac-Moody currents $\Jc^\pm$ are valued in the real Lie algebra $\g$.
If we have a non-split R-matrix, \textit{i.e.} $c=i$, then the poles $z_\pm$ and the levels $\ell_\pm$ form complex conjugate pairs.
The Kac-Moody currents $\Jc^\pm$ are valued in the complexification $\g^\Complex$ of $\g$ and are also complex conjugate to each other.

\paragraph{Homogeneous case -- Takiff currents and double pole of the twist function.}
In this paragraph, we take $R$ to be a homogeneous R-matrix, \textit{i.e.} $c=0$.
The twist function~\eqref{Eq:Twist} then has a double pole at
\begin{equation}
z_0 = \frac{\kay}{2\xi},
\end{equation}
and we define the residues
\begin{equation}\label{Eq:LevelsHom}
\ell_0 = \res_{z=z_0} \vp(z)\, dz = -2\kay \qquad \text{ and } \qquad \ell_1 = \res_{z=z_0} (z-z_0)\vp(z)\, dz = \frac{2\kay}{\sinh\chi}.
\end{equation}
Using the following partial fraction decomposition to extract the currents $\Jc_{[0]}$ and $\Jc_{[1]}$ from the Lax matrix
\begin{equation}
\vp(z)\Lc(z) = \frac{\Jc_{[0]}}{z-z_0} + \frac{\Jc_{[1]}}{(z-z_0)^2},
\end{equation}
we find that
\begin{equation}\label{Eq:Takiff}
\Jc_{[0]} = Z, \qquad \Jc_{[1]} = \frac{(\Ergm - \cosh{c\rho})Z + 2\kay\,\Ergm\,j }{\sinh{\chi}}.
\end{equation}
From the Maillet bracket~\eqref{Eq:Maillet} it follows that these are Takiff currents of multiplicity two
\begin{subequations}\begin{align}
\big\lbrace \Jc_{[0]}\null\ti{1}(x),\Jc_{[0]}\null\ti{2}(y) \big\rbrace &= \big[C\ti{12},\Jc_{[0]}\null\ti{1}(x) \big]\delta_{xy} - \ell_0 \,C\ti{12} \,\delta'_{xy}, \\
\big\lbrace \Jc_{[0]}\null\ti{1}(x),\Jc_{[1]}\null\ti{2}(y) \big\rbrace &= \big[C\ti{12},\Jc_{[1]}\null\ti{1}(x) \big]\delta_{xy} - \ell_1 \,C\ti{12} \,\delta'_{xy}, \\
\big\lbrace \Jc_{[1]}\null\ti{1}(x),\Jc_{[1]}\null\ti{2}(y) \big\rbrace &= 0.
\end{align}\end{subequations}
These brackets can again be checked directly using techniques similar to those described in \appref{sec:Maillet}.
Let us note for completeness that these currents are real, \textit{i.e.} valued in the real Lie algebra $\g$.

\paragraph{Hamiltonian and zeroes of the twist function.}
The zeroes $+1$ and $-1$ of the twist function \eqref{Eq:Twist} are also simple poles of the Lax matrix \eqref{Eq:LaxLC}.
The local charge
\begin{equation}
\mathsf{Q}(z) = - \frac{\vp(z)}{2} \int d x\;\kappa\big(\Lc(z),\Lc(z)\big),
\end{equation}
which is rational in the spectral parameter $z$, has poles at the zeroes $\pm 1$ of the twist function.
We define $\mathsf{Q}_\pm$ as the corresponding residues
\begin{equation}
\mathsf{Q}_\pm = \res_{z=\pm 1} \mathsf{Q}(z) \, dz.
\end{equation}
From the expression \eqref{Eq:LaxLC} of $\Lc(z)$ we find
\begin{equation}
\mathsf{Q}_\pm = - \frac{\vp'(\pm 1)}{8} \int dx \; \kappa\Big( g^{-1}\p_\pm g, \big( \mathcal{Q}^\pm_g \; \tra \mathcal{Q}^\pm_g \big)^{-1} \, g^{-1}\p_\pm g \Big).
\end{equation}
Using eqs.~\eqref{Eq:OExp}, \eqref{Eq:QSolv} and \eqref{Eq:Twist} it follows that
\begin{equation}
-\frac{\vp'(\pm 1)}{8} \big( \mathcal{Q}^\pm \; \tra \mathcal{Q}^\pm \big)^{-1} = \pm \frac{\kay(e^{2\chi}-1)e^{\rho R}}{4(e^\chi-e^{\rho R})(e^\chi e^{\rho R}-1)} = \pm\frac{\Oc + \tra \Oc}{4},
\end{equation}
and hence
\begin{equation}
\mathsf{Q}_\pm = \pm\frac{1}{4} \int dx \; \kappa\big(g^{-1}\p_\pm g,(\Oc_g + \tra \Oc_g)g^{-1}\p_\pm g).
\end{equation}
The Hamiltonian \eqref{Eq:HamLC} can then be written as
\begin{equation}\label{Eq:HQ}
\mathsf{H} = \mathsf{Q}_+ - \mathsf{Q}_-,
\end{equation}
while substituting in for the light-cone currents in terms of phase space variables using~\eqref{eq:lc1} and using the definition \eqref{Eq:Z} of $Z$ and the orthogonality relation \eqref{Eq:Wj} we find that
\begin{equation}
\mathsf{P} = \mathsf{Q}_+ + \mathsf{Q}_- = \int dx \; \kappa(X,j).
\end{equation}
The Poisson bracket of $\mathsf{P}$ with the canonical fields $g(x)$ and $X(x)$ generates their spatial derivatives, and hence $\mathsf{P}$ defines the spatial momentum of the model.

As we have just observed, it is possible to extract the Hamiltonian $\mathsf{H}$ and spatial momentum $\mathsf{P}$, both of which are local conserved charges, from the zeroes of the twist function.
In fact, this is a general statement for models with twist functions~\cite{Lacroix:2017isl}.
More precisely, it was shown in~\cite{Lacroix:2017isl} that an infinite tower of local conserved charges in involution can be constructed from each zero of the twist function.
These charges are integrals of polynomials of increasing degree in the currents appearing in the Lax matrix.
In particular, the first charges in each infinite tower are always quadratic and in the present case correspond to the charges $\mathsf{Q}_\pm$.
It follows that for the YB deformation of the PCM plus WZ term there exist an infinite number of local conserved charges in involution, whose density are well chosen polynomials of the currents $\mathcal{Q}_g^{\pm\,-1}\,g^{-1}\p_\pm g$.

\paragraph{Affine Gaudin model structure.}
The model considered here is such that
\begin{itemize}
\item[(i)] the Lax matrix satisfies a Maillet bracket \eqref{Eq:Maillet} with twist function,
\item[(ii)] the Hamiltonian is given by a linear combination \eqref{Eq:HQ} of the quadratic charges $\mathsf{Q}_\pm$.
\end{itemize}
These properties ensure that the model can be interpreted as a realisation of an affine Gaudin model in the formalism proposed in~\cite{Vicedo:2017cge} and developed further in~\cite{Delduc:2019bcl}.

The affine Gaudin model either has two real sites of multiplicity one in the split case, $c=1$, two complex conjugate sites of multiplicity one in the non-split case, $c=i$, or one real site of multiplicity two in the homogeneous case, $c=0$.
\unskip\footnote{Technically, the model also possesses a site of multiplicity two at infinity, which is treated in a slightly different way (see~\cite{Vicedo:2017cge,Delduc:2019bcl}).}
This defines the structure of the underlying formal affine Gaudin model.
The $\sigma$-model of interest is then obtained as a realisation of this formal theory in the algebra of canonical fields on $T^\star_{\vphantom{g}} \grp{G}$.
This realisation is given concretely by the expression \eqref{Eq:KM} of the Kac-Moody currents in the inhomogeneous case and by the expression \eqref{Eq:Takiff} of the Takiff currents in the homogeneous case.
That these currents form a realisation of the formal affine Gaudin model Poisson structure is ensured by point (i) above.
Point (ii) then implies that the Hamiltonian of the model is the image in this realisation of the Hamiltonian of the formal affine Gaudin model.

\section{Relation to alternative formulations}
\label{sec:altform}

In this section we explain how the action \eqref{Eq:SolvS} can be found from three alternative formulations.
In \ssecref{ssec:emodel} we describe its origin as an $\mathcal{E}$-model and in \ssecref{ssec:4dcs} we outline how it can be obtained from 4-dimensional Chern-Simons theory.
Finally, in \ssecref{ssec:NATD} we explain how it is equivalent to a non-abelian T-dual model when $R$ is a homogeneous R-matrix.

\subsection{\texorpdfstring{$\mathcal{E}$}{E}-models}
\label{ssec:emodel}

It has been shown~\cite{Klimcik:2017ken,Severa:2017kcs} that when $R$ is the standard Drinfel'd-Jimbo R-matrix the action \eqref{Eq:SolvS} follows from an $\mathcal{E}$-model, a first-order model on the Drinfel'd double.
Also using the results of~\cite{Klimcik:2019kkf,Klimcik:2020fhs}, the generalisation to any R-matrix with solvable $\alg{h}_\pm$ is largely straightforward.
For completeness, we present a brief overview of this construction, which holds for all three cases, $c=1$, $c=i$ and $c=0$.

\subsubsection{Structure of the Drinfel'd double}
\label{ssec:drindoub}

Let $\alg{g}$ be a simple Lie algebra with Lie bracket $[\cdot,\cdot]$ and normalised Killing form $\kappa(\cdot,\cdot)$~\eqref{kappa}.
We introduce a vector space $\tilde{\alg{g}}$ such that $\dim\tilde{\alg{g}} = \dim\alg{g}$, their direct sum (as vector spaces)
\begin{equation}\label{eq:dd}
\alg{d} = \alg{g} \dotp \tilde{\alg{g}},
\end{equation}
and an invertible linear map $\sigma : \alg{g} \to \tilde{\alg{g}}$.
We denote an element of $\alg{d}$ as $X + \sigma Y$, $X,Y\in\alg{g}$.

Given a skew-symmetric R-matrix on $\alg{g}$, \textit{i.e.} $R: \alg{g} \to \alg{g}$, the vector space $\alg{d}$ can be understood as the Lie algebra of an associated Drinfel'd double.
The Lie bracket
\begin{equation}\begin{split}\label{eq:lb1}
[X_1+\sigma Y_1,X_2+\sigma Y_2] = [X_1,X_2] + \sigma [Y_1,Y_2]_R & + [X_1, R Y_2] - R[X_1,Y_2]+\sigma[X_1,Y_2]
\\ & \quad + [RY_1, X_2] - R[Y_1,X_2]+\sigma[Y_1,X_2]
\end{split}\end{equation}
satisfies the Jacobi identity as a consequence of the (m)cYBE \eqref{cybe},
and the subalgebras $\alg{\alg{g}}$ and $\tilde{\alg{g}} \cong \alg{g}_R$ are Lagrangian with respect to the following invariant bilinear form
\begin{equation}\label{eq:bl1}
\langle X_1+\sigma Y_1,X_2+\sigma Y_2\rangle = \kappa(X_1,Y_2) + \kappa(Y_1,X_2) .
\end{equation}

Defining $\iota X = \sigma X - RX$, the Lie bracket \eqref{eq:lb1} and invariant bilinear form \eqref{eq:bl1} can be equivalently written as
\begin{equation}\label{Eq:ComIota}
\begin{gathered}\hspace{0pt}
[X_1+\iota Y_1,X_2+\iota Y_2] = [X_1,X_2] + c^2[Y_1,Y_2] + \iota([X_1,Y_2]+[Y_1,X_2]),
\\
\langle X_1+\iota Y_1,X_2+\iota Y_2\rangle = \kappa(X_1,Y_2) + \kappa(Y_1,X_2) ,
\end{gathered}
\end{equation}
and we recover the standard result that $\alg{d}$ is isomorphic to the real double $\alg{g} \oplus \alg{g}$, the complex double, $\alg{g}^\Complex$ or the semi-abelian double, $\alg{g} \ltimes \alg{g}^{\text{ab}}$, for $c = 1$, $c=i$ and $c=0$ respectively (see \ssecref{ssec:doubles} for more details).

Given this isomorphism it follows that we have a second invariant bilinear form on $\alg{d}$
\begin{equation}
\langle X_1+\iota Y_1,X_2+\iota Y_2\rangle' = \kappa(X_1,X_2) + c^2 \kappa(Y_1,Y_2),
\end{equation}
or equivalently
\begin{equation}
\langle X_1+\sigma Y_1,X_2+\sigma Y_2\rangle' = \kappa(X_1,X_2) + \kappa(X_1,RY_2) + \kappa(RY_1,X_2) + \kappa(R_\pm Y_1,R_\pm Y_2).
\end{equation}

Let us now consider
\begin{equation}\label{eq:mblmodel}
\llangle \cdot,\cdot \rrangle = \cosh c\rho\, \langle\cdot,\cdot\rangle + \frac{\sinh c\rho}{c} \,\langle\cdot,\cdot\rangle' = \langle\cdot,\cdot\rangle + O(\rho),
\end{equation}
and ask when $\tilde{\alg{g}}$ can be deformed to $\tilde{\alg{g}}_\rho$ such that it remains a Lagrangian subalgebra with respect to this new invariant bilinear form.
Limiting ourselves to the case that $\alg{h}_\pm = \im R_\pm$ is solvable, we recall that the operator
\begin{equation}\label{eq:rhatr}
\hat{R} = \frac{c}{\sinh c\rho} (e^{\rho R} - \cosh c\rho) = R + O(\rho) ,
\end{equation}
introduced in eq.~\eqref{eq:defrhat}, solves the (m)cYBE~\eqref{cybe} as a consequence of the identity~\eqref{Eq:IdExp}.
Defining $\hat{\sigma} = \iota + \hat R$, this motivates the following definition of $\tilde{\alg{g}}_\rho$:
\begin{equation}\label{eq:defg}
\tilde{\alg{g}}_\rho = \{ \hat{\sigma} X, \, X \in \alg{g} \} ,
\end{equation}
\textit{i.e.} $\hat{R}$ identifies the Lagrangian subalgebra $\tilde{\alg{g}}_\rho$ within the semi-abelian, real or complex double in the same way that $R$ identifies $\tilde{\alg{g}}$.
It is also worth noting that since a general element of $\alg{d}$, $X + \sigma Y$, $X,Y \in\alg{g}$, can be uniquely written in the form $\hat{X} + \hat{\sigma} \hat{Y}$, $\hat{X}, \hat{Y} \in \alg{g}$ (to be precise we have $\hat{X} = X + (R - \hat{R})Y$ and $\hat{Y} = Y$), it follows that
\begin{equation}\label{eq:dsgr}
\alg{d} = \alg{g} \dotp \tilde{\alg{g}}_\rho,
\end{equation}
and $\dim \tilde{\alg{g}}_\rho = \dim\alg{g}$.

It is easy to see that $\tilde{\alg{g}}_\rho$ is a deformation of $\alg{g}$ with $\hat\sigma X = \sigma X + O(\rho)$.
To check that \eqref{eq:defg} has the remaining required properties, we first verify the isotropy condition:
\begin{equation}\begin{split}
\llangle\hat \sigma X,\hat\sigma Y \rrangle
= \kappa\left(X, \left(\cosh c\rho\,(\trap\hat{R}+\hat{R}) + \frac{\sinh c\rho}{c} (\trap\hat{R} \hat{R} + c^2 ) \right)Y \right) = 0,
\end{split}\end{equation}
using the symmetry property~\eqref{eq:transposer}, and hence $\tilde{\alg{g}}_\rho$ is indeed isotropic with respect to the invariant bilinear form \eqref{eq:mblmodel}.
Secondly, $\tilde{\alg{g}}_\rho$ should form an algebra.
For any linear operator $\mathcal{O}:\alg{g}\to\alg{g}$ we have
\begin{equation}\begin{split}
[(\iota + \mathcal{O}) X , (\iota + \mathcal{O}) Y] & = (\iota+\mathcal{O})[ X, Y]_{\mathcal{O}}
+ [\mathcal{O} X, \mathcal{O} Y] - \mathcal{O}[ X,Y]_{\mathcal{O}} + c^2[X,Y] , \qquad X,Y\in\alg{g}.
\end{split}\end{equation}
Therefore, if $\mathcal{O}$ solves the (m)cYBE this bracket closes and $\{(\iota +\mathcal{O})X:X\in \alg{g}\}$ is a subalgebra of $\alg{d}$.
This is indeed the case for our choice of $\mathcal{O} = \hat{R}$~\eqref{eq:defg}.

Therefore, for solvable $\alg{h}_\pm$, we have constructed $\tilde{\alg{g}}_\rho$, a deformation of $\tilde{\alg{g}}$ that is a Lagrangian subalgebra with respect to the invariant bilinear form \eqref{eq:mblmodel}.

\subsubsection{Formulation as an \texorpdfstring{$\mathcal{E}$}{E}-model.}
\label{ssec:emodelform}

To formulate the action~\eqref{Eq:SolvS} as an $\mathcal{E}$-model, our starting point is the first-order action~\cite{Klimcik:1995dy,Klimcik:1996nq}
\begin{equation}\begin{split}\label{eq:doubact}
\mathsf{S} & = N \Big[\int d^2 x \; \llangle \gdsl^{-1} \partial_t \gdsl, \gdsl^{-1}\partial_x \gdsl \rrangle
- \frac{1}{6} \int d^3 x \; \epsilon^{abc} \llangle \gdsl^{-1}\partial_a \gdsl,[\gdsl^{-1}\partial_b \gdsl, \gdsl^{-1} \partial_c \gdsl] \rrangle
\\ & \hspace{75pt}
- \int d^2 x\; \llangle \gdsl^{-1} \partial_x \gdsl, \mathcal{E} \gdsl^{-1} \partial_x \gdsl \rrangle \Big],
\end{split}\end{equation}
where $\gdsl$ is a field valued in the Drinfel'd double $\grp{D}$, whose Lie algebra is $\alg{d}$, and
$\llangle\cdot,\cdot\rrangle$ is an invariant bilinear form on $\alg{d}$.
$\mathcal{E} : \alg{d} \to \alg{d}$ is a constant linear operator that squares to the identity, $\mathcal{E}^2 = 1$, and
is symmetric with respect to the bilinear form, $\llangle \mathcal{E}\mathscr{X},\mathscr{Y} \rrangle = \llangle \mathscr{X},\mathcal{E}\mathscr{Y} \rrangle$, $\mathscr{X},\mathscr{Y} \in \alg{d}$.

The invariant bilinear form $\llangle\cdot,\cdot\rrangle$ does not need to be the one that defines $\grp{D}$ as a Drinfel'd double; however, we do require that $\alg{d}$ has at least one Lagrangian subalgebra, which we denote $\alg{b}$, \textit{i.e.} $\dim\alg{b} = \frac12\dim\alg{d}$ and $\llangle\alg{b},\alg{b}\rrangle = 0$, where $\alg{b} = \Lie \grp{B}$.
Redefining
\begin{equation}\label{eq:redefgb}
\gdsl \to b \gdsl, \qquad b \in \grp{B},
\end{equation}
it is an immediate consequence of the Polyakov-Wiegmann identity and the isotropy of $\alg{b}$ that the action~\eqref{eq:doubact} only depends on $b$ through $b^{-1}\partial_x b \in \alg{b}$.
If $\mathcal{E}$ is such that $\Ad_\gdsl^{-1} \alg{b}$ and $\mathcal{E} \Ad_\gdsl^{-1} \alg{b}$ have trivial intersection, then we can integrate out the degrees of freedom in $b$ to obtain the action
\begin{equation}\begin{split}\label{eq:doubact1}
\mathsf{S} & = N \Big[\frac12\int d^2 x \; \big(\llangle \gdsl^{-1} \partial_+ \gdsl, \mathcal{E}\mathcal{P}(\mathcal{E}+1) \gdsl^{-1}\partial_- \gdsl \rrangle
- \llangle \gdsl^{-1} \partial_- \gdsl, \mathcal{E}\mathcal{P}(\mathcal{E}-1) \gdsl^{-1}\partial_+ \gdsl \rrangle\big)
\\ & \hspace{75pt}
- \frac{1}{6} \int d^3 x \; \epsilon^{abc} \llangle \gdsl^{-1}\partial_a \gdsl,[\gdsl^{-1}\partial_b \gdsl, \gdsl^{-1} \partial_c \gdsl] \rrangle\Big],
\end{split}\end{equation}
where $\mathcal{P}$ is the projector with $\im \mathcal{P} = \ker \llangle\mathcal{E} \Ad_\gdsl^{-1} \alg{b} ,\cdot\rrangle = \mathcal{E} \Ad_\gdsl^{-1} \alg{b}$ and $\ker \mathcal{P} = \Ad_\gdsl^{-1} \alg{b}$.
\unskip\footnote{To reach this form it is useful to use the identities $\mathcal{E} \mathcal{P} + \mathcal{P} \mathcal{E} = \mathcal{E}$ and $\llangle \mathcal{P}\mathscr{X},\mathcal{P}\mathscr{Y}\rrangle = 0$, $\mathscr{X},\mathscr{Y} \in \alg{d}$.}
It follows that the operators $\mathcal{E}\mathcal{P}(\mathcal{E}\pm1)$ are projectors with $\im \mathcal{E}\mathcal{P}(\mathcal{E}\pm1) = \Ad_\gdsl^{-1} \alg{b}$ and $\ker \mathcal{E}\mathcal{P}(\mathcal{E}\pm1) = \alg{e}_\mp$ where $\alg{e}_\pm$ are the eigenspaces of $\mathcal{E}$ with eigenvalues $\pm 1$.
To compensate the additional degrees of freedom that the redefinition \eqref{eq:redefgb} introduces, the action~\eqref{eq:doubact1} has a $\grp{B}$ gauge symmetry
\begin{equation}\label{eq:gsym}
\gdsl \to b' \gdsl , \qquad b' \in \grp{B},
\end{equation}
and hence describes a relativistic second-order model on $\grp{B} \backslash \grp{D}$.

Let us now turn to the model of interest~\eqref{Eq:SolvS}.
In particular, we use the algebraic structures introduced in \ssecref{ssec:drindoub}.
We identify the bilinear form with that in eq.~\eqref{eq:mblmodel} and set
\begin{equation}
\grp{B} = \widetilde{\grp{G}}_\rho, \qquad \alg{b} = \tilde{\alg{g}}_\rho,
\end{equation}
where $\tilde{\alg{g}}_\rho$ is defined in eq.~\eqref{eq:defg}.
Writing a general element of the Drinfel'd double $\alg{d}$ \eqref{eq:dd} as $X + \iota Y$, $X,Y\in\alg{g}$,
the operator $\mathcal{E}$ is defined as
\begin{equation}\begin{gathered}
(\mathcal{E}\pm 1) (X + \iota Y) = (\mathrm{s}_\pm X + \mathrm{c}_+ Y) - \iota (\mathrm{s}_\mp Y + \mathrm{c}_- X) , \\
\mathrm{s}_\pm = \frac{\sinh c\rho \pm \sinh c\nu}{\sinh c\nu}, \qquad
\mathrm{c}_\pm = c^{\pm 1}\frac{\cosh c\rho \pm \cosh c\nu}{\sinh c\nu},
\end{gathered}\end{equation}
where $\nu$ is a free parameter that will eventually be related to the parameter $\chi$ of the action~\eqref{Eq:SolvS}.

Assuming that the decomposition \eqref{eq:dsgr} lifts to the group, \textit{i.e.} the quotient $\widetilde{\grp{G}}_\rho\backslash\grp{D}$ can be identified with $\grp{G}$, we parametrise
\begin{equation}
\gdsl = \tilde{g}_\rho \, g, \qquad g \in \grp{G}, \quad \tilde{g}_\rho \in \widetilde{\grp{G}}_\rho,
\end{equation}
and use the gauge symmetry \eqref{eq:gsym} to fix $\tilde{g}_\rho = 1$, \textit{i.e.}
\begin{equation}
\gdsl = g \in \grp{G}.
\end{equation}
Let us now determine the action of the projectors $\mathcal{E}\mathcal{P}(\mathcal{E}\pm1)$, which are defined by their image and kernel.
In the current setup these are given by $\im \mathcal{E} \mathcal{P}(\mathcal{E}\pm 1) = \Ad_g^{-1} \tilde{\alg{g}}_\rho$ and $\ker \mathcal{E} \mathcal{P}(\mathcal{E}\pm 1) = \alg{e}_\mp$.
We start by writing
\begin{equation}
\mathcal{E}\mathcal{P}(\mathcal{E}\pm1) = \Ad_g^{-1} \hat{\mathcal{P}} \Ad_g^{\vphantom{-1}} (\mathcal{E} \pm 1),
\end{equation}
where $\im \hat{\mathcal{P}} = \tilde{\alg{g}}_\rho$.
This automatically means that we have the required image and that $\alg{e}_\mp$ lies in the kernel.
From the commutation relations \eqref{Eq:ComIota} it follows that $\Ad_g^{\vphantom{-1}}$ commutes with $\iota$ and hence $\mathcal{E}$.
Therefore, we have
\begin{equation}
\mathcal{E}\mathcal{P}(\mathcal{E}\pm1) = \Ad_g^{-1} \hat{\mathcal{P}} (\mathcal{E} \pm 1) \Ad_g^{\vphantom{-1}}.
\end{equation}
The requirement that this is a projector (and that the full kernel is $\alg{e}_\mp$) can then be written in the simple form
\begin{equation}\label{eq:projcond}
\hat{\mathcal{P}} (\mathcal{E} \pm 1) \hat{\mathcal{P}} (\mathcal{E} \pm 1) = \hat{\mathcal{P}} (\mathcal{E} \pm 1) .
\end{equation}
Parametrising
\begin{equation}
\hat{\mathcal{P}} (X + \iota Y) = \hat{\sigma} \big(f_0(\hat{R}) X + f_1(\hat{R}) Y\big),
\qquad X,Y \in \alg{g},
\end{equation}
we find that the condition \eqref{eq:projcond} yields
\begin{equation}
f_0(\hat R) = \frac{1}{\mathrm{c}_+ + \mathrm{c}_- \hat{R}^2 + (\mathrm{s}_+ + \mathrm{s}_-)\hat{R}}, \qquad
f_1(\hat{R}) = -\frac{\hat{R}}{\mathrm{c}_+ + \mathrm{c}_- \hat{R}^2 + (\mathrm{s}_+ + \mathrm{s}_-)\hat{R}}.
\end{equation}
Finally, we arrive at the following expression for the action of the projectors $\mathcal{E}\mathcal{P}(\mathcal{E}\pm1)$
\begin{equation}\label{eq:actprojepe}
\mathcal{E}\mathcal{P}(\mathcal{E}\pm1) (X + \iota Y) = \Ad_g^{-1} \hat\sigma \frac{1}{\hat{R} + c \coth \frac{c}{2} (\rho \pm \nu)}
\Ad_g^{\vphantom{-1}} \Big(X + c\coth \frac{c}{2} (\rho \pm \nu) Y\Big)
, \qquad X,Y \in \alg{g}.
\end{equation}

To conclude, we fix $\gdsl = g$ in the action~\eqref{eq:doubact1}.
Using the action of the projectors in eq.~\eqref{eq:actprojepe} and the bilinear form~\eqref{eq:mblmodel} we find
\begin{equation}\begin{split}\label{eq:doubact2}
\mathsf{S} & = N \Big[\frac12\int d^2 x \; \kappa( g^{-1} \partial_+ g, \Big(\frac{\cosh c\rho + c^{-1}\sinh c\rho \hat{R}_g}{\hat{R}_g + c \coth \frac{c}{2}(\rho+\nu)} - \frac{\cosh c\rho + c^{-1}\sinh c\rho \trap\hat{R}_g}{\trap\hat{R}_g + c \coth \frac{c}{2}(\rho-\nu)} \Big) g^{-1}\partial_- g)
\\ & \hspace{75pt}
- \frac{1}{6} \frac{\sinh c\rho}{c} \int d^3 x \; \epsilon^{abc} \kappa( g^{-1}\partial_a g,[g^{-1}\partial_b g, g^{-1} \partial_c g] )\Big],
\end{split}\end{equation}
where $\hat{R}_g = \Ad_g^{-1} \hat{R} \Ad_g^{\vphantom{-1}}$.
Recalling that $\hat{R}$ is defined in terms of the skew-symmetric R-matrix $R$ in eq.~\eqref{eq:rhatr} and setting
\begin{equation}
N = - \frac{\kay c}{\sinh c\rho}, \qquad e^{c\nu} = \frac{e^\chi - e^{c\rho}}{e^\chi e^{c\rho} -1},
\end{equation}
it is straightforward to check that eq.~\eqref{eq:doubact2} indeed reproduces the action~\eqref{Eq:SolvS} as claimed.

\subsection{4-dimensional Chern-Simons theory}
\label{ssec:4dcs}

The models constructed in \secref{sec:solvable} can also be obtained from the 4-dimensional Chern-Simons theory proposed in~\cite{Costello:2013zra}.
Here we will explain how this is done following~\cite{Costello:2019tri,Vicedo:2019dej,Delduc:2019whp}.

\subsubsection{Real, complex and semi-abelian doubles}
\label{ssec:doubles}

We start by discussing the structure of the Drinfel'd double $\alg{d}$ introduced in \ssecref{ssec:drindoub} in more detail.
Let us recall that any element of $\alg{d}$ can be written as $X + \iota Y$ with $X,Y\in\g$ and that the Lie bracket of $\alg{d}$ in this parametrisation is given by \eqref{Eq:ComIota}.

\paragraph{Real double.} In the case $c=1$ the Drinfel'd double $\alg{d}$ is isomorphic to the real double $\g \oplus \g$.
This identification is given explicitly by the map
\begin{equation}
\phi_1:
\begin{array}{ccc}
\alg{d} & \longrightarrow & \g \oplus \g \\
X + \iota Y & \longmapsto & (X-Y,X+Y)
\end{array},
\end{equation}
which sends the Lie bracket \eqref{Eq:ComIota} to that of the direct sum $\g\oplus\g$.

Under the isomorphism $\phi_1$ the subalgebra $\g$ is identified with the diagonal subalgebra of $\g\oplus\g$, $\phi_1(\g) = \g_{\text{diag}} = \big\lbrace (X,X), \, X\in\g \big\rbrace$.
Moreover, the subalgebra $\tilde{\g}$ is mapped to
\begin{equation}
\phi_1(\tilde{\g}) = \big\lbrace (R_-X,R_+X), \, X\in\g \big\rbrace,
\end{equation}
while the image of the subalgebra $\tilde{\g}_\rho$, introduced in \ssecref{ssec:drindoub} as a deformation of $\tilde{\g}$, is
\begin{equation}\label{Eq:gtildeReal}
\phi_1(\tilde{\g}_\rho) = \big\lbrace (\hat{R}_-X,\hat{R}_+X), \, X\in\g \big\rbrace = \bigg\lbrace \Big( \frac{e^{\rho R} - e^{\rho}}{\sinh \rho} X,\frac{e^{\rho R} - e^{-\rho}}{\sinh \rho}X\Big), \, X\in\g \bigg\rbrace.
\end{equation}
The subalgebra $\tilde{\g}_\rho$ is Lagrangian with respect to the deformed invariant bilinear form $\llangle \cdot,\cdot \rrangle$ defined in eq.~\eqref{eq:mblmodel}.
This induces an invariant bilinear form on $\g\oplus\g$ through the isomorphism $\phi_1$, which, rescaling by $-2\kay/\sinh\rho$, reads
\begin{equation}
\llangle (X_1,Y_1), (X_2,Y_2) \rrangle_1 = \kay(\coth{\rho} - 1 ) \kappa(X_1,X_2) - \kay ( \coth{\rho} + 1 ) \kappa(Y_1,Y_2).
\end{equation}
By construction, the subalgebra $\phi_1(\tilde{\g}_\rho)$ of $\g\oplus\g$ is Lagrangian with respect to $\llangle \cdot,\cdot \rrangle_1$.
From eq.~\eqref{Eq:LevelsI} we see that, in terms of the levels $\ell_\pm$ of the Kac-Moody currents $\mathcal{J}_\pm$ introduced in \ssecref{ssec:HamInt}, this bilinear form can be written as
\begin{equation}\label{Eq:FormLevelsReal}
\llangle (X_1,Y_1), (X_2,Y_2) \rrangle_1 = \ell_+\, \kappa(X_1,X_2) + \ell_- \, \kappa(Y_1,Y_2).
\end{equation}

\paragraph{Complex double.} In the case $c=i$ the Drinfel'd double can be identified with the complex double $\g^\Complex$ through the isomorphism
\begin{equation}
\phi_i:
\begin{array}{ccc}
\alg{d} & \longrightarrow & \g^\Complex \\
X + \iota Y & \longmapsto & X-i Y
\end{array},
\end{equation}
which sends the Lie bracket \eqref{Eq:ComIota} to that of the complexification $\g^\Complex$.
We let $\tau$ denote the antilinear involutive automorphism of $\g^\Complex$ defined by $\tau: X+i Y \mapsto X-i Y$.

Under the isomorphism $\phi_i$, the subalgebra $\g$ is identified with the real form $\g$ in $\g^\Complex$, \textit{i.e.} the set of fixed-points of the automorphism $\tau$.
Moreover, the subalgebra $\tilde{\g}$ is mapped to
\begin{equation}
\phi_i(\tilde{\g}) = \lbrace R_-X, \; X\in\g \rbrace,
\end{equation}
while the image of the deformed subalgebra $\tilde{\g}_\rho$ is
\begin{equation}
\phi_i(\tilde{\g}_\rho) = \lbrace \hat{R}_-X, \, X\in\g \rbrace = \bigg\lbrace \frac{e^{\rho R} - e^{i\rho}}{\sin \rho} X, \, X\in\g \bigg\rbrace.
\end{equation}
This deformed subalgebra is Lagrangian with respect to the pullback by $\phi_i$ of the bilinear form $\llangle \cdot,\cdot \rrangle$, which, after rescaling by $-2\kay/\sin\rho$, is given by
\begin{equation}
\llangle X_1+i Y_1, X_2+ i Y_2 \rrangle_i = 2\kay \big( \kappa(Y_1,Y_2)-\kappa(X_1,X_2) \big) + 2\kay \cot\rho \big( \kappa(X_1,Y_2) - \kappa(Y_1,X_2) \big).
\end{equation}
In terms of the level $\ell_+$ of the complex Kac-Moody current $\mathcal{J}_+$ introduced in \ssecref{ssec:HamInt}, this invariant bilinear form on $\g^\Complex$ is given by
\begin{equation}\label{Eq:FormLevelComp}
\llangle X, Y \rrangle_i = 2 \operatorname{Re} \big( \ell_+\, \kappa(X,Y) \big), \qquad X,Y \in \g^\Complex.
\end{equation}
where $\kappa$ has been extended from $\g$ to $\g^\Complex$ by $\Complex$-bilinearity.
For comparison with eq.~\eqref{Eq:FormLevelsReal} in the split case, we note that eq.~\eqref{Eq:FormLevelComp} can also be written as
\begin{equation}
\llangle X, Y \big\rrangle_i = \ell_+\, \kappa(X,Y) + \ell_-\, \kappa(\tau X, \tau Y),
\end{equation}
where $\ell_-=\overline{\ell_+}$ is the level of the conjugate Kac-Moody current $\mathcal{J}_-=\tau(\mathcal{J}_+)$.

\paragraph{Semi-abelian double.} Finally, in the case $c=0$ the Drinfel'd double $\alg{d}$ is then isomorphic to the semi-abelian double $\g \ltimes \g^{\text{ab}}$, where $\g^{\text{ab}}$ denotes the vector space $\g$ equipped with the trivial Lie bracket (making it an abelian Lie algebra) and $\g$ acts on $\g^{\text{ab}}$ by the adjoint action.
This isomorphism is given by the map
\begin{equation}
\phi_0:
\begin{array}{ccc}
\alg{d} & \longrightarrow & \g \ltimes \g^{\text{ab}} \\
X + \iota Y & \longmapsto & (X,-\rho^{-1} \sinh\chi \, Y)
\end{array},
\end{equation}
which sends the Lie bracket \eqref{Eq:ComIota} to that of the semi-direct product $\g \ltimes \g^{\text{ab}}$.
Note that we have introduced the factor of $-\rho^{-1}\sinh\chi$ using the automorphism $(X,Y) \mapsto (X,aY)$ of the semi-abelian double.
Under the isomorphism $\phi_0$, the subalgebra $\g$ is identified with the subalgebra $\g \ltimes \lbrace 0 \rbrace$ of $\g \ltimes \g^{\text{ab}}$.
Moreover, the subalgebra $\tilde{\g}$ is mapped to
\begin{equation}
\phi_0(\tilde{\g}) = \bigg\lbrace \Big( RX,-\frac{\sinh\chi}{\rho}X \Big), \, X\in\g \bigg\rbrace,
\end{equation}
while the image of the deformed subalgebra $\tilde{\g}_\rho$ is
\begin{equation}
\phi_0(\tilde{\g}_\rho) = \bigg\lbrace \Big( \hat{R}X,-\frac{\sinh\chi}{\rho}X \Big), \, X\in\g \bigg\rbrace = \bigg\lbrace \Big( \frac{e^{\rho R}-1}{\rho}X, -\frac{\sinh\chi}{\rho}X \Big), \, X\in\g \bigg\rbrace.
\end{equation}
The pullback of the bilinear form $\llangle \cdot,\cdot \rrangle$ by $\phi_0$ defines an invariant bilinear form on $\g\ltimes\g^{\text{ab}}$ with respect to which $\phi_0(\tilde{\g}_\rho)$ is Lagrangian.
After rescaling by $-2\kay/\rho$, this bilinear form reads
\begin{equation}
\llangle (X_1,Y_1), (X_2,Y_2) \rrangle_0 = - 2\kay\, \kappa(X_1,X_2) + \frac{2\kay}{\sinh\chi} \big( \kappa(X_1,Y_2) + \kappa(Y_1,X_2) \big).
\end{equation}
In terms of the levels $\ell_0$ and $\ell_1$~\eqref{Eq:LevelsHom} of the two Takiff currents $\Jc_0$ and $\Jc_1$ characterising the integrable structure of the model in the homogeneous case, the above bilinear form can be rewritten as
\begin{equation}\label{Eq:FormLevelSemi}
\llangle (X_1,Y_1), (X_2,Y_2) \rrangle_0 = \ell_0\, \kappa(X_1,X_2) + \ell_1 \Big( \kappa(X_1,Y_2) + \kappa(Y_1,X_2) \Big).
\end{equation}
\noindent

\paragraph{Summary.} Let us summarise the results of this subsection.
In the three cases $c=1$, $c=i$ and $c=0$, the Drinfel'd double is mapped through the isomorphism $\phi_c$ to the real double $\g\oplus\g$, the complex double $\g^\Complex$ and the semi-abelian double $\g\ltimes\g^{\text{ab}}$ respectively.
This double can be written as the direct sum (as a vector space)
\begin{equation}\label{Eq:DoubleSum}
\phi_c(\alg{d}) = \phi_c(\g) \dotplus \phi_c(\tilde{\g}_\rho)
\end{equation}
of a subalgebra $\phi_c(\g)$, isomorphic to $\g$, and another subalgebra $\phi_c(\tilde{\g}_\rho)$.
The latter is Lagrangian with respect to an invariant bilinear form $\llangle\cdot,\cdot\rrangle_c$ on $\phi_c(\alg{d})$, which can be expressed in terms of the levels characterising the integrable structure of the model.
We will assume that in all three cases the decomposition \eqref{Eq:DoubleSum} lifts to the group and hence that $\phi_c(\grp{D})$ possesses the factorisation
\begin{equation}\label{Eq:DoubleFact}
\phi_c(\grp{D}) = \phi_c(\widetilde{\grp{G}}_\rho) \cdot \phi_c(\grp{G}).
\end{equation}
When $\grp{G}$ is a compact group and $c=i$, this factorisation corresponds to the Iwasawa decomposition of the complex double $\grp{G}^\Complex$ in the limit without WZ term.

\subsubsection{4-dimensional Chern-Simons theory}\label{ssec:2d}

In this subsection we review the 4-dimensional variant of Chern-Simons theory ($\mathrm{CS}_4$) initially proposed in~\cite{Costello:2013zra}, from which we will obtain the YB deformation of the PCM plus WZ term~\eqref{Eq:SolvS}.
This 4-dimensional theory is related to both integrable lattice models~\cite{Costello:2013sla,Witten:2016spx,Costello:2017dso,Costello:2018gyb} as well as integrable field theories in 2 dimensions~\cite{Costello:2019tri}.
The undeformed PCM plus WZ term was first obtained from $\mathrm{CS}_4$ in~\cite{Costello:2019tri} by introducing disorder defects.
Subsequently it was shown in~\cite{Vicedo:2019dej} that all integrable field theories obtained in this way satisfy a Maillet bracket with twist function and can be related to affine Gaudin models.
This was developed further in~\cite{Delduc:2019whp}\footnote{See also~\cite{Benini:2020skc} for a recent analysis of $\mathrm{CS}_4$ and its relation to 2-dimensional integrable models using homotopy theory.}, where various other integrable field theories were constructed from $\mathrm{CS}_4$, including the YB deformation of the PCM plus WZ term in the case where $R$ is the standard Drinfel'd-Jimbo R-matrix.
Later we will extend these results to the more general case of an R-matrix with solvable $\alg{h}_\pm$.

$\mathrm{CS}_4$ is defined on $\Real^2 \times \Projective^1$, where the real plane $\Real^2$ is described by coordinates $(x,t)$ and the Riemann sphere $\Projective^1$ by a complex coordinate $z$ and its conjugate $\bar z$.
It depends on a gauge field $A$, a $\g^{\Complex}$-valued 1-form on $\Real^2 \times \Projective^1$, restricted such that $A=A_x\, \dd x + A_t\, \dd t + A_{\bar z}\, \dd \bar z$.
The theory is further specified by the choice of a meromorphic 1-form $\omega = \vp(z)\dd z$.
Its action reads
\begin{equation}\label{Eq:CSAction}
\mathsf{S}_{4d} = \frac{i}{4\pi} \int_{\Real^2 \times \Projective^1} \omega \wedge C\!S(A),
\end{equation}
where $C\!S(A)=\kappa \big( A \; \overset{\wedge}{,} \; \dd A + \frac{2}{3} A \wedge A \big)$ is the standard Chern-Simons 3-form.
To ensure that the action~\eqref{Eq:CSAction} is real, we impose reality conditions on the gauge field $A$ and the 1-form $\omega$~\cite{Delduc:2019whp}.
More precisely, we ask that their pullback under complex conjugation $z\mapsto\bar z$ on $\Projective^1$ gives their complex conjugate in $\g^{\Complex}$ and $\Complex$ respectively.

\paragraph{Parametrisation of the gauge field.} We parametrise the component $A_{\bar z}$ of the gauge field as
\begin{equation}\label{Eq:Ghat}
A_{\bar z} = - \p_{\bar z} \gh \gh^{-1},
\end{equation}
where $\gh$ is a $\grp{G}^{\Complex}$-valued field on $\Real^2 \times \Projective^1$.
Let us note that $\gh$ is not uniquely determined; it is defined up to $\gh \to \gh h$, where $h$ is an arbitrary $\grp{G}^{\Complex}$-valued field on $\Real^2$, but independent of $z$ and $\bar z$.
We then parametrise the other components of the gauge field as
\begin{equation}\label{Eq:LaxCS}
A_x = \gh \Lc \gh^{-1} - \p_x \gh \gh^{-1}, \qquad A_t = \gh \mathcal{M} \gh^{-1} - \p_t \gh \gh^{-1} ,
\end{equation}
in terms of two $\g^{\Complex}$-valued fields $\Lc$ and $\mathcal{M}$ on $\Real^2 \times \Projective^1$.

\paragraph{Lax connection and twist function.} A 2-dimensional integrable structure naturally arises from $\mathrm{CS}_4$ when we parametrise the gauge field as in eqs.~\eqref{Eq:Ghat} and \eqref{Eq:LaxCS}.
Varying the gauge field $A$ in the action \eqref{Eq:CSAction}, the bulk equation of motion of $\mathrm{CS}_4$ is simply given by $\omega \wedge F(A)=0$, where $F(A)=dA + A \wedge A$ is the curvature of the gauge field.
In the parametrisation~\eqref{Eq:Ghat} and \eqref{Eq:LaxCS}, this equation of motion translates into three equations on $\Lc$ and $\mathcal{M}$
\begin{equation}
\vp(z) \, \p_{\bar z} \Lc = \vp(z) \, \p_{\bar z} \mathcal{M} = 0, \qquad \p_t \Lc - \p_x \mathcal{M} + \big[ \mathcal{M}, \Lc \big]=0,
\end{equation}
where we recall that $\omega=\vp(z)\dd z$.
The first two equations tell us that $\Lc$ and $\mathcal{M}$ are meromorphic functions of $z$, with poles at the zeroes of $\vp(z)$.
The last equation imposes the flatness of the connection $(\p_x+\Lc,\p_t+\mathcal{M})$.
Therefore, we find a 2-dimensional connection on $\Real^2$ that depends meromorphically on a complex parameter $z$ and which is flat on-shell.
These are the defining characteristics of a Lax connection of a 2-dimensional integrable field theory, with $(x,t)$ becoming the space-time coordinates of the 2-dimensional model, and the complex coordinate $z$ playing the role of the spectral parameter.
In this framework, the poles of $\Lc$ and $\mathcal{M}$ and thus of $A$, situated at the zeroes of $\vp(z)$, are referred to as disorder defects~\cite{Costello:2019tri}.

The proof of the integrability of these models was completed in~\cite{Vicedo:2019dej}, where it was shown that the Lax matrix $\Lc$ satisfies a Maillet bracket with twist function, with the latter given by the meromorphic function $\vp(z)$ parametrising $\omega$.
It then follows that the conserved charges extracted from the monodromy of this Lax matrix are in involution.
Since our goal is to construct the YB deformation of the PCM plus WZ term from $\mathrm{CS}_4$ we use this result as a guide and fix $\omega$ in terms of the twist function \eqref{Eq:Twist} to be
\begin{equation}\label{Eq:Omega}
\omega = \frac{2\xi (1-z^2)}{\left(z-\dfrac{\kay}{2\xi}\right)^2 - \dfrac{\sinh^2{\rho c}}{\sinh^2{\chi}}} \dd z.
\end{equation}
The zeroes of this twist function are located at $z=+1$ and $z=-1$, and hence the Lax connection $(\p_x+\Lc,\p_t+\mathcal{M})$ is meromorphic in $z$ with poles at these points.
We will take the poles at $z=+1$ and $z=-1$ to be in the light-cone components $\Lc_+ = \mathcal{M} + \Lc$ and $\Lc_- = \mathcal{M} - \Lc$ respectively.
This completely specifies the $z$-dependence of the Lax connection
\begin{equation}\label{Eq:LaxCSz}
\Lc_\pm(z,x,t) = \frac{V_\pm(x,t)}{1 \mp z} + U_\pm(x,t).
\end{equation}

\paragraph{Boundary conditions.}
In addition to the bulk equation of motion, varying the action \eqref{Eq:CSAction} with respect to $A$ also gives a boundary equation of motion on $A$ and its variation $\delta A$, which comes from the presence of poles in the 1-form $\omega$~\eqref{Eq:Omega}.
To deal with the double pole at $z = \infty$, the following Dirichlet boundary conditions are imposed (see~\cite{Costello:2019tri,Delduc:2019whp} for further details)
\begin{equation}\label{Eq:BoundInf}
A_x\big|_\infty = A_t\big|_\infty = 0.
\end{equation}
The nature of the other poles depends on the choice of the parameter $c$:
\begin{itemize}
\item if $c=1$, $\omega$ has a pair of real simple poles $z_\pm=\frac{\kay}{2\xi} \pm \frac{\sinh{\rho}}{\sinh{\chi}}$, with residues $\ell_\pm = \kay \left( \pm \coth{\rho} - 1 \right)$;
\item if $c=i$, $\omega$ has a pair of complex conjugate simple poles $z_\pm=\frac{\kay}{2\xi} \pm i\frac{\sin{\rho}}{\sinh{\chi}}$, with residues $\ell_\pm = \kay \left( \mp i \cot{\rho} - 1 \right)$;
\item if $c=0$, $\omega$ has a double pole at $z_0 = \kay/2\xi$, with coefficients $\ell_0 = \res_{z_0} \omega = -2\kay $ and $\ell_1 = \res_{z_0} (z-z_0)\omega = \frac{2\kay}{\sinh\chi}$.
\end{itemize}
These three cases give rise to different boundary equations on the components $A_\mu$, $\mu=x,t$, of the gauge field~\cite{Delduc:2019whp}.
Here we treat them in a uniform way using the formalism developed in \ssecref{ssec:doubles}.
This is achieved by introducing $\mathbb{A}_\mu$ defined as:
\unskip\footnote{In the cases $c=1$ and $c=0$, $A_\mu|_{z_\pm}$, $A_\mu|_{z_0}$ and $\p_z A_\mu|_{z_0}$ are valued in the real form $\g$ due to the reality condition imposed on $A$ and the reality of the poles $z_\pm$ and $z_0$.
Similarly, in the case $c=i$ we have $z_-=\overline{z_+}$ and $A_\mu|_{z_-}$ is the complex conjugate of $A_\mu|_{z_+}$.}
\begin{itemize}
\item for $c=1$ we let $\mathbb{A}_\mu = ( A_\mu|_{z_+}, A_\mu|_{z_-} )$, which belongs to the real double $\g\oplus\g$;
\item for $c=i$ we let $\mathbb{A}_\mu = A_\mu|_{z_+}$, which belongs to the complex double $\g^{\Complex}$;
\item for $c=0$ we let $\mathbb{A}_\mu = ( A_\mu|_{z_0}, \p_z A_\mu|_{z_0} )$, which belongs to the semi-abelian double $\g \ltimes \g^{\text{ab}}$.
\end{itemize}
Therefore, $\mathbb{A}_i$ belongs to the realisation $\phi_c(\alg{d})$ of the Drinfel'd double $\alg{d}$.
Recalling that $\phi_c(\alg{d})$ admits an invariant bilinear form $\llangle \cdot,\cdot \rrangle_c$, given in terms of the residues of $\omega$ by the eqs.~\eqref{Eq:FormLevelsReal}, \eqref{Eq:FormLevelComp} and \eqref{Eq:FormLevelSemi} for $c=1$, $c=i$ and $c=0$ respectively, the boundary equation of motion can then be written as
\begin{equation}
\epsilon^{\mu\nu}\,\llangle \mathbb{A}_\mu, \delta\mathbb{A}_\nu \rrangle_c = 0,
\end{equation}
where $\epsilon^{\mu\nu}$ denotes the 2-dimensional Levi-Civita symbol, with $\epsilon^{tx}=-\epsilon^{xt}=1$ and $\epsilon^{tt}=\epsilon^{xx}=0$.
This equation can be solved by demanding that $\mathbb{A}_\mu$ belongs to an isotropic subspace of $\phi_c(\alg{d})$.
In what follows we will impose the boundary condition
\begin{equation}\label{Eq:BoundDouble}
\mathbb{A}_x, \mathbb{A}_t \in \phi_c(\tilde{\g}_\rho).
\end{equation}

\paragraph{Gauge symmetry and 2-dimensional degrees of freedom.}
To identify the dynamical fields of the 2-dimensional integrable model we observe that the bulk equation of motion $F(A)=0$ is invariant under the local transformation
\begin{equation}\label{Eq:Gauge}
d+A \longmapsto u(d+A)u^{-1} = d + A^u,
\end{equation}
for any $u: \Real^2 \times \Projective^1 \rightarrow \grp{G}^{\Complex}$, with the curvature $F(A)$ transforming covariantly, \textit{i.e.} $F(A^u)=uF(A)u^{-1}$.
Not all such transformations are admissible gauge symmetries since, in addition to preserving the bulk equation of motion, they also need to preserve the boundary conditions imposed on the gauge field.

Under the gauge transformation \eqref{Eq:Gauge} the field $\gh$ parametrising $A_{\bar z}$~\eqref{Eq:Ghat} transforms as $\gh \mapsto u\gh$.
This allows us to eliminate almost all degrees of freedom in $\gh$.
In particular, if $y\in\Projective^1$ is not a pole of $\omega$, it is always possible to find a gauge transformation that sets the evaluation $\gh|_y$ to the identity of $\grp{G}$.
Moreover, this can be done while preserving the boundary conditions since these only involve fields evaluated at the poles of $\omega$.
Similarly, one can also eliminate the evaluations of all derivatives $\p_z^k \gh|_y$ at points $y$ which are not poles of $\omega$.

Schematically, this tells us that the physical degrees of freedom contained in $\gh$ are ``located'' at the poles $p$ of $\omega$ and can be extracted from the evaluations $\gh|_p$ and $\p_z^k\gh|_p$.
Some of these derivatives can also be ``gauged'' away by gauge transformations~\eqref{Eq:Gauge} that are non-trivial in a neighbourhood of $p$ but respect the boundary condition at $p$.
For instance, the boundary condition \eqref{Eq:BoundInf} imposed at infinity is preserved by any gauge transformation $A \mapsto A^u$ such that $u|_\infty$ is a constant field on $\Real^2$.
This allows us to bring $\gh$ to a form such that it is constant in a neighbourhood of $\infty$~\cite{Costello:2019tri} (see also~\cite{Delduc:2019whp}) and hence the only physical degree of freedom located at infinity is the evaluation $\gh|_\infty$.

A similar analysis for the other poles $z_\pm$ or $z_0$ of $\omega$ was performed in~\cite{Delduc:2019whp}.
To summarise, we first introduce the following notation:
\begin{itemize}
\item if $c=1$, we define $\gdsl = \big( \gh|_{z_+}, \gh|_{z_-} \big)$, which is valued in the real double $\grp{G} \times \grp{G}$;
\item if $c=i$, we define $\gdsl = \gh|_{z_+}$, which is valued in the complex double $\grp{G}^{\Complex}$;
\item if $c=0$, we define $\gdsl = \big( \gh|_{z_0}, \p_z \gh \gh^{-1}|_{z_0} \big)$, which is valued in the semi-abelian double $\grp{G} \ltimes \g$.
\unskip\footnote{In the semi-abelian double $\grp{G} \ltimes \g$, $\g$ is seen as an abelian group equipped with addition, on which $\grp{G}$ acts by the adjoint action.}
\end{itemize}
The field $\gdsl$ is valued in the realisation $\phi_c(\grp{D})$ of the Drinfel'd double $\grp{D}$.
In the same way, we also define $\udsl: \Real^2 \rightarrow \phi_c(\grp{D})$ in terms of $u$, such that under gauge transformations~\eqref{Eq:Gauge}, $\mathbb{A}$ and $\gdsl$ transform as
\begin{equation}\label{Eq:GaugeDouble}
\gdsl \longmapsto \udsl \gdsl,\qquad
\mathbb{A}_\mu \longmapsto \udsl\, \mathbb{A}_\mu \udsl^{-1} - \p_\mu \udsl \udsl^{-1},
\end{equation}
where the product is understood in $\phi_c(\grp{D})$ and the adjoint action is that of $\phi_c(\grp{D})$ on its Lie algebra $\phi_c(\alg{d})$.
In order to preserve the boundary condition \eqref{Eq:BoundDouble}, it is clear that $\udsl$ should belong to the subgroup $\phi_c(\widetilde{\grp{G}}_\rho)$ corresponding to the Lagrangian subalgebra $\phi_c(\tilde{\g}_\rho)$.
\unskip\footnote{Note that it is for this reason that $\mathbb{A}_\mu$ should belong to an isotropic subalgebra of $\phi_c(\alg{d})$ and not any isotropic subset. In particular, the subset should be stable under transformations of the form \eqref{Eq:GaugeDouble} for some well-chosen $\udsl$.}
The allowed gauge transformations on $\gdsl$ are thus $\gdsl \mapsto \udsl \gdsl$, for $\udsl \in \phi_c(\widetilde{\grp{G}}_\rho)$, and the physical degrees of freedom in $\gdsl$ are valued in the quotient $\phi_c(\widetilde{\grp{G}}_\rho)\backslash \phi_c(\grp{D})$.
Assuming the factorisation \eqref{Eq:DoubleFact}, this quotient can be parametrised by fixing $\gdsl \in \phi_c(\grp{G})$.
The physical field extracted from $\gdsl$ is thus simply a $\grp{G}$-valued field $g$, such that:
\begin{itemize}
\item if $c=1$, $\gdsl=(g,g)$ is in the diagonal subgroup $\phi_1(\grp{G})=\grp{G}^{\text{diag}}$ of $\phi_1(\grp{D})=\grp{G} \times \grp{G}$;
\item if $c=i$, $\gdsl=g$ is in the real form $\phi_i(\grp{G})=\grp{G}$ of $\phi_i(\grp{D})=\grp{G}^{\Complex}$;
\item if $c=0$, $\gdsl=(g,0)$ is in the subgroup $\phi_0(\grp{G})=\grp{G} \ltimes \lbrace 0 \rbrace$ of $\phi_0(\grp{D}) = \grp{G} \ltimes \g$.
\end{itemize}
In the first two cases, we have $g=\gh|_{z_+}=\gh|_{z_-}$ and in the third, $g=\gh|_{z_0}$.
We can then choose a gauge where $\gh$ is constant equal to $g$ in a neighbourhood of $z_\pm$ or $z_0$, such that the archipelago conditions are satisfied~\cite{Delduc:2019whp}.

Thus far, we have seen that the physical degrees of freedom of the 4-dimensional field $\gh : \Real^2 \times \Projective^1 \to \grp{G}^{\Complex}$ are two 2-dimensional fields $g: \Real^2 \to \grp{G}$ and $\gh|_\infty: \Real^2 \to \grp{G}$, attached to the poles of $\omega$.
However, let us recall that $\gh$ is only defined~\eqref{Eq:Ghat} up to $\gh \mapsto \gh h$, where $h: \Real^2 \to \grp{G}^{\Complex}$ is independent of $z$ and $\bar{z}$.
This freedom can be used to eliminate one of the two fields.
In what follows, we will choose to fix $\gh|_\infty=\Id$, such that we are left with a single $\grp{G}$-valued field $g$.

\subsubsection{Equivalence with the YB deformation of the PCM plus WZ term}

To conclude, we demonstrate that the model following from $\mathrm{CS}_4$ is equivalent to the YB deformation of the PCM plus WZ term constructed in \secref{sec:solvable}.

\paragraph{Determining the Lax connection.} Having discussed the physical degrees of freedom of $\gh$, or equivalently $A_{\bar{z}}$, let us now turn to the remaining components of the gauge field, $A_x$ and $A_t$.
In eq.~\eqref{Eq:LaxCS} we parametrised these components in terms of $\gh$ and the $\g^\Complex$-valued fields $\Lc$ and $\mathcal{M}$, which determine the Lax connection of the 2-dimensional integrable field theory.
Under the gauge transformation~\eqref{Eq:Gauge} the components $(\Lc,\mathcal{M})$ of the Lax connection are invariant.
As we will now explain, these components can be expressed in terms of the field $g$, \textit{i.e.} the only physical degree of freedom of $\gh$ that cannot be eliminated by gauge transformations.

To determine the light-cone components of the Lax connection $\Lc_\pm=\mathcal{M}\pm \Lc$ we start from the form derived in eq.~\eqref{Eq:LaxCSz}, which makes the meromorphic dependence on $z$ manifest.
First we consider the boundary condition at infinity~\eqref{Eq:BoundInf}, which can be written $A_\pm|_{\infty}=0$.
Evaluating eq.~\eqref{Eq:LaxCS} at $z=\infty$ and using that we have fixed $\gh|_\infty=\Id$, we find that $A_\pm|_\infty = \Lc_\pm|_{\infty}$.
From the form~\eqref{Eq:LaxCSz}, it then follows that $A_\pm|_\infty = U_\pm$ and hence $U_\pm=0$.

To determine $V_\pm$ we use the second boundary condition \eqref{Eq:BoundDouble}.
Considering the case $c=1$, this boundary condition can be written as $\mathbb{A}_\mu = \big( A_\mu|_{z_+}, A_\mu|_{z_-} \big) \in \phi_1(\tilde{\g}_\rho)$.
From eq.~\eqref{Eq:LaxCS} and using $\gh|_{z_+} = \gh|_{z_-} = g$, we have that
\begin{equation}
A_\mu|_{z_\pm} = g\,\Lc_\mu|_{z_\pm}g^{-1} - \p_\mu g g^{-1} = \Lc^g_\mu |_{z_\pm},
\end{equation}
where $\Lc^g_\mu$ denotes the formal gauge transformation of the Lax connection.
Substituting in eq.~\eqref{Eq:LaxCSz} with $U_\pm = 0$ gives the expressions
\begin{equation}\label{Eq:APoles}
A_+|_{z_\pm} = \frac{g V_+ g^{-1}}{1 - z_\pm} - \p_+g g^{-1}, \qquad A_-|_{z_\pm} = \frac{g V_- g^{-1}}{1 + z_\pm} - \p_-g g^{-1}.
\end{equation}
Now demanding that $( A_\mu|_{z_+}, A_\mu|_{z_-} )$ belongs to $\phi_1(\tilde{\g}_\rho)$~\eqref{Eq:gtildeReal}, \textit{i.e.}
\begin{equation}
\big( e^{\rho R} - e^{-\rho}\big) A_\mu|_{z_+} = \big( e^{\rho R} - e^{\rho} \big)A_\mu|_{z_-},
\end{equation}
we find the following equations for $V_\pm$
\begin{equation}
\frac{1}{2\sinh \rho} \Big( \frac{e^{\rho R_g}-e^{-\rho}}{1 \mp z_+} - \frac{e^{\rho R_g}-e^{\rho}}{1 \mp z_-} \Big) V_\pm = j_\pm,
\end{equation}
where $j_\pm = g^{-1}\p_\pm g$.
Using $z_\pm=\frac{\kay}{2\xi} \pm \frac{\sinh{\rho}}{\sinh{\chi}}$ it is straightforward to see that this implies $\mathcal{Q}_g^\pm V_\pm = j_\pm$, with $\mathcal{Q}_\pm$ defined in eq.~\eqref{Eq:QSolv}, and hence that $V_\pm=K_\pm$, using eq.~\eqref{intq}.

A similar analysis can be carried out for the $c=i$ and $c=0$ cases, also leading to $V_\pm=K_\pm$.
In all three cases, the boundary condition~\eqref{Eq:BoundDouble} can be interpreted as a condition on the gauge transformed Lax connection $\Lc_\pm^g$ evaluated at the poles of $\omega$.
More precisely, we find that $\big(\Lc_\pm^g|_{z_+},\Lc_\pm^g|_{z_-}\big)$ for $c=1$, $\Lc_\pm^g|_{z_+}$ for $c=i$ and $\big(\Lc_\pm^g|_{z_0}, \p_z\Lc_\pm^g|_{z_0} \big)$ for $c=0$ belong to the Lagrangian subalgebra $\phi_c(\tilde{\g}_\rho)$.
To conclude, the Lax connection is given by $\Lc_\pm = \frac{K_\pm}{1\mp z}$, which agrees with that of the YB deformed PCM plus WZ term constructed in \secref{sec:solvable}.

\paragraph{2-dimensional action.} The final step is to show that the action following from $\mathrm{CS}_4$ coincides with that of the YB deformed PCM plus WZ term.
In~\cite{Delduc:2019whp} it was shown in general how to recast the 4-dimensional action \eqref{Eq:CSAction} as a 2-dimensional action under the assumption that the field $\gh$ satisfies the archipelago conditions.
Denoting the set of finite poles of $\omega$ by $P$, \textit{i.e.} $P=\lbrace z_+, z_- \rbrace$ if $c=1$ or $c=i$ and $P=\lbrace z_0 \rbrace$ if $c=0$, and recalling that in all three cases, $\gh|_p=g$ for all $p\in P$, the action of the 2-dimensional model is given by~\cite{Delduc:2019whp}
\unskip\footnote{This follows from Theorem 3.2 of~\cite{Delduc:2019whp}. Note that the contribution from the pole at $\infty$ vanishes as we have chosen to set $\gh|_\infty = \Id$.}
\begin{equation}\begin{split}
\mathsf{S} &= \frac{1}{4}\sum_{p\in P} \int \dd^2 x \; \big( \kappa ( \res_{p} \omega\Lc_+, j_- ) - \kappa ( \res_p \omega\Lc_-, j_+ ) \big) \\
& \qquad - \frac{1}{12}\Big( \sum_{p\in P} \res_p \omega \Big) \int d^3 x \; \epsilon^{abc} \kappa(g^{-1}\partial_a g, [g^{-1}\partial_b g , g^{-1} \partial_c g]).
\end{split}\end{equation}
A direct computation shows that this coincides with the action \eqref{Eq:SolvS} of the YB deformed PCM plus WZ term as claimed.

\subsection{Homogeneous R-matrices and non-abelian T-duality}
\label{ssec:NATD}

YB deformations of the PCM (as well as the symmetric space and semi-symmetric space sigma models) based on homogeneous R-matrices ($c=0$) are known to be equivalent to the addition of a closed B-field term and non-abelian T-duality~\cite{Hoare:2016wsk,Borsato:2016pas,Borsato:2017qsx}.
This relation has been used to generalise homogeneous YB deformations to other sigma models including the WZW model~\cite{Borsato:2018idb,Borsato:2018spz}.
\unskip\footnote{This construction can be implemented without complications when the metric and B-field of the sigma model are invariant under the action of the algebra $\alg{h}$ used in the non-abelian T-duality transformation~\cite{Hull:1989jk,Jack:1989ne}.
In the presence of the WZ term this leads to the condition that $\kappa([X,Y],Z)$ with $X,Y,Z \in \alg{h}$ is exact in the $\alg{h}$-cohomology, which is trivially satisfied if $\alg{h}$ is solvable.}$^{,}$
\unskip\footnote{To construct non-trivial deformations of the WZW model one needs to consider deformations that mix the left and right symmetries \cite{Borsato:2018spz}. This is consistent with the fact that the YB deformation of the PCM plus WZ term~\eqref{Eq:SolvS}, which preserves the right-acting $\grp{G}$-symmetry, is trivial in the limit $\chi\to+\infty$.}.
Here we will demonstrate that this prescription coincides with the YB deformation of the PCM plus WZ term for solvable $\alg{h}$ defined by the action~\eqref{Eq:SolvS}, \textit{i.e.} for homogeneous R-matrices this model is equivalent to the addition of a closed B-field term and non-abelian T-duality.

\paragraph{Alternative R-matrix.}
Since we are working with homogeneous R-matrices we have $c=0$ and $\alg{h}_+ = \alg{h}_- = \alg{h}$.
The action \eqref{Eq:SolvS} can be simplified by noting that if $R$ is a skew-symmetric solution of the cYBE with $\im R = \alg{h}$ solvable, then
\begin{equation}\label{eq:rb}
\frac{1}{2\hayb} \etab \Rb = \frac{1}{\kay} \frac{e^{\rho R} - 1}{e^{\rho R} +1},
\end{equation}
is also a skew-symmetric solution of the cYBE with $\im \Rb = \im R = \alg{h}$ solvable and $\ker \Rb = \ker R$.
\unskip\footnote{It follows that \eqref{eq:rb} defines a map between two R-matrices in the same subspace of skew-symmetric solutions to the cYBE, where this subspace is specified by the image and kernel of the R-matrices. Therefore, for a rank-2 r-matrix $R$ and $\Rb$ are proportional, while for higher ranks the relation will be more involved.}
This can be seen by using eq.~\eqref{Eq:IdExp} to derive the identity
\begin{equation}\begin{split}
\Big[\frac{e^{\rho R} - 1}{e^{\rho R} + 1}X,\frac{e^{\rho R} - 1}{e^{\rho R} + 1} Y\Big]
& - \frac{e^{\rho R} - 1}{e^{\rho R} + 1} \Big(\Big[X,\frac{e^{\rho R} - 1}{e^{\rho R} + 1}Y\Big] + \Big[\frac{e^{\rho R} - 1}{e^{\rho R} + 1}X,Y\Big]\Big)
\\ & = - \frac{(e^{\rho R} - 1)^2}{e^{\rho R} + 1}\Big[\frac{1}{e^{\rho R} + 1} X,\frac{1}{e^{\rho R} + 1}Y\Big]_{e^{\rho R} - 1}
\qquad \forall ~ X,Y \in \alg{g},
\end{split}\end{equation}
and then showing that the right-hand side is zero using the identity~\eqref{Eq:RRExp} with $c=0$.

The skew-symmetry, image and kernel of $\Rb$ follow straightforwardly from the analogous properties of $R$.
Also defining
\begin{equation}\label{eq:hay}
\hayb = \frac{\kay}{2} \coth \frac{\chi}{2},
\end{equation}
we can then rewrite the action \eqref{Eq:SolvS} in terms of $\etab \Rb$ and $\hayb$
\unskip\footnote{The parameters $\etab$ and $\hayb$ defined in eqs.~\eqref{eq:rb} and \eqref{eq:hay} coincide with $\eta$ and $\hay$ of eq.~\eqref{eq:chirho} in the $\kay \to 0$ limit.}
\begin{equation}\label{Eq:SolvSH}
\mathsf{S} = \int d^2x \; \kappa \Big(g^{-1}\partial_+ g, \frac{\hayb^2 - \frac{\kay^2}{4}\etab \Rb_g}{\hayb(1-\etab\Rb_g)} g^{-1}\partial_- g \Big) + \frac{\kay}{6} \int d^3 x \; \epsilon^{abc} \kappa(g^{-1}\partial_a g, [g^{-1}\partial_b g , g^{-1} \partial_c g]),
\end{equation}
where $\Rb_g = \Ad_g^{-1} \Rb \Ad_g^{\vphantom{-1}}$.

\paragraph{2-cocycle.} To see that this action can be found by non-abelian T-duality, we start by recapping the relation between solutions of the cYBE and non-degenerate 2-cocycles on $\alg{h}$~\cite{Belavin:1982,Belavin:1984,Stolin:1991n}.
Introducing $\alg{h}^\star$, the dual vector space to $\alg{h}$, we identify it with a subspace of $\alg{g}$ using the Killing form, \textit{i.e.} $\alg{h}^\star \subset \alg{g}$ such that $\kappa(\alg{h}^\star,\alg{h})$ is non-degenerate.
This identification is not unique and the following construction works for any choice.
Let us denote $\alg{p}^\star = \ker \Rb = \ker \kappa(\alg{h},\cdot)$ such that $\alg{g} = \alg{h}^\star \dotp \alg{p}^\star$, and the corresponding projector onto $\alg{h}^\star$ as $\trap P$.
It will also be useful to introduce $\alg{p} = \ker \kappa(\alg{h}^\star,\cdot)$ such that $\alg{g} = \alg{h} \dotp \alg{p}$, and $P$, the corresponding projector onto $\alg{h}$.

Recalling that $\im \Rb = \alg{h}$ and $\ker \Rb = \alg{p}^\star$, the restriction of $\Rb$ to $\alg{h}^\star$ has an inverse, which we denote
\begin{equation}\label{ommap}
\bar\omega = (P\,\Rb \,\trap P)^{-1} : \alg{h} \to \alg{h}^\star ,
\end{equation}
with $\tra\bar\omega = - \bar\omega$.
Defining the corresponding 2-cochain on $\alg{h}$
\begin{equation}
\omega(X,Y) = \kappa(X,\bar\omega Y) , \qquad \forall ~ X,Y\in\alg{h},
\end{equation}
the condition that this is a 2-cocycle, \text{i.e.} $d\omega = 0$, is equivalent to the cYBE for $\Rb$.
Note that, while the map $\bar\omega: \alg{h} \to \alg{h}^\star$ depends on the choice of $\alg{h}^\star$, the 2-cocycle itself does not.
We extend the map $\bar\omega$ to act on $\alg{g}$ by setting $\ker \bar\omega = \alg{p}$, and hence
\begin{equation}\label{rominv}
\bar R \bar\omega = P , \qquad \bar\omega \bar R = \trap P .
\end{equation}

Given that $\alg{h}$ is a subalgebra of $\alg{g}$
we have that for any linear operator $\bar{\mathcal{O}}:\alg{h} \to \alg{h}$
\begin{equation}\begin{aligned}\label{psx}
P\bar{\mathcal{O}} P & = \bar{\mathcal{O}} P, \qquad & \trap P \bar{\mathcal{O}} \trap P & = \trap P \bar{\mathcal{O}}.
\end{aligned}\end{equation}
This includes, in particular, taking $\bar{\mathcal{O}}$ to be $\ad_X$, $\Ad_h$, $\Rb$ or $\Rb_h$ where $X\in \alg{h}$ and $ h\in \grp{H}$ with the Lie group $\grp{H}$ is defined via the exponential map, \textit{i.e.} $\exp:\alg{h}\to\grp{H}$.
Together with the cocycle condition written in the form
\begin{equation}
\bar\omega[X,Y] = \trap P ([\bar\omega X,Y] + [X,\bar\omega Y]), \qquad X,Y \in \alg{h},
\end{equation}
this implies that $\bar\omega$ acts as a derivative followed by a projection onto $\alg{h}^\star$ when acting on commutators of $\alg{h}$.
This can be used to make sense of expressions such as $\trap P (h^{-1} \bar\omega h)$ and $\trap P(\bar\omega h h^{-1})$ where $h \in \grp{H}$~\cite{Borsato:2018idb}.
In particular, parametrising $h = \exp \lh$, $\lh \in \alg{h}$, we have
\begin{equation}\label{eq:omcdef}
\trap P(h^{-1} \bar\omega h) = \trap P\frac{1-e^{-\ad_\lh}}{\ad_\lh} \bar\omega \lh ,
\qquad
\trap P(\bar\omega h h^{-1}) = \trap P \frac{e^{\ad_\lh -1}}{\ad_\lh} \bar\omega \lh .
\end{equation}
Furthermore, noting that $[\p_\pm,\bar\omega] = 0$ acting on $\alg{h}$-valued fields, we have the following identities
\begin{equation}\begin{gathered}\label{eq:omegaid}
\bar\omega (h^{-1}\partial_\pm h) - \trap P \partial_\pm (h^{-1} \bar\omega h) + \trap P[h^{-1}\bar\omega h,h^{-1}\partial_\pm h] = 0 ,
\\
\bar\omega (\partial_\pm hh^{-1}) - \trap P \partial_\pm (\bar\omega h h^{-1}) - \trap P[\bar\omega h h^{-1} , \partial_\pm h h^{-1}] = 0 ,
\\
\trap P \Ad_h^{-1} \bar\omega \Ad_h^{\vphantom{-1}} P = \trap P(\bar\omega + \ad_{h^{-1}\bar\omega h}) P ,
\\
\trap P \Ad_h^{\vphantom{-1}} \bar\omega \Ad_h^{-1} P = \trap P (\bar\omega - \ad_{\bar\omega h h^{-1}}) P ,
\end{gathered}\end{equation}
where we have left the projector $\trap P$ acting on $\bar\omega h h^{-1}$ or $h^{-1} \bar\omega h$ implicit when it follows from the relations \eqref{psx}.

\paragraph{Non-abelian T-duality.}
Adapting the results of~\cite{Borsato:2016pas,Borsato:2017qsx} to the presence of the WZ term, let us now outline the non-abelian T-duality transformation that can be used to find the action \eqref{Eq:SolvSH}.
Our starting point is the action for the PCM plus WZ term
\begin{equation}\begin{split}\label{eq:act1}
\mathsf{S}_0 & = \hayb \int d^2x \; \kappa (g^{-1}\partial_+ g, g^{-1}\partial_- g ) + \frac{\kay}{6} \int d^3 x \; \epsilon^{abc} \kappa(g^{-1}\partial_a g, [g^{-1}\partial_b g , g^{-1} \partial_c g]) .
\end{split}\end{equation}
As the action \eqref{Eq:SolvSH} is invariant under right multiplication, we will non-abelian T-dualise in the left-acting $\grp{H}$-symmetry.
To this end we redefine $g \to h g$ with $h \in \grp{H}$ and $g \in \grp{G}$.
To compensate for the additional degrees of freedom this introduces a $\grp{H}$ gauge symmetry
\begin{equation}\label{eq:hgauge}
h \longmapsto h h', \qquad g\longmapsto h'^{-1} g, \qquad h' \in \grp{H}.
\end{equation}

Given that $h^{-1}\p_\pm h \in \alg{h}$ and $\omega$ is a 2-cocycle on $\alg{h}$, adding
\begin{equation}
- \hayb\etab^{-1} \int d^2x \; \omega(h^{-1}\p_+h,h^{-1}\p_-h),
\end{equation}
to \eqref{eq:act1} contributes a closed B-field term, which is locally a total derivative and hence does not modify the equations of motion.
Introducing $l_\pm = h^{-1}\p_\pm h$ and $k_\pm = \p_\pm g g^{-1}$ we arrive at
\begin{align}\nn
\mathsf{S}_0 & = \int d^2x \; \big[\hayb\kappa(l_+,l_-) + (\hayb-\frac{\kay}{2})\kappa(l_+,k_-) + (\hayb+\frac{\kay}{2}) \kappa(k_+,l_-)
+\hayb\kappa(k_+,k_-) - \hayb \etab^{-1} \omega(l_+,l_-) \big]
\\ & \qquad + \frac{\kay}{6} \int d^3 x \; \epsilon^{abc} \kappa(k_a,[k_b,k_c]) ,\label{eq:act2}
\end{align}
where we have used that $\kappa(l_a,[l_b,l_c]) = 0$ since $l_a \in \alg{h}$ and $\alg{h}$ is solvable by assumption.

To non-abelian T-dualise we gauge the left-acting $\grp{H}$-symmetry and fix $h = 1$.
To compensate we introduce a Lagrange multiplier, $v \in \alg{h}^\star$, imposing that the gauge field, $A_\pm \in \alg{h}$, has vanishing field strength
\begin{align}\nn
\bar{\mathsf{S}} & = \int d^2x \; \big[\hayb\kappa(A_+,A_-) + (\hayb-\frac{\kay}{2})\kappa(A_+,k_-) + (\hayb+\frac{\kay}{2}) \kappa(k_+,A_-)
+\hayb\kappa(k_+,k_-) -\hayb \etab^{-1} \omega(A_+,A_-) \big]
\\ & \qquad + \frac{\kay}{6} \int d^3 x \; \epsilon^{abc} \kappa(k_a,[k_b,k_c]) + \int d^2x \;
\kappa(v , \partial_+ A_- - \partial_- A_+ + [A_+,A_-]) .\label{eq:act3}
\end{align}
Integrating out the Lagrange multiplier we recover the action \eqref{eq:act2}.
On the other hand the non-abelian T-dual model is found by integrating out the gauge field
\begin{equation}\begin{split}\label{eq:act4}
\bar{\mathsf{S}} & =
\int d^2 x\; \big[\hayb \kappa(k_+,k_-) +
\kappa\big(\partial_+v - (\hayb+\frac{\kay}{2}) k_+, \mathcal{M}_-^{-1} (\partial_-v + (\hayb-\frac{\kay}{2}) k_-)\big)\big]
\\ & \qquad + \frac{\kay}{6} \int d^3 x \; \epsilon^{abc} \kappa(k_a,[k_b,k_c]),
\end{split}\end{equation}
where
\begin{equation}
\mathcal{M}_\pm = \trap P(\hay \pm \hay\etab^{-1}\bar\omega \pm \ad_v) P : \alg{g} \to \alg{h}^\star, \qquad \ker \mathcal{M}_\pm = \alg{p}, \qquad \mathcal{M}_+ = \tra \mathcal{M}_- ,
\end{equation}
and their inverses are defined such that
\begin{equation}
\mathcal{M}_\pm^{-1} : \alg{g} \to \alg{h} , \qquad \ker \mathcal{M}_\pm^{-1} = \alg{p}^\star, \qquad \mathcal{M}_\pm^{-1}\mathcal{M}_\pm = P, \qquad \mathcal{M}_\pm\mathcal{M}_\pm^{-1} = \trap P.
\end{equation}
Let us note that the actions \eqref{eq:act3} and \eqref{eq:act4} are invariant under the gauge symmetry \eqref{eq:hgauge}, up to a closed B-field term, if $v$ transforms as
\begin{equation}\label{eq:hgauge2}
v \longmapsto \trap P\Ad_{h'}\big(v+ \hayb\etab^{-1} (\bar\omega h' h'^{-1})\big),
\end{equation}
where $\trap P (\bar\omega h' h'^{-1})$ is defined in eq.~\eqref{eq:omcdef}.

\paragraph{Demonstration of equivalence.}
To demonstrate that \eqref{eq:act4} is equal to \eqref{Eq:SolvSH} up to a closed B-field term we first redefine $g \to h g$ in \eqref{Eq:SolvSH} to give
\begin{equation}\begin{split}\label{Eq:SolvSHex}
\mathsf{S} & = \int d^2x \; \Big[ \kappa \Big( (l_+ + k_+), \frac{\hayb^2 - \frac{\kay^2}{4}\etab \Rb_h}{\hayb(1-\etab\Rb_h)} (l_-+k_-) \Big)
+ \frac{\kay}{2}(\kappa(k_+,l_-) - \kappa(l_+,k_-) ) \Big]
\\ & \qquad + \frac{\kay}{6} \int d^3 x \; \epsilon^{abc} \kappa(k_a,[k_b,k_c]),
\end{split}\end{equation}
where $\Rb_h = \Ad_h^{-1} \Rb \Ad_h^{\vphantom{-1}}$.
Comparing to \eqref{eq:act4} we see that if
\begin{subequations}
\label{eqi}
\begin{gather}
\label{eqi1}
\frac{\hayb^2 - \frac{\kay^2}{4}\etab \Rb_h}{\hayb(1-\etab\Rb_h)} = \hayb - (\hayb^2 - \frac{\kay^2}{4}) \mathcal{M}_-^{-1} ,
\quad \
\Big(\frac{\hayb^2 \pm \frac{\kay^2}{4}\etab \Rb_h}{\hayb(1\pm\etab\Rb_h)} \mp \frac{\kay}{2}\Big)l_\pm = \pm (\hayb \mp \frac{\kay}{2}) \mathcal{M}_\pm^{-1}\partial_\pm v,\ \
\\
\label{eqi3}
\kappa\Big(l_+, \frac{\hayb^2 - \frac{\kay^2}{4}\etab \Rb_h}{\hayb(1-\etab\Rb_h)}l_-\Big) - \kappa(\partial_+ v ,\mathcal{M}_-^{-1}\partial_-v) = \text{closed B-field term},
\end{gather}\end{subequations}
then we have the desired result.

To solve the system of equations \eqref{eqi}, we first rearrange \eqref{eqi1} to give
\unskip\footnote{Intermediate steps in these rearrangements are given by
\begin{equation*}\begin{gathered}
\text{first eq.~in \eqref{eqi1}}
\ \ \Leftrightarrow \ \
\mathcal{M}_\pm^{-1} = \pm \frac{\etab \Rb_h}{\hayb(1\pm \etab \Rb_h)}
\ \ \Leftrightarrow \ \
\mathcal{M}_\pm = \trap P(\hayb \pm \hayb \etab^{-1}\bar\omega_h) P
\ \ \Leftrightarrow \ \
\text{first eq.~in \eqref{eqi22}},
\\
\text{second eq.~in \eqref{eqi1}}
\ \ \Leftrightarrow \ \
\etab \Rb_h \partial_\pm v = (\hayb - \frac{\kay}{2}\etab \Rb_h)l_\pm
\ \ \Leftrightarrow \ \
\text{second eq.~in \eqref{eqi22}}.
\end{gathered}\end{equation*}}
\begin{equation}\label{eqi22}
\trap P \ad_v P = \hayb \etab^{-1} \trap P (\bar\omega_h - \bar\omega) P , \qquad
\partial_\pm v = \trap P (\hayb\etab^{-1} \bar\omega_h - \frac{\kay}{2})l_\pm,
\end{equation}
where we have used eqs.~\eqref{rominv} and \eqref{psx}, and $\bar\omega_h = \Ad_h^{-1} \bar \omega \Ad_h^{\vphantom{-1}}$.
Parametrising $h = \exp \lh$, $\lh \in \alg{h}$, these equations are solved by
\begin{equation}\label{eqv}
v = \trap P \big(\hayb \etab^{-1} (h^{-1}\bar\omega h) - \frac{\kay}{2} \lh\big) ,
\end{equation}
where $\trap P(h^{-1}\bar\omega h)$ is defined in eq.~\eqref{eq:omcdef}.
For $\kay = 0$ this agrees with the solution found in~\cite{Borsato:2016pas,Borsato:2017qsx} and it is straightforward to see that it solves \eqref{eqi22} using the identities \eqref{psx} and \eqref{eq:omegaid} together with $\trap P\ad_{\trap P X} P = \trap P \ad_X P$ for $X \in \alg{g}$.
\unskip\footnote{Using \eqref{psx} we have $\trap P\ad_{\trap P X} P Y = - \trap P \ad_{PY} \trap P X = - \trap P \ad_{PY} X = \trap P \ad_X PY$ for $X,Y\in \alg{g}$.}
For $\kay \neq 0$, eq.~\eqref{eqv} will typically not solve the system of equations \eqref{eqi22};
however in the case of interest, \textit{i.e.} when $\alg{h}$ is solvable, it does.
In particular, for solvable $\alg{h}$ we have $\trap P l_\pm = \trap P \partial_\pm \lh$ and $\trap P \ad_{\trap P X} P = \trap P \ad_{X} P = 0$ for $X \in \alg{h}$.
These relations follow from the property $\trap P [X,Y] = 0$ for $X,Y \in \alg{h}$, a consequence of the fact that $\kappa([X,Y],PZ) = 0$ for $Z \in \alg{g}$ if $\alg{h}$ is solvable.

It remains to check that eq.~\eqref{eqi3} is satisfied.
Using the relations \eqref{eqi22} we find that the left-hand side equals $-\hayb \etab^{-1} \omega(\partial_+ h h^{-1},\partial_- h h^{-1})$, which is indeed a closed B-field term by virtue of the fact that $d\omega = 0$.
Therefore, when $R$ is a solution of the cYBE with solvable $\alg{h}$ we have, as claimed, that
\begin{equation}
\mathsf{S} = \bar{\mathsf{S}} - \int d^2x \;\hayb \etab^{-1} \omega(\partial_+ h h^{-1},\partial_- h h^{-1}),
\end{equation}
where $\mathsf{S}$ is the YB deformation of the PCM plus WZ term \eqref{Eq:SolvS} and $\bar{\mathsf{S}}$ is the non-abelian T-dual model \eqref{eq:act4}.


\section{Concluding comments}
\label{sec:comments}

In this article we have investigated YB deformations of the PCM plus WZ term.
The admissibility of such a deformation at first order in $\kay/\hay$ is governed by the cohomological constraint~\eqref{diffeq}, assuming that the Lax connection is still based on the existence of a flat and conserved current.
Since all the R-matrices that we considered solving this condition satisfy $\Omega_R = 0$, we focused on this class of solutions, \textit{i.e.} deformations based on R-matrices with solvable $\alg{h}_\pm$.
Having proposed the action~\eqref{Eq:SolvS} motivated by the results of~\cite{Klimcik:2019kkf}, we proved its classical integrability by constructing a Lax connection and showing that the Lax matrix satisfies a Maillet bracket with twist function.
We also demonstrated that this model follows from various alternative formulations used to study integrable $\sigma$-models, including affine Gaudin models, $\mathcal{E}$-models, 4-dimensional Chern-Simons theory and, in the case of homogeneous R-matrices, non-abelian T-duality.

In \appref{app:gendef} we argued that any integrable deformation of the PCM plus WZ term whose Lax connection remains of the form \eqref{Eq:Lax} falls into the class of models constructed in \secref{sec:solvable}.
It is therefore natural to ask if it can be proven using cohomological arguments that the solvability of $\alg{h}_\pm$ is both a necessary and sufficient condition to construct the YB deformation of the PCM plus WZ term.
It would also be insightful to investigate the space of solutions to the identity~\eqref{Eq:IdExp} more fully; in particular, whether there are additional isolated solutions for fixed $\rho$.
The resulting theories would not be deformations of the PCM plus WZ term, but may lead to new examples of integrable $\sigma$-models.
Another interesting question is to ask whether the assumption that the Lax connection is of the form~\eqref{Eq:Lax} can be relaxed and if a more general setup could allow for the construction of integrable deformations based on R-matrices that do not satisfy the cohomological constraint \eqref{diffeq}.
However, it is not immediately clear what a natural generalisation of our ansatz, which could still be analysed systematically as in \secref{sec:general}, would be.

Using the formulation of the YB deformation of the PCM plus WZ term as an $\mathcal{E}$-model, it would also be interesting to study the space of Poisson-Lie T-duals~\cite{Klimcik:1995dy,Klimcik:1996nq}.
For the case of the standard Drinfel'd-Jimbo R-matrix, examples of such dualities were considered in~\cite{Demulder:2017zhz}.
Finally, an important next step would be to study the quantum properties of these models, including, for example, their quantum integrability, their renormalisability and renormalisation group flow, and their infrared degrees of freedom and scattering matrices.
Again for the standard non-split Drinfel'd-Jimbo R-matrix, the one-loop renormalisability has been studied in \cite{Demulder:2017zhz,Klimcik:2019kkf} where it was shown that in terms of the parameters $\kay$, $\chi$ and $\rho$ of the action~\eqref{Eq:SolvS} only $\chi$ runs.
Up to a convention-dependent normalisation, the flow is given by $\dot\chi \propto h^\vee \kay \xi^{-2}$ where $h^\vee$ is the dual Coxeter number and $\xi$ is defined in eq.~\eqref{Eq:XiSolv} with $c=i$.
We expect that this relation holds for general values of $c$ and for any R-matrix with solvable $\alg{h}_\pm$.
If this is indeed the case, then, in addition to the fixed points at $\chi \to \pm \infty$ corresponding the WZW model, there are additional fixed points at $\chi = \pm c\rho$ for $c=0$ and $c=1$.
It would be interesting to investigate the resulting models.

\medskip

The unifying framework for YB deformations of the PCM plus WZ term developed in this article should be readily applicable to a number of closely related models.
A defining property of the model we have considered is that the right-acting $\grp{G}$-symmetry is preserved.
This can be relaxed by recasting the PCM for $\grp{G}$ as the symmetric space $\sigma$-model for the $\Integer_2$ permutation coset $\frac{\grp{G} \times \grp{G}}{\grp{G}}$.
This allows deformations breaking the full $\grp{G} \times \grp{G}$ symmetry to be constructed, including the bi-Yang-Baxter deformation, for which the left and right symmetries are independently deformed.
The bi-Yang-Baxter deformation has been extensively studied for the standard Drinfel'd-Jimbo R-matrix both with and without WZ term, for $\grp{SU}(2)$ in~\cite{Fateev:1996ea,Lukyanov:2012zt,Hoare:2014pna} and for arbitrary $\grp{G}$ in~\cite{Klimcik:2008eq,Klimcik:2014bta,Delduc:2015xdm,Delduc:2017fib,Klimcik:2019kkf,Klimcik:2020fhs}.

Models with the deformation mixing the left and right symmetries have also been considered.
Without the WZ term this just amounts to the YB deformation of the symmetric space $\sigma$-model~\cite{Delduc:2013fga}.
Examples including the WZ term include those based on TsT transformations of the bi-Yang-Baxter deformation~\cite{Lukyanov:2012zt,Delduc:2017fib} and homogeneous YB deformations of the $\grp{SL}(2,\Real)$ WZW model~\cite{Borsato:2018spz}.
Formulating these models in a unified framework would provide a better understanding of the space of integrable deformations of the PCM plus WZ term.
A related direction is to investigate, along similar lines, the YB deformation of $\Integer_4$ permutation supercosets with WZ term~\cite{Cagnazzo:2012se,Delduc:2018xug}, generalising the case without WZ term of~\cite{Babichenko:2009dk,Hoare:2014oua}.
It would also be interesting to study if the new YB deformations of the PCM constructed in~\cite{Fukushima:2020kta} can be applied to the PCM plus WZ term, as well as the classical integrability of the action~\eqref{Eq:SolvS} with local couplings~\cite{Hoare:2020fye}.

Finally, we may also ask what happens if we consider semi-simple Lie groups.
Of course, we can take a copy of the PCM plus WZ term for each simple normal subgroup and independently deform each one; however, we may also explore what happens if we allow these models to mix.
In particular, the results of~\cite{Bassi:2019aaf} can be used to construct YB deformations of the model defined in~\cite{Delduc:2018hty,Delduc:2019bcl}, which couples together an arbitrary number of copies of the PCM plus WZ term for the same Lie group $\grp{G}$.
Such deformations have been considered in~\cite{Bassi:2019aaf} for the standard Drinfel'd-Jimbo R-matrix.
The results in this article open up the possibility for constructing new integrable deformations of this coupled model, based on more general R-matrices with solvable $\alg{h}_\pm$.

\section*{Acknowledgements}
SL would like to thank F.~Delduc, M.~Magro and B.~Vicedo for useful discussions.
BH was supported by the Swiss National Science Foundation through the NCCR SwissMAP.
The work of SL is funded by the Deutsche Forschungsgemeinschaft (DFG, German Research Foundation) under Germany’s Excellence Strategy – EXC 2121 ``Quantum Universe'' – 390833306.

\appendix

\section{Proof of the integrability identity for \texorpdfstring{$e^{\rho R}$}{exp(\unichar{"03C1}R)}}
\label{sec:identity}

In this appendix we prove the integrability identity \eqref{Eq:IdExp} obeyed by $e^{\rho R}$ when $R$ is a skew-symmetric R-matrix such that $\alg{h}_\pm$ is solvable.
Let us first introduce some notation.
For a linear operator $\mathcal{O}$ on $\g$ we let
\unskip\footnote{Note that, in general, $[X,Y]_{\mathcal{O}}$ is not a Lie bracket.}
\begin{equation}
[X,Y]_{\mathcal{O}} = [\mathcal{O}X,Y] + [X,\mathcal{O}Y].
\end{equation}
We also define the functions
\begin{equation}
f(\rho) = \cosh(c\rho), \qquad g(\rho) = \frac{\sinh(c\rho)}{c},
\end{equation}
and the differential operator
\begin{equation}
\square = \frac{\p^2\;}{\p\rho^2} - c^2,
\end{equation}
such that $f(\rho)$ and $g(\rho)$ are solutions of $\square f(\rho) = \square g(\rho) = 0$ with initial conditions $f(0)=g'(0)=1$ and $f'(0)=g(0)=0$.
Note that the exponential operator $\er$ satisfies
\begin{equation}\label{Eq:BoxExp}
\square \er = (R^2-c^2)\er = R_+ R_- \er.
\end{equation}

\paragraph{An intermediate result.} We start by proving an intermediate result.
For fixed $X,Y\in\g$, consider the function
\begin{equation}\label{Eq:F}
\mathcal{F}(\rho) = R_+ R_- [X,Y]_{\er}.
\end{equation}
Acting with $\square$ on $\mathcal{F}(\rho)$ and using eq.~\eqref{Eq:BoxExp} gives
\begin{equation}
\square\mathcal{F}(\rho) = R_+ R_- [X,Y]_{R_+ R_- \er} = R_+R_- [R_+R_- \er X, Y] + R_+ R_- [X,R_+R_- \er Y].
\end{equation}
Applying the identity \eqref{Eq:IdentityR2} we find
\begin{equation}
\square\mathcal{F}(\rho) = [R_+R_- \er X, R_+R_-Y] + [R_+R_-X, R_+R_-\er Y],
\end{equation}
and hence, since $\im R_+R_-$ is abelian~\eqref{Eq:IdentityR3}, we have that $\square \mathcal{F}(\rho) = 0$.
Therefore, $\mathcal{F}(\rho)$ is a linear combination of the functions $f(\rho)$ and $g(\rho)$ introduced above.
More precisely, we have
\begin{equation}
\mathcal{F}(\rho) = f(\rho) \mathcal{F}(0) + g(\rho) \mathcal{F}'(0).
\end{equation}
It is clear from the definition \eqref{Eq:F} that $\mathcal{F}(0)=2 R_+R_-[X,Y]$ and $\mathcal{F}'(0)=R_+R_-[X,Y]_R$, which vanishes by eq.~\eqref{Eq:IdentityR1}.
Therefore, we have the identity
\begin{equation}\label{Eq:RRExp}
R_+ R_- [X,Y]_{\er} = 2f(\rho) R_+R_- [X,Y].
\end{equation}

\paragraph{Proof of the identity.} Let us now consider the function
\begin{equation}\label{eq:E}
\mathcal{E}(\rho) = [\er X,\er Y] - \er[X,Y]_{\er} - [X,Y] + (e^{\rho R_+} + e^{\rho R_-}) [X,Y] ,
\end{equation}
for fixed $X,Y \in \g$.
Our goal is to prove that $\mathcal{E}(\rho) = 0$.
Acting with $\square$ on $\mathcal{E}(\rho)$ and using eq.~\eqref{Eq:BoxExp} gives
\begin{equation}\begin{split}
\square \mathcal{E}(\rho) & = [R_+R_-\er X,\er Y] + [\er X, R_+R_- \er Y] - \er [X,Y]_{R_+R_-\er}
\\ & \quad - R_+R_- \er [X,Y]_{\er} + R_+R_- (e^{\rho R_+} + e^{\rho R_-}) [X,Y] .
\end{split}\end{equation}
The first line of the right-hand side vanishes as a consequence of the identity~\eqref{Eq:IdentityR2}, while the second line vanishes due to eq.~\eqref{Eq:RRExp}.
We thus have that $\square \mathcal{E} = 0$.
The solution to this differential equation is
\begin{equation}
\mathcal{E}(\rho) = f(\rho) \mathcal{E}(0) + g(\rho) \mathcal{E}'(0) .
\end{equation}
It is straightforward to see from eq.~\eqref{eq:E} that $\mathcal{E}(0) = 0$, while
\begin{equation}
\mathcal{E}'(\rho) = [\er X,\er Y]_R - \er[X,Y]_{R\er} - R\er[X,Y]_{\er} + (R_+ e^{\rho R_+} + R_- e^{\rho R_-}) [X,Y],
\end{equation}
and hence we also have $\mathcal{E}'(0) = 0$.
Therefore, it follows that
\begin{equation}
\mathcal{E}(\rho) = [\er X,\er Y] - \er[X,Y]_{\er} - [X,Y] + (e^{\rho R_+} + e^{\rho R_-}) [X,Y] = 0 ,
\end{equation}
as claimed.

\section{Integrable deformations of the PCM plus WZ term.}\label{app:gendef}

In \ssecref{ssec:ybdef} we defined the YB deformation of the PCM plus WZ term by the action~\eqref{Eq:SolvS}, which depends on $R$, a skew-symmetric solution of the (m)cYBE with solvable $\alg{h}_\pm$.
Such R-matrices satisfy the identity~\eqref{Eq:IdExp}.
As discussed in \ssecref{ssec:ybdef}, for a Lax connection to exist it is sufficient for $R$ to be skew-symmetric and solve this identity.
However, if this holds for all $\rho \in \Real$, then we can expand~\eqref{Eq:IdExp} for small $\rho$ to see that this implies that $R$ solves the (m)cYBE and that $\alg{h}_\pm$ is solvable, and hence we return to our original setup.

In this appendix we investigate the reverse logic and show that the identity~\eqref{Eq:IdExp} and the skew-symmetry of $R$ are also necessary conditions for integrability under certain assumptions, including that the Lax connection remains of the form~\eqref{Eq:Lax}.
In particular, these conditions follow from the integrability condition \eqref{Eq:IntCond}.

\paragraph{Parametrisation of $\mathcal{O}$.} The form of the action \eqref{Eq:SolvS} suggests a natural parametrisation for exploring generalisations of the setup we have considered thus far.
We take the operator $\mathcal{O}$ in the general action~\eqref{ansatz} to be
\begin{equation}\begin{gathered}\label{eq:ooqqgam2}
\mathcal{O} = \frac{\kay}{2} \frac{1 + \gamma^{-1} S}{1 - \gamma^{-1} S} , \qquad
\tra \mathcal{O} = - \frac{\kay}{2} \frac{1 + \gamma\,\tra S^{-1}}{1 - \gamma\, \tra S^{-1}} ,
\end{gathered}\end{equation}
where $\gamma$ is a free parameter and $S:\alg{g}\to\alg{g}$ is a constant invertible linear operator, which we assume is independent of $\kay$ and $\gamma$.
In this ansatz we have the freedom to rescale both $\gamma$ and $S$, or, equivalently, to fix the normalisation of $S$.
The undeformed limit, in which we recover the PCM plus WZ term, is $S \to 1$, while taking $\gamma \to + \infty$ gives the WZW model.
The equations of motion are equivalent to the conservation equation~\eqref{eom} of the Noether current
\begin{equation}\begin{gathered}
K_\pm = (\mathcal{Q}_g^\pm)^{-1} j_\pm , \\
\mathcal{Q}^- = \frac{\xi}{\kay} (1-\gamma^{-1} S),\qquad
\mathcal{Q}^+ = - \frac{\xi}{\kay} (1 - \gamma \, \tra S^{-1}) ,
\end{gathered}\end{equation}
where the constant $\xi$ parametrises the freedom in normalising $K_\pm$.

\paragraph{Integrability.}
Substituting $\mathcal{Q}^\pm$ into \eqref{zeq}, the integrability condition~\eqref{Eq:IntCond} becomes
\begin{equation}\begin{split}\label{eq:genint}
[\tra S^{-1}X,SY] & - \tra{S}^{-1}[X,SY] - S[\tra S^{-1}X,Y] \\ & = \bigg(1-\Big(1-\frac{\kay}{2\xi}\Big)\gamma\,\tra S^{-1} -\Big(1+\frac{\kay}{2\xi}\Big)\gamma^{-1} S\bigg)[X,Y],
\qquad \forall ~ X,Y \in \g.
\end{split}\end{equation}
Using the ad-invariance of $\kappa$, this identity can be rewritten as
\begin{equation}\begin{split}\label{eq:genint2}
[SX,SY] & - S[X,SY] - S[SX,Y] \\ & = \bigg(1- \Big(1+\frac{\kay}{2\xi}\Big)\gamma^{-1} S\bigg)[X,\tra S S Y] - \Big(1-\frac{\kay}{2\xi}\Big)\gamma S[X,Y],
\qquad \forall ~ X,Y \in \g,
\end{split}\end{equation}
which is a form that is particularly useful for computation since it is polynomial in $S$ and its transpose.
Summing \eqref{eq:genint2} with itself with $X$ and $Y$ interchanged we find
\begin{equation}
\bigg(1 - \Big(1+\frac{\kay}{2\xi}\Big)\gamma^{-1} S\bigg)\big([X,\tra S S Y] - [\tra S S X, Y] \big) = 0,
\qquad \forall ~ X,Y \in \g.
\end{equation}
Assuming that $1 - (1+\frac{\kay}{2\xi}) \gamma^{-1}S$ is invertible this implies that
\begin{equation}
[X,\tra S S Y] = \tra S S [X,Y], \qquad \forall ~ X,Y \in \g,
\end{equation}
and hence, by Schur's lemma, $\tra S S$ is proportional to the identity.
The freedom to fix the normalisation of $S$ in the ansatz~\eqref{eq:ooqqgam2} allows us to choose $\tra S S = 1$, \textit{i.e.} $S$ to be orthogonal.
Using this result in the identity~\eqref{eq:genint2}, and parametrising $\xi$ in terms of the new parameter $\hat\rho$ as
\begin{equation}\label{eq:gamxi}
\xi = \frac{\kay}{2} \frac{\gamma - \gamma^{-1}}{\gamma+\gamma^{-1} - 2 \sqrt{1+c^2\hat\rho^2}},
\end{equation}
where $c=1$, $i$ or $0$, we find that solutions to
\begin{equation}\begin{aligned}\label{eq:genint2a}
& [SX,SY] - S[X,SY] - S[SX,Y] = (1-2\sqrt{1+c^2\hat\rho^2} \, S)[X,Y] , \qquad
\forall ~ X,Y \in \g,
\\ & \qquad \tra S S = 1,
\end{aligned}\end{equation}
with $S \to 1$ as $\hat{\rho} \to 0$, define integrable deformations of the PCM plus WZ term.

If we parametrise
\begin{equation}\label{eq:sgamparam}
S = e^{\rho R}, \qquad \gamma = e^\chi , \qquad \hat\rho = \frac{\sinh c\rho}{c},
\end{equation}
where $R$ is skew-symmetric by the orthogonality of $S$, the identity~\eqref{eq:genint2a} becomes~\eqref{Eq:IdExp}, and we find the YB deformation of the PCM plus WZ term as defined by the action~\eqref{Eq:SolvS}.
Alternatively, introducing $\hat{R} = \hat\rho^{-1}( S - \sqrt{1+c^2\hat\rho^2})$, we see that the equations~\eqref{eq:genint2a} are equivalent to
\begin{equation}\begin{aligned}\label{eq:genint2b}
& [\hat R X,\hat RY] - \hat R[X,\hat RY] - \hat R[\hat RX,Y] + c^2 \,[X,Y] = 0 , \qquad
\forall ~ X,Y \in \g,
\\
& \qquad \hat\rho\,\big(\trap \hat{R} \hat{R} +c^2\big) + \sqrt{1+c^2\hat\rho^2} \, (\trap \hat{R} + \hat{R})= 0,
\end{aligned}\end{equation}
\textit{i.e.} $\hat{R}$ is an asymmetric solution of the (m)cYBE satisfying the symmetry property \eqref{eq:transposer}.
Therefore, we find the YB deformation of the PCM plus WZ term in the form~\eqref{Eq:SolvSasym}.

\paragraph{Limit without WZ term.}
Starting from the operators~\eqref{eq:ooqqgam2}, to take the limit without WZ term we parametrise $S$ and $\gamma$ as in eq.~\eqref{eq:sgamparam} without assuming that $R$ is skew-symmetric, and take $\kay,\rho,\chi \to 0$ with their ratios fixed as
\begin{equation}
\frac{\rho}{\kay} = \frac{\eta}{\hay} , \qquad \frac{\chi}{\kay} = \frac{1}{\hay}, \qquad \frac{\rho}{\chi} = \eta .
\end{equation}
In this limit the operator $\mathcal{O}$~\eqref{eq:ooqqgam2} takes the form
\begin{equation}\label{eq:opo}
\mathcal{O} = \frac{\hay}{1-\eta R}.
\end{equation}
In this ansatz we have the freedom to rescale both $\eta^{-1}$ and $R$, or, equivalently, to fix the normalisation of $R$.
We may also shift $R$ by the identity, which together with a compensating rescaling of $\hay$ and $\eta$, also leaves the form of~\eqref{eq:opo} unchanged.
In the limit without WZ term, the identity \eqref{eq:genint} becomes
\begin{equation}\begin{split}\label{eq:genint3}
[\trap R X, R Y] & - \trap R[X,RY] - R[\trap R X, Y]
\\ & = \frac{1}{\eta^2}\Big(1-\frac{\hay}{\xi}\Big) [X,Y] - \frac{1}{\eta}\Big(1-\frac{\hay}{2\xi}\Big) (\trap R + R) [X,Y] , \qquad \forall ~ X,Y \in \g.
\end{split}\end{equation}
Using the ad-invariance of $\kappa$, this identity can be rewritten as
\begin{equation}\begin{split}\label{eq:genint4}
[R X, R Y] & - R[X,RY] + R[\trap R X, Y]
\\ & = - \frac{1}{\eta^2}\Big(1-\frac{\hay}{\xi}\Big) [X,Y] + \frac{1}{\eta}\Big(1-\frac{\hay}{2\xi}\Big)[(\trap R + R)X,Y], \qquad \forall ~ X,Y\in \g.
\end{split}\end{equation}
Again summing with itself with $X$ and $Y$ interchanged we find
\begin{equation}
\bigg(R - \frac{1}{\eta}\Big(1-\frac{\hay}{2\xi}\Big)\bigg)\big([(\trap R + R) X, Y] - [X,(\trap R + R)Y] \big) = 0, \qquad \forall ~ X,Y\in \g.
\end{equation}
Following a similar logic to above, this implies that, assuming that $R - \frac{1}{\eta}(1-\frac{\hay}{2\xi})$ is invertible, $\trap R + R$ is proportional to the identity.
The freedom to shift $R$ in the ansatz~\eqref{eq:opo}, together with a compensating rescaling of $\hay$ and $\eta$, allows us to choose $\trap R + R = 0$, \textit{i.e.} $R$ to be skew-symmetric.
Using this result in the identity~\eqref{eq:genint4}, and parametrising $\xi$ in terms of a new parameter, $c$,
\begin{equation}
\xi = \frac{\hay}{1-c^2\eta^2},
\end{equation}
we find that solutions to
\begin{equation}\begin{aligned}\hspace{0pt}\label{eq:eqseqseqs2}
& [R X, R Y] - R[X,RY] - R[ R X, Y] + c^2 [X,Y] = 0, \qquad \forall ~ X,Y\in\g,
\\
& \qquad \trap R + R = 0,
\end{aligned}\end{equation}
define integrable deformations of the PCM.
Furthermore, we can use the freedom to fix the normalisation of $R$ in the ansatz~\eqref{eq:opo} to set $c$ equal to $1$, $i$ or $0$,
recovering the standard YB deformation of the PCM~\cite{Klimcik:2002zj,Klimcik:2008eq}.

\section{Maillet bracket with twist function}
\label{sec:Maillet}

In this appendix we compute the Poisson bracket of the Lax matrix \eqref{Eq:LaxHam} with itself and show that it satisfies a Maillet bracket~\eqref{Eq:Maillet} with twist function.

\paragraph{A technical result.}
To do so we first prove the following technical result.
Let $A(x)$ and $B(x)$ be $\g$-valued fields satisfying the brackets
\begin{equation}\begin{gathered}\label{eq:pbassump}
\left\lbrace A\ti{1}(x), B\ti{2}(y) \right\rbrace = \big[ C\ti{12}, D^{AB}\ti{1}(x) \big] \delta_{xy} - \ell_{AB}\, C\ti{12} \delta'_{xy},\\
\left\lbrace A\ti{1}(x), g\ti{2}(y) \right\rbrace = p_A\, g\ti{2}(x) C\ti{12} \delta_{xy}, \qquad
\left\lbrace B\ti{1}(x), g\ti{2}(y) \right\rbrace = p_B\, g\ti{2}(x) C\ti{12} \delta_{xy},
\end{gathered}\end{equation}
for a $\g$-valued field $D^{AB}(x)$ and numbers $\ell_{AB}$, $p_A$ and $p_B$, and let $\mathcal{M}$ and $\mathcal{N}$ be constant linear operators on $\g$ with $\Mg = \Ad_g^{-1}\mathcal{M} \Ad_g^{\vphantom{-1}}$ and $\Ng = \Ad_g^{-1} \mathcal{N} \Ad_g^{\vphantom{-1}}$.
It then follows that
\begin{equation}\begin{split}\label{eq:idpb}
\left\lbrace (\Mg A)\ti{1}(x), (\Ng B)\ti{2}(y) \right\rbrace &= \Mgx\null\ti{1} \big[ \tra \Ngx\null\ti{1}C\ti{12}, D^{AB}\ti{1}(x) - p_A\,B\ti{1}(x) - p_B\,A\ti{1}(x) - \ell_{AB}\,j\ti{1}(x) \big]\delta_{xy} \\
& \qquad + p_A\, \Mgx\null\ti{1} \big[ C\ti{12}, (\Ng B)\ti{1}(x) \big]\delta_{xy} + p_B\, \big[ \tra \Ngx\null\ti{1}C\ti{12}, (\Mg A)\ti{1}(x) \big]\delta_{xy} \\
& \qquad + \ell_{AB} \, \Mgx\null\ti{1} \,\tra \Ngx\null\ti{1} \Big( \big[ C\ti{12}, j\ti{1}(x) \big] \delta_{xy} - C\ti{12}\,\delta'_{xy} \Big).
\end{split}\end{equation}
Note that, using the identity
\begin{equation}\label{Eq:dDelta}
f(y)\delta'_{xy}=f(x)\delta'_{xy}+f'(x)\delta_{xy},
\end{equation}
which holds for a general function $f$, we have written the right-hand side of eq.~\eqref{eq:idpb} with all fields evaluated at the point $x$.
Furthermore, all operators act on and every field appears in the first tensor space. This can be achieved using the identity
\begin{equation}\label{Eq:TranspCas}
\mathcal{O}\ti{2}C\ti{12} = \tra \mathcal{O}\ti{1}C\ti{12},
\end{equation}
which holds for any linear operator $\mathcal{O}$ by the definition of the quadratic split Casimir~\eqref{eq:splitcas}.
In particular, it is useful to note that taking $\mathcal{O} = \ad_X$, $X\in \g$, eq.~\eqref{Eq:TranspCas} gives us the standard identity
\begin{equation}\label{Eq:ComCas}
[ C\ti{12},X\ti{2} ] = -[ C\ti{12}, X\ti{1} ], \qquad \forall ~ X\in\g.
\end{equation}

To prove eq.~\eqref{eq:idpb} we start by recalling the following result.
If $\delta$ is a derivation of the algebra of observables of the model (\textit{e.g.} the Poisson bracket with a fixed observable or a space-time derivative)
then
\begin{equation}\label{Eq:DerOg}
\delta(\Oc_g L) = \Oc_g\delta L + \Oc_g [g^{-1}\delta g,L] - [ g^{-1}\delta g, \Oc_gL ],
\end{equation}
where $\Oc$ is a constant linear operator on $\g$ and $L$ is a $\g$-valued observable.
This is straightforward to show using the properties of a derivation.

Consider the identity~\eqref{Eq:DerOg} with $\delta=\left\lbrace (\Mg A)\ti{1}(x), \cdot \right\rbrace$, $\Oc=\mathcal{N}\ti{2}$ and $L=B\ti{2}(y)$.
This gives
\begin{equation}\begin{split}\label{eqvv1}
\left\lbrace (\Mg A)\ti{1}(x), (\Ng B)\ti{2}(y) \right\rbrace
& = \Ngy\null\ti{2} \left\lbrace (\Mg A)\ti{1}(x), B\ti{2}(y) \right\rbrace
\\ & \qquad + p_A \Mgx\null\ti{1}\Big(\Ngy\null\ti{2} \big[ C\ti{12}, B\ti{2}(y) \big]
- \big[ C\ti{12} , (\Ng B)\ti{2}(y) \big]\Big) \delta_{xy} ,
\end{split}\end{equation}
where we have used that
\begin{equation*}
g\ti{2}(y)^{-1}\left\lbrace (\Mg A)\ti{1}(x), g\ti{2}(y) \right\rbrace
= \Mgx\null\ti{1} \big( g\ti{2}(y)^{-1}\left\lbrace A\ti{1}(x), g\ti{2}(y) \right\rbrace \!\big)
= p_A \Mgx\null\ti{1} C\ti{12}\delta_{xy},
\end{equation*}
since $g$ Poisson commutes with itself.

Now taking the identity~\eqref{Eq:DerOg} with $\delta = \lbrace \cdot, B\ti{2}(y) \rbrace$, $\Oc = \mathcal{M}\ti{1}$ and $L = A\ti{1}(x)$, we find
\begin{equation}\begin{split}\label{eqvv2}
\left\lbrace (\Mg A)\ti{1}(x), B\ti{2}(y) \right\rbrace
&= \Mgx\null\ti{1} \left\lbrace A\ti{1}(x), B\ti{2}(y) \right\rbrace
\\ & \qquad - p_B \Big(\Mgx\null\ti{1} \big[ C\ti{12}, A\ti{1}(x) \big]
- \big[ C\ti{12}, (\Mg A)\ti{1}(x) \big]\Big) \delta_{xy} ,
\end{split}\end{equation}
where we have used that
\begin{equation*}
g\ti{1}(x)^{-1}\left\lbrace g\ti{1}(x), B\ti{2}(y)\right\rbrace = - p_B C\ti{12}\delta_{xy} .
\end{equation*}

Substituting \eqref{eqvv2} into \eqref{eqvv1} and using the Poisson bracket of $A$ with $B$~\eqref{eq:pbassump} we arrive at
\begin{equation}\begin{aligned}
\left\lbrace (\Mg A)\ti{1}(x), (\Ng B)\ti{2}(y) \right\rbrace
& = \Mgx\null\ti{1} \Ngy\null\ti{2}
\Big(\big[ C\ti{12}, D^{AB}\ti{1}(x) \big] \delta_{xy} - \ell_{AB}\, C\ti{12} \delta'_{xy}\Big)
\\ & \quad + p_A \Mgx\null\ti{1}\Big(\Ngy\null\ti{2} \big[ C\ti{12}, B\ti{2}(y) \big]
- \big[ C\ti{12} , (\Ng B)\ti{2}(y) \big]\Big) \delta_{xy}
\\ & \quad - p_B \Ngy\null\ti{2} \Big(\Mgx\null\ti{1} \big[ C\ti{12}, A\ti{1}(x) \big]
- \big[ C\ti{12}, (\Mg A)\ti{1}(x) \big]\Big) \delta_{xy} .
\end{aligned}\end{equation}
To bring this expression into the required form~\eqref{eq:idpb} we first use the identities~\eqref{Eq:TranspCas} and \eqref{Eq:ComCas} so that all operators act on and every field appears in the first tensor space
\begin{equation}\begin{aligned}\label{eq:subsub}
\left\lbrace (\Mg A)\ti{1}(x), (\Ng B)\ti{2}(y) \right\rbrace
& = \Mgx\null\ti{1}\big[\tra \Ngy\null\ti{1}
C\ti{12}, D^{AB}\ti{1}(x) - p_A B\ti{1}(y) - p_B A\ti{1}(x) \big] \delta_{xy}
\\ & \quad + p_A \Mgx\null\ti{1} \big[ C\ti{12} , (\Ng B)\ti{1}(y) \big] \delta_{xy}
+ p_B \big[\tra\Ngy\null\ti{2} C\ti{12}, (\Mg A)\ti{1}(x) \big] \delta_{xy}
\\ & \quad - \ell_{AB} \Mgx\null\ti{1} \tra \Ngy\null\ti{1} C\ti{12} \delta'_{xy}.
\end{aligned}\end{equation}
Finally, to have all fields evaluated at the same point $x$, we use eq.~\eqref{Eq:dDelta} to write
\begin{equation}\begin{split}
\tra\Ngy\null\ti{1} C\ti{12} \delta'_{xy} & = \tra\Ngx\null\ti{1} C\ti{12} \delta'_{xy} + \p_x(\tra\Ngx\null\ti{1} C\ti{12})\delta_{xy}
\\ & = \tra\Ngx\null\ti{1} C\ti{12} \delta'_{xy} - \tra \Ngx\null\ti{1} \big[ C\ti{12}, j\ti{1}(x) \big]\delta_{xy} + \big[\tra \Ngx\null\ti{1} C\ti{12}, j\ti{1}(x) \big]\delta_{xy}.
\end{split}\end{equation}
Substituting into eq.~\eqref{eq:subsub} and evaluating all fields at the point $x$ we find eq.~\eqref{eq:idpb} as claimed.

\paragraph{Poisson bracket of the Lax matrix.}

To compute the Poisson bracket of the Lax matrix with itself we first define the operators
\begin{equation}\begin{aligned}\label{Eq:UV}
\Uc(z) & = \alpha(z)\,e^{-\rho R} + \beta(z),
\qquad &
\tra \Uc(z) & = \alpha(z) e^{\rho R} + \beta(z),
\\ \Vc(z) & = 2\kay\,\alpha(z)\,e^{-\rho R},
\qquad &
\tra \Vc(z) & = 2\kay\,\alpha(z) e^{\rho R},
\end{aligned}\end{equation}
on $\g$, such that the Lax matrix \eqref{Eq:LaxHam} is given by
\begin{equation}\label{Eq:LaxUV}
\Lc(z) = \Uc(z)_g\,Z + \Vc(z)_g\,j,
\end{equation}
where $\trab \Uc(z)_g=\Ad_g^{-1}\trab \Uc(z) \Ad_g^{\vphantom{-1}}$ and $\trab \Vc(z)_g=\Ad_g^{-1}\trab \Vc(z) \Ad_g^{\vphantom{-1}}$.

The Poisson brackets between the fields $Z$, $j$ and $g$, given in eqs.~\eqref{Eq:PBgj}, \eqref{Eq:PbZ1} and \eqref{Eq:PbZ2}, are of the form \eqref{eq:pbassump} with
\begin{equation}\begin{aligned}
& D^{ZZ}= Z, && \ell^{ZZ} = -2\kay, && p_Z = 1,
\\
& D^{jj} = 0, && \ell^{jj} = 0, && p_j = 0,
\\
& D^{Zj} = D^{jZ} = j, &\quad & \ell^{Zj} = \ell^{jZ} = 1. &\quad&
\end{aligned}\end{equation}
Now using eq.~\eqref{eq:idpb} we find
\begin{equation}\begin{split}\label{Eq:PbL1}
\big\lbrace \Lc\ti{1}(z,x), \Lc\ti{2}(w,y) \big\rbrace &=
\Big(\big[\tra \Uc_g\null\ti{1}(w) C\ti{12}, (\Uc_g(z)Z)\ti{1} \big]
+ \Uc_g\null\ti{1}(z)\big[ C\ti{12}, (\Uc_g(w)Z)\ti{1} \big]
\\ & \hspace{190pt}
- \Uc_g\null\ti{1}(z)\big[\tra \Uc_g\null\ti{1} (w)C\ti{12}, Z\ti{1} \big]\Big)\delta_{xy}
\\ & \quad
+ \Big(\big[ \tra \Uc_g\null\ti{1}(w)C\ti{12}, (\Vc_g(z) j)\ti{1} \big]
+ \Uc_g\null\ti{1}(z) \big[ C\ti{12}, (\Vc_g(w)j)\ti{1} \big]
\\ & \hspace{70pt}
- \Vc_g\null\ti{1}(z) \big[ \tra \Uc_g\null\ti{1}(w)C\ti{12}, j\ti{1} \big]
- \Uc_g\null\ti{1}(z) \big[ \tra \Vc_g\null\ti{1}(w)C\ti{12}, j\ti{1} \big]
\\ & \hspace{180pt}
+ 2\kay \, \Uc_g\null\ti{1}(z) \big[ \tra \Uc_g\null\ti{1}(w)C\ti{12}, j\ti{1} \big] \Big)\delta_{xy}
\\ & \quad
+ W\ti{1}(z,w) \Big( \big[ C\ti{12}, j\ti{1} \big] \delta_{xy} - C\ti{12}\,\delta'_{xy} \Big),
\end{split}\end{equation}
where all fields on the right-hand side are evaluated at the point $x$ and we have defined
\begin{equation}\begin{split}
W(z,w) & = \Uc_g(z)\,\tra \Vc_g(w) + \Vc_g(z)\,\tra \Uc_g(w) - 2\kay\,\Uc_g(z)\,\tra \Uc_g(w)
\\ & = 2\kay\,\big(\alpha(z)\alpha(w)-\beta(z)\beta(w)\big).
\end{split}\end{equation}
The identity~\eqref{Eq:IdExp} can be rewritten in the form
\begin{equation}
\big[ \Er X, \Emr Y \big] - \Emr \big[ \Er X, Y \big] + \Emr \big[ X, \Emr Y \big] + \big[X,Y\big] - 2\cosh c \rho\, [X,\Emr Y\big] = 0 ,
\quad \forall ~ X,Y \in \g,
\end{equation}
and used to obtain the following identities for the operators $\Uc(z)$ and $\Vc(z)$
\begin{equation}\begin{split}
& \big[\tra \Uc_g(w) X,\Uc_g(z)Y\big] + \Uc_g(z) \big[X,\Uc_g(w)Y\big] - \Uc_g(z) \big[\tra \Uc_g(w) X, Y\big] \\
& \ \ = \big[ X, \big( \beta(z)\beta(w) - \alpha(z)\alpha(w) \big)Y + \big( \alpha(z)\beta(w)+\beta(z)\alpha(w)+2\alpha(z)\alpha(w) \cosh{c\rho}\big) \Ergm Y \big],
\end{split}\end{equation}
and
\begin{equation}\begin{split}
& \big[ \tra \Uc_g(w)X, \Vc_g(z) Y \big]
+ \Uc_g(z) \big[ X, \Vc_g(w)Y \big]
\\ & \hspace{42pt}
- \Vc_g(z) \big[ \tra \Uc_g(w)X, Y \big]
- \Uc_g(z) \big[ \tra \Vc_g(w)X, Y \big]
+ 2 \kay \, \Uc_g(z) \big[ \tra \Uc_g(w)X, Y \big]
\\
& \ \ = 2\kay \big[ X, \big( \beta(z)\beta(w) - \alpha(z)\alpha(w) \big)Y + \big( \alpha(z)\beta(w)+\beta(z)\alpha(w)+2\alpha(z)\alpha(w)\cosh c\rho \big) \Ergm Y \big].
\end{split}\end{equation}
These can then be used to simplify the Poisson bracket~\eqref{Eq:PbL1} to give
\begin{equation}\begin{split}\label{Eq:PbL2}
\big\lbrace \Lc\ti{1}(z,x), \Lc\ti{2}(w,y) \big\rbrace
& = \big( \alpha(z)\beta(w)+\beta(z)\alpha(w)+2\alpha(z)\alpha(w)\cosh c\rho \big) \big[ C\ti{12}, \big(e^{-\rho R_g}(Z+2\kay\,j)\big)\ti{1}(x) \big]\delta_{xy} \\
& \quad + \big( \beta(z)\beta(w) - \alpha(z)\alpha(w) \big) \big[ C\ti{12}, Z\ti{1}(x) \big]\delta_{xy} - 2\kay \big( \alpha(z)\alpha(w)-\beta(z)\beta(w) \big)C\ti{12}\delta'_{xy}.
\end{split}\end{equation}

\paragraph{Comparison with the Maillet bracket with twist function.}

Let us now substitute the $\Rc$-matrix~\eqref{Eq:RMat} and Lax matrix~\eqref{Eq:LaxHam} into the right-hand side of the Maillet bracket~\eqref{Eq:Maillet} and use the identity~\eqref{Eq:ComCas} to write it in the form
\begin{equation}\begin{split}\label{Eq:PbLc}
\big\lbrace \Lc\ti{1}(z,x), \Lc\ti{2}(w,y) \big\rbrace & =
\frac{\vp(z)^{-1}\alpha(w)-\vp(w)^{-1}\alpha(z)}{z-w} \big[ C\ti{12}, \big(e^{-\rho R_g}(Z+2\kay\,j)\big)\ti{1}(x) \big]\delta_{xy} \\
& \quad + \frac{\vp(z)^{-1}\beta(w)-\vp(w)^{-1}\beta(z)}{z-w} \big[ C\ti{12}, Z\ti{1}(x) \big]\delta_{xy} - \frac{\vp(z)^{-1}-\vp(w)^{-1}}{z-w} C\ti{12}\,\delta'_{xy}.
\end{split}\end{equation}
From the definition of the twist function~\eqref{Eq:Twist} and the expression for $\xi$~\eqref{Eq:XiSolv} we find that
\begin{equation}\begin{split}
&\frac{\vp(z)^{-1}\alpha(w)-\vp(w)^{-1}\alpha(z)}{z-w} = \alpha(z)\beta(w)+\beta(z)\alpha(w)+2\alpha(z)\alpha(w)\cosh c\rho, \\
&\frac{\vp(z)^{-1}\beta(w)-\vp(w)^{-1}\beta(z)}{z-w} = \beta(z)\beta(w) - \alpha(z)\alpha(w), \\
&\frac{\vp(z)^{-1}-\vp(w)^{-1}}{z-w} = 2\kay \big( \alpha(z)\alpha(w)-\beta(z)\beta(w) \big).
\end{split}\end{equation}
Using these to compare eqs.~\eqref{Eq:PbL2} and \eqref{Eq:PbLc} we immediately see that the Poisson bracket of the Lax matrix with itself indeed satisfies a Maillet bracket~\eqref{Eq:Maillet} with twist function, with the twist function given by eq.~\eqref{Eq:Twist}, as claimed.

\linespread{1.05}

\begin{bibtex}[\jobname]

@article{Delduc:2017fib,
author = "Delduc, Francois and Hoare, Ben and Kameyama, Takashi and Magro, Marc",
title = "{Combining the bi-Yang-Baxter deformation, the Wess-Zumino term and TsT transformations in one integrable $\sigma$-model}",
eprint = "1707.08371",
archivePrefix = "arXiv",
primaryClass = "hep-th",
doi = "10.1007/JHEP10(2017)212",
journal = "JHEP",
volume = "10",
pages = "212",
year = "2017"
}

@article{Delduc:2013fga,
author = "Delduc, Francois and Magro, Marc and Vicedo, Benoit",
title = "{On classical $q$-deformations of integrable $\sigma$-models}",
eprint = "1308.3581",
archivePrefix = "arXiv",
primaryClass = "hep-th",
doi = "10.1007/JHEP11(2013)192",
journal = "JHEP",
volume = "11",
pages = "192",
year = "2013"
}

@article{Klimcik:2017ken,
author = "Klim\v{c}\'{i}k, Ctirad",
title = "{Yang-Baxter $\sigma$-model with WZNW term as $\mathcal{E}$-model}",
eprint = "1706.08912",
archivePrefix = "arXiv",
primaryClass = "hep-th",
doi = "10.1016/j.physletb.2017.07.051",
journal = "Phys. Lett. B",
volume = "772",
pages = "725--730",
year = "2017"
}

@article{Klimcik:2019kkf,
author = "Klim\v{c}\'{i}k, Ctirad",
title = "{Dressing cosets and multi-parametric integrable deformations}",
eprint = "1903.00439",
archivePrefix = "arXiv",
primaryClass = "hep-th",
doi = "10.1007/JHEP07(2019)176",
journal = "JHEP",
volume = "07",
pages = "176",
year = "2019"
}

@article{Klimcik:2020fhs,
author = "Klim\v{c}\'{i}k, Ctirad",
title = "{Strong integrability of the bi-YB-WZ model}",
eprint = "2001.05466",
archivePrefix = "arXiv",
primaryClass = "hep-th",
doi = "10.1007/s11005-020-01300-1",
month = "1",
year = "2020"
}

@article{Klimcik:2002zj,
author = "Klim\v{c}\'{i}k, Ctirad",
title = "{Yang-Baxter $\sigma$-models and dS/AdS T-duality}",
eprint = "hep-th/0210095",
archivePrefix = "arXiv",
doi = "10.1088/1126-6708/2002/12/051",
journal = "JHEP",
volume = "12",
pages = "051",
year = "2002"
}

@article{Klimcik:2008eq,
author = "Klim\v{c}\'{i}k, Ctirad",
title = "{On integrability of the Yang-Baxter $\sigma$-model}",
eprint = "0802.3518",
archivePrefix = "arXiv",
primaryClass = "hep-th",
doi = "10.1063/1.3116242",
journal = "J. Math. Phys.",
volume = "50",
pages = "043508",
year = "2009"
}

@article{Delduc:2014uaa,
author = "Delduc, Francois and Magro, Marc and Vicedo, Benoit",
title = "{Integrable double deformation of the principal chiral model}",
eprint = "1410.8066",
archivePrefix = "arXiv",
primaryClass = "hep-th",
doi = "10.1016/j.nuclphysb.2014.12.018",
journal = "Nucl. Phys. B",
volume = "891",
pages = "312--321",
year = "2015"
}

@article{Delduc:2018hty,
author = "Delduc, F. and Lacroix, S. and Magro, M. and Vicedo, B.",
title = "{Integrable Coupled $\sigma$ Models}",
eprint = "1811.12316",
archivePrefix = "arXiv",
primaryClass = "hep-th",
doi = "10.1103/PhysRevLett.122.041601",
journal = "Phys. Rev. Lett.",
volume = "122",
number = "4",
pages = "041601",
year = "2019"
}

@article{Delduc:2019bcl,
author = "Delduc, F. and Lacroix, S. and Magro, M. and Vicedo, B.",
title = "{Assembling integrable $\sigma$-models as affine Gaudin models}",
eprint = "1903.00368",
archivePrefix = "arXiv",
primaryClass = "hep-th",
doi = "10.1007/JHEP06(2019)017",
journal = "JHEP",
volume = "06",
pages = "017",
year = "2019"
}

@article{Maillet:1985fn,
author = "Maillet, Jean Michel",
title = "{Kac-Moody algebra and extended Yang-Baxter relations in the $O(N)$ non-linear $\sigma$-model}",
doi = "10.1016/0370-2693(85)91075-5",
journal = "Phys. Lett. B",
volume = "162",
pages = "137--142",
year = "1985"
}

@article{Maillet:1985ek,
author = "Maillet, Jean Michel",
title = "{New integrable canonical structures in two-dimensional models}",
doi = "10.1016/0550-3213(86)90365-2",
journal = "Nucl. Phys. B",
volume = "269",
pages = "54--76",
year = "1986"
}

@article{Maillet:1985ec,
author = "Maillet, Jean Michel",
title = "{Hamiltonian structures for integrable classical theories from graded Kac-Moody algebras}",
doi = "10.1016/0370-2693(86)91289-X",
journal = "Phys. Lett. B",
volume = "167",
pages = "401--405",
year = "1986"
}

@article{Reyman:1988sf,
author = "Reyman, A.G. and Semenov-Tian-Shansky, M.A.",
title = "{Compatible Poisson structures for Lax equations: an r-matrix approach}",
doi = "10.1016/0375-9601(88)90707-4",
journal = "Phys. Lett. A",
volume = "130",
pages = "456--460",
year = "1988"
}

@article{Sevostyanov:1995hd,
author = "Sevostyanov, Alexei",
title = "{The classical r matrix method for the nonlinear sigma model}",
eprint = "hep-th/9509030",
archivePrefix = "arXiv",
doi = "10.1142/S0217751X96001978",
journal = "Int. J. Mod. Phys. A",
volume = "11",
pages = "4241--4254",
year = "1996"
}

@article{Vicedo:2010qd,
author = "Vicedo, Benoit",
title = "{The Classical R-Matrix of AdS/CFT and its Lie Dialgebra Structure}",
eprint = "1003.1192",
archivePrefix = "arXiv",
primaryClass = "hep-th",
doi = "10.1007/s11005-010-0446-9",
journal = "Lett. Math. Phys.",
volume = "95",
pages = "249--274",
year = "2011"
}

@article{Vicedo:2015pna,
author = "Vicedo, Benoit",
title = "{Deformed integrable $\sigma$-models, classical R-matrices and classical exchange algebra on Drinfel'd doubles}",
eprint = "1504.06303",
archivePrefix = "arXiv",
primaryClass = "hep-th",
doi = "10.1088/1751-8113/48/35/355203",
journal = "J. Phys. A",
volume = "48",
number = "35",
pages = "355203",
year = "2015"
}

@article{Lacroix:2018njs,
author = "Lacroix, Sylvain",
title = "{Integrable models with twist function and affine Gaudin models}",
eprint = "1809.06811",
archivePrefix = "arXiv",
primaryClass = "hep-th",
school = "Lyon, Ecole Normale Superieure",
year = "2018"
}

@article{Bassi:2019aaf,
author = "Bassi, Cristian and Lacroix, Sylvain",
title = "{Integrable deformations of coupled $\sigma$-models}",
eprint = "1912.06157",
archivePrefix = "arXiv",
primaryClass = "hep-th",
doi = "10.1007/JHEP05(2020)059",
journal = "JHEP",
volume = "05",
pages = "059",
year = "2020"
}

@article{Lacroix:2017isl,
author = "Lacroix, Sylvain and Magro, Marc and Vicedo, Benoit",
title = "{Local charges in involution and hierarchies in integrable sigma-models}",
eprint = "1703.01951",
archivePrefix = "arXiv",
primaryClass = "hep-th",
doi = "10.1007/JHEP09(2017)117",
journal = "JHEP",
volume = "09",
pages = "117",
year = "2017"
}

@article{Vicedo:2017cge,
author = "Vicedo, Benoit",
title = "{On Integrable Field Theories as Dihedral Affine Gaudin Models}",
eprint = "1701.04856",
archivePrefix = "arXiv",
primaryClass = "hep-th",
doi = "10.1093/imrn/rny128",
journal = "International Mathematics Research Notices",
volume = "15",
pages = "4513-4601",
year = "2018",
}

@article{Stolin:1991a,
author = "Stolin, A",
title = "{Constant solutions of Yang-Baxter equation for sl(2) and sl(3)}",
doi = "10.7146/math.scand.a-12370",
journal = "Math. Scand.",
volume = "69",
pages = "81",
year = "1991",
}

@article{Stolin:1991n,
author = "Stolin, A",
title = "{On rational solutions of Yang-Baxter equation for sl(n)}",
doi = "10.7146/math.scand.a-12369",
journal = "Math. Scand.",
volume = "69",
pages = "57",
year = "1991",
}

@article{Belavin:1982,
author = "Belavin, A. A. and Drinfel'd, V. G.",
title = "{Solutions of the classical Yang-Baxter equation for simple Lie algebras}",
doi = "10.1007/BF01081585",
journal = "Funct. Anal. Appl.",
volume = "16",
pages = "159--180",
year = "1982",
}

@article{Belavin:1984,
author = "Belavin, A. A. and Drinfel'd, V. G.",
title = "{Triangle equations and simple Lie algebras}",
journal = "Sov. Sci. Rev.",
volume = "C4",
pages = "93",
year = "1984",
}

@article{Hoare:2016wsk,
author = "Hoare, B. and Tseytlin, A.A.",
title = "{Homogeneous Yang-Baxter deformations as non-abelian duals of the $AdS_5$ $\sigma$-model}",
eprint = "1609.02550",
archivePrefix = "arXiv",
primaryClass = "hep-th",
doi = "10.1088/1751-8113/49/49/494001",
journal = "J. Phys. A",
volume = "49",
number = "49",
pages = "494001",
year = "2016"
}

@article{Borsato:2016pas,
author = "Borsato, Riccardo and Wulff, Linus",
title = "{Integrable Deformations of $T$-Dual $\sigma$ Models}",
eprint = "1609.09834",
archivePrefix = "arXiv",
primaryClass = "hep-th",
doi = "10.1103/PhysRevLett.117.251602",
journal = "Phys. Rev. Lett.",
volume = "117",
number = "25",
pages = "251602",
year = "2016"
}

@article{Borsato:2017qsx,
author = "Borsato, Riccardo and Wulff, Linus",
title = "{On non-abelian T-duality and deformations of supercoset string sigma-models}",
eprint = "1706.10169",
archivePrefix = "arXiv",
primaryClass = "hep-th",
doi = "10.1007/JHEP10(2017)024",
journal = "JHEP",
volume = "10",
pages = "024",
year = "2017"
}

@article{Borsato:2018idb,
author = "Borsato, Riccardo and Wulff, Linus",
title = "{Non-abelian T-duality and Yang-Baxter deformations of Green-Schwarz strings}",
eprint = "1806.04083",
archivePrefix = "arXiv",
primaryClass = "hep-th",
doi = "10.1007/JHEP08(2018)027",
journal = "JHEP",
volume = "08",
pages = "027",
year = "2018"
}

@article{Borsato:2018spz,
author = "Borsato, Riccardo and Wulff, Linus",
title = "{Marginal deformations of WZW models and the classical Yang-Baxter equation}",
eprint = "1812.07287",
archivePrefix = "arXiv",
primaryClass = "hep-th",
doi = "10.1088/1751-8121/ab1b9c",
journal = "J. Phys. A",
volume = "52",
number = "22",
pages = "225401",
year = "2019"
}

@article{Drinfeld:1985rx,
author = "Drinfeld, V.G.",
title = "{Hopf algebras and the quantum Yang-Baxter equation}",
journal = "Sov. Math. Dokl.",
volume = "32",
pages = "254--258",
year = "1985"
}

@article{Jimbo:1985zk,
author = "Jimbo, Michio",
title = "{A q-Difference Analogue of U(g) and the Yang-Baxter Equation}",
doi = "10.1007/BF00704588",
journal = "Lett. Math. Phys.",
volume = "10",
pages = "63--69",
year = "1985"
}

@article{Matsumoto:2014nra,
author = "Matsumoto, Takuya and Yoshida, Kentaroh",
title = "{Lunin-Maldacena backgrounds from the classical Yang-Baxter equation -- towards the gravity/CYBE correspondence}",
eprint = "1404.1838",
archivePrefix = "arXiv",
primaryClass = "hep-th",
doi = "10.1007/JHEP06(2014)135",
journal = "JHEP",
volume = "06",
pages = "135",
year = "2014"
}

@article{Osten:2016dvf,
author = "Osten, David and van Tongeren, Stijn J.",
title = "{Abelian Yang-Baxter deformations and TsT transformations}",
eprint = "1608.08504",
archivePrefix = "arXiv",
primaryClass = "hep-th",
doi = "10.1016/j.nuclphysb.2016.12.007",
journal = "Nucl. Phys. B",
volume = "915",
pages = "184--205",
year = "2017"
}

@article{Reshetikhin:1990ep,
author = "Reshetikhin, N.",
title = "{Multiparameter Quantum Groups and Twisted Quasitriangular Hopf Algebras}",
doi = "10.1007/BF00626530",
journal = "Lett. Math. Phys.",
volume = "20",
pages = "331--335",
year = "1990"
}

@article{Ogievetsky:1992ph,
author = "Ogievetsky, O.",
title = "{Hopf structures on the Borel subalgebra of sl(2)}",
editor = "Bure\v{s}, J. and Sou\v{c}ek, J.",
booktitle = "{Proceedings of the 13th Winter School `Geometry and Physics', Zd\'{i}kov, Czech Republic, 1993}",
publisher = "Circolo Matematico di Palermo",
address = "Palermo",
journal = "Suppl. Rend. Circ. Mat. Palermo, II. Ser.",
volume = "37",
year = "1994",
pages = "185",
}

@article{Kulish:1998be,
author = "Kulish, P.P. and Lyakhovsky, V.D. and Mudrov, A.I.",
title = "{Extended Jordanian twists for Lie algebras}",
eprint = "math/9806014",
archivePrefix = "arXiv",
doi = "10.1063/1.532987",
journal = "J. Math. Phys.",
volume = "40",
pages = "4569",
year = "1999"
}

@article{Kulish:1999ua,
author = "Kulish, Peter P. and Lyakhovsky, Vladimir D. and del Olmo, Mariano A.",
title = "{Chains of twists for classical Lie algebras}",
eprint = "math/9908061",
archivePrefix = "arXiv",
doi = "10.1088/0305-4470/32/49/308",
journal = "J. Phys. A",
volume = "32",
pages = "8671",
year = "1999"
}

@article{Tolstoy,
author = "Tolstoy, V. N.",
title = "{Chains of extended Jordanian twists for Lie superalgebras}",
editor = "Ivanov, E. and Pashnev, A.",
booktitle = "{Proceedings of the 5th International Workshop `Supersymmetries and Quantum Symmetries', Dubna, Russia, 2003}",
address = "Dubna",
eprint = "math/0402433",
archivePrefix = "arXiv",
primaryClass = "math",
pages = "242",
year = "2004"
}

@article{vanTongeren:2016eeb,
author = "van Tongeren, Stijn J.",
title = "{Almost abelian twists and AdS/CFT}",
eprint = "1610.05677",
archivePrefix = "arXiv",
primaryClass = "hep-th",
doi = "10.1016/j.physletb.2016.12.002",
journal = "Phys. Lett. B",
volume = "765",
pages = "344--351",
year = "2017"
}

@article{Borsato:2016ose,
author = "Borsato, Riccardo and Wulff, Linus",
title = "{Target space supergeometry of $\eta$ and $\lambda$-deformed strings}",
eprint = "1608.03570",
archivePrefix = "arXiv",
primaryClass = "hep-th",
doi = "10.1007/JHEP10(2016)045",
journal = "JHEP",
volume = "10",
pages = "045",
year = "2016"
}

@article{Costello:2013zra,
author = "Costello, Kevin",
title = "{Supersymmetric gauge theory and the Yangian}",
eprint = "1303.2632",
archivePrefix = "arXiv",
primaryClass = "hep-th",
year = "2013"
}

@article{Costello:2019tri,
author = "Costello, Kevin and Yamazaki, Masahito",
title = "{Gauge Theory And Integrability, III}",
eprint = "1908.02289",
archivePrefix = "arXiv",
primaryClass = "hep-th",
year = "2019"
}

@article{Vicedo:2019dej,
author = "Vicedo, Benoit",
title = "{Holomorphic Chern-Simons theory and affine Gaudin models}",
eprint = "1908.07511",
archivePrefix = "arXiv",
primaryClass = "hep-th",
year = "2019"
}

@article{Delduc:2019whp,
author = "Delduc, Francois and Lacroix, Sylvain and Magro, Marc and Vicedo, Benoit",
title = "{A unifying 2d action for integrable $\sigma$-models from 4d Chern-Simons theory}",
eprint = "1909.13824",
archivePrefix = "arXiv",
primaryClass = "hep-th",
doi = "10.1007/s11005-020-01268-y",
journal = "Lett. Math. Phys.",
volume = "110",
pages = "1645--1687",
year = "2019"
}

@article{Costello:2013sla,
author = "Costello, Kevin",
editor = "Donagi, Ron and Douglas, Michael R. and Kamenova, Ljudmila and Rocek, Martin",
title = "{Integrable lattice models from four-dimensional field theories}",
eprint = "1308.0370",
archivePrefix = "arXiv",
primaryClass = "hep-th",
doi = "10.1090/pspum/088/01483",
journal = "Proc. Symp. Pure Math.",
volume = "88",
pages = "3--24",
year = "2014"
}

@article{Witten:2016spx,
author = "Witten, Edward",
title = "{Integrable lattice models from gauge theory}",
eprint = "1611.00592",
archivePrefix = "arXiv",
primaryClass = "hep-th",
doi = "10.4310/ATMP.2017.v21.n7.a10",
journal = "Adv. Theor. Math. Phys.",
volume = "21",
pages = "1819--1843",
year = "2017"
}

@article{Costello:2017dso,
author = "Costello, Kevin and Witten, Edward and Yamazaki, Masahito",
title = "{Gauge Theory and Integrability, I}",
eprint = "1709.09993",
archivePrefix = "arXiv",
primaryClass = "hep-th",
reportNumber = "IPMU17-0136",
doi = "10.4310/ICCM.2018.v6.n1.a6",
month = "9",
year = "2017"
}

@article{Costello:2018gyb,
author = "Costello, Kevin and Witten, Edward and Yamazaki, Masahito",
title = "{Gauge Theory and Integrability, II}",
eprint = "1802.01579",
archivePrefix = "arXiv",
primaryClass = "hep-th",
reportNumber = "IPMU18-0025",
doi = "10.4310/ICCM.2018.v6.n1.a7",
month = "2",
year = "2018"
}

@article{Benini:2020skc,
author = "Benini, Marco and Schenkel, Alexander and Vicedo, Benoit",
title = "{Homotopical analysis of 4d Chern-Simons theory and integrable field theories}",
eprint = "2008.01829",
archivePrefix = "arXiv",
primaryClass = "hep-th",
month = "8",
year = "2020"
}

@article{Hull:1989jk,
author = "Hull, C.M. and Spence, Bill J.",
title = "{The Gauged Nonlinear $\sigma$-Model With Wess-Zumino Term}",
doi = "10.1016/0370-2693(89)91688-2",
journal = "Phys. Lett. B",
volume = "232",
pages = "204--210",
year = "1989"
}

@article{Jack:1989ne,
author = "Jack, I. and Jones, D.R.T. and Mohammedi, N. and Osborn, H.",
title = "{Gauging the General $\sigma$-Model With a Wess-Zumino Term}",
doi = "10.1016/0550-3213(90)90099-Y",
journal = "Nucl. Phys. B",
volume = "332",
pages = "359",
year = "1990"
}

@article{Klimcik:1995dy,
author = "Klim\v{c}\'{i}k, C. and \v{S}evera, P.",
title = "{Poisson-Lie T-duality and loop groups of Drinfeld doubles}",
eprint = "hep-th/9512040",
archivePrefix = "arXiv",
doi = "10.1016/0370-2693(96)00025-1",
journal = "Phys. Lett. B",
volume = "372",
pages = "65--71",
year = "1996"
}

@article{Klimcik:1996nq,
author = "Klim\v{c}\'{i}k, C. and \v{S}evera, P.",
title = "{Non-abelian momentum winding exchange}",
eprint = "hep-th/9605212",
archivePrefix = "arXiv",
doi = "10.1016/0370-2693(96)00755-1",
journal = "Phys. Lett. B",
volume = "383",
pages = "281--286",
year = "1996"
}

@article{Kawaguchi:2011mz,
author = "Kawaguchi, Io and Orlando, Domenico and Yoshida, Kentaroh",
title = "{Yangian symmetry in deformed WZNW models on squashed spheres}",
eprint = "1104.0738",
archivePrefix = "arXiv",
primaryClass = "hep-th",
doi = "10.1016/j.physletb.2011.06.007",
journal = "Phys. Lett. B",
volume = "701",
pages = "475--480",
year = "2011"
}

@article{Kawaguchi:2013gma,
author = "Kawaguchi, Io and Yoshida, Kentaroh",
title = "{A deformation of quantum affine algebra in squashed Wess-Zumino-Novikov-Witten models}",
eprint = "1311.4696",
archivePrefix = "arXiv",
primaryClass = "hep-th",
doi = "10.1063/1.4880341",
journal = "J. Math. Phys.",
volume = "55",
pages = "062302",
year = "2014"
}

@article{Hoare:2014pna,
author = "Hoare, B. and Roiban, R. and Tseytlin, A.A.",
title = "{On deformations of $AdS_n \times S^n$ supercosets}",
eprint = "1403.5517",
archivePrefix = "arXiv",
primaryClass = "hep-th",
doi = "10.1007/JHEP06(2014)002",
journal = "JHEP",
volume = "06",
pages = "002",
year = "2014"
}

@article{Novikov:1982ei,
author = "Novikov, S.P.",
title = "{The Hamiltonian formalism and a many-valued analogue of Morse theory}",
doi = "10.1070/RM1982v037n05ABEH004020",
journal = "Usp. Mat. Nauk",
volume = "37N5",
number = "5",
pages = "3--49",
year = "1982"
}

@article{Witten:1983ar,
author = "Witten, Edward",
title = "{Non-abelian bosonization in two dimensions}",
doi = "10.1007/BF01215276",
journal = "Commun. Math. Phys.",
volume = "92",
pages = "455--472",
year = "1984"
}

@article{Witten:1983tw,
author = "Witten, Edward",
title = "{Global aspects of current algebra}",
doi = "10.1016/0550-3213(83)90063-9",
journal = "Nucl. Phys. B",
volume = "223",
pages = "422--432",
year = "1983"
}

@article{Fateev:1996ea,
author = "Fateev, V.A.",
title = "{The sigma model (dual) representation for a two-parameter family of integrable quantum field theories}",
doi = "10.1016/0550-3213(96)00256-8",
journal = "Nucl. Phys. B",
volume = "473",
pages = "509--538",
year = "1996"
}

@article{Lukyanov:2012zt,
author = "Lukyanov, Sergei L.",
title = "{The integrable harmonic map problem versus Ricci flow}",
eprint = "1205.3201",
archivePrefix = "arXiv",
primaryClass = "hep-th",
doi = "10.1016/j.nuclphysb.2012.08.002",
journal = "Nucl. Phys. B",
volume = "865",
pages = "308--329",
year = "2012"
}

@article{Cherednik:1981df,
author = "Cherednik, I.V.",
title = "{Relativistically invariant quasiclassical limits of integrable two-dimensional quantum models}",
doi = "10.1007/BF01086395",
journal = "Theor. Math. Phys.",
volume = "47",
pages = "422--425",
year = "1981"
}

@article{Klimcik:2014bta,
author = "Klimcik, Ctirad",
title = "{Integrability of the Bi-Yang-Baxter $\sigma$-Model}",
eprint = "1402.2105",
archivePrefix = "arXiv",
primaryClass = "math-ph",
doi = "10.1007/s11005-014-0709-y",
journal = "Lett. Math. Phys.",
volume = "104",
pages = "1095--1106",
year = "2014"
}

@article{Delduc:2015xdm,
author = "Delduc, Francois and Lacroix, Sylvain and Magro, Marc and Vicedo, Benoit",
title = "{On the Hamiltonian integrability of the bi-Yang-Baxter $\sigma$-model}",
eprint = "1512.02462",
archivePrefix = "arXiv",
primaryClass = "hep-th",
doi = "10.1007/JHEP03(2016)104",
journal = "JHEP",
volume = "03",
pages = "104",
year = "2016"
}

@article{Babichenko:2009dk,
author = "Babichenko, A. and Stefa\'{n}ski~jr., B. and Zarembo, K.",
title = "{Integrability and the AdS$_3$/CFT$_2$ correspondence}",
eprint = "0912.1723",
archivePrefix = "arXiv",
primaryClass = "hep-th",
doi = "10.1007/JHEP03(2010)058",
journal = "JHEP",
volume = "03",
pages = "058",
year = "2010"
}

@article{Hoare:2014oua,
author = "Hoare, Ben",
title = "{Towards a two-parameter $q$-deformation of $AdS_3 \times S^3 \times M^4$ superstrings}",
eprint = "1411.1266",
archivePrefix = "arXiv",
primaryClass = "hep-th",
doi = "10.1016/j.nuclphysb.2014.12.012",
journal = "Nucl. Phys. B",
volume = "891",
pages = "259--295",
year = "2015"
}

@article{Cagnazzo:2012se,
author = "Cagnazzo, A. and Zarembo, K.",
title = "{B-field in AdS$_3$/CFT$_2$ correspondence and integrability}",
eprint = "1209.4049",
archivePrefix = "arXiv",
primaryClass = "hep-th",
doi = "10.1007/JHEP11(2012)133",
journal = "JHEP",
volume = "11",
pages = "133",
year = "2012",
note = "[Erratum: \href{http://dx.doi.org/10.1007/JHEP04(2013)003}{\textsf{JHEP 04, 003 (2013)}}]"
}

@article{Delduc:2018xug,
author = "Delduc, F. and Hoare, B. and Kameyama, T. and Lacroix, S. and Magro, M.",
title = "{Three-parameter integrable deformation of $\Integer_4$ permutation supercosets}",
eprint = "1811.00453",
archivePrefix = "arXiv",
primaryClass = "hep-th",
doi = "10.1007/JHEP01(2019)109",
journal = "JHEP",
volume = "01",
pages = "109",
year = "2019"
}

@article{Severa:2017kcs,
author = "\v{S}evera, Pavol",
title = "{On integrability of 2-dimensional $\sigma$-models of Poisson-Lie type}",
eprint = "1709.02213",
archivePrefix = "arXiv",
primaryClass = "hep-th",
doi = "10.1007/JHEP11(2017)015",
journal = "JHEP",
volume = "11",
pages = "015",
year = "2017"
}

@article{Demulder:2017zhz,
author = "Demulder, Saskia and Driezen, Sibylle and Sevrin, Alexander and Thompson, Daniel C.",
title = "{Classical and quantum aspects of Yang-Baxter Wess-Zumino models}",
eprint = "1711.00084",
archivePrefix = "arXiv",
primaryClass = "hep-th",
doi = "10.1007/JHEP03(2018)041",
journal = "JHEP",
volume = "03",
pages = "041",
year = "2018"
}

@article{Fukushima:2020kta,
author = "Fukushima, Osamu and Sakamoto, Jun-ichi and Yoshida, Kentaroh",
title = "{Comments on $\eta$-deformed principal chiral model from 4D Chern-Simons theory}",
eprint = "2003.07309",
archivePrefix = "arXiv",
primaryClass = "hep-th",
doi = "10.1016/j.nuclphysb.2020.115080",
journal = "Nucl. Phys. B",
volume = "957",
pages = "115080",
year = "2020"
}

@article{Hoare:2020fye,
author = "Hoare, Ben and Levine, Nat and Tseytlin, Arkady A.",
title = "{Sigma models with local couplings: a new integrability--RG flow connection}",
eprint = "2008.01112",
archivePrefix = "arXiv",
primaryClass = "hep-th",
month = "8",
year = "2020"
}

@article{Lichnerowicz,
author = "Lichnerowicz, Andr\'{e} and Medina, Alberto",
title = "{On Lie Groups with Left-Invariant Symplectic or K\"{a}hlerian Structures}",
doi = "10.1007/BF00398959",
journal = "Lett. Math. Phys.",
volume = "16",
pages = "225-235",
year = "1988"
}

@article{Chu,
author = "Chu, Bon-Yao",
title = "{Symplectic Homogeneous Spaces}",
doi = "10.1090/S0002-9947-1974-0342642-7 ",
journal = "Trans. Amer. Math. Soc.",
volume = "197",
pages = "145-159",
year = "1974"
}

\end{bibtex}

\bibliographystyle{nb}
\bibliography{\jobname}

\begin{thebibliography}{10}
\providecommand{\href}[2]{#2}
\providecommand{\arxivref}[2]{\href{http://arxiv.org/abs/#1}{#2}}
\providecommand{\doiref}[2]{\href{http://dx.doi.org/#1}{#2}}
\providecommand{\nbbstauthor}[1]{#1}
\providecommand{\nbbstjournal}[1]{\textsf{#1}}
\providecommand{\nbbsttitle}[1]{\textit{#1}}
\providecommand{\nbbsturl}[1]{\texttt{#1}}
\providecommand{\nbbsteprint}[1]{\texttt{#1}}
\providecommand{\nbbststyle}{\raggedright\small\parskip0pt}
\nbbststyle

\bibitem{Klimcik:2002zj}
\nbbstauthor{C.~Klim\v{c}\'{i}k},
\nbbsttitle{``{Yang-Baxter $\sigma$-models and dS/AdS T-duality}''},
\nbbstjournal{\doiref{10.1088/1126-6708/2002/12/051}{JHEP~0212,~051~(2002)}},
\nbbsteprint{\arxivref{hep-th/0210095}{hep-th/0210095}}.

\bibitem{Klimcik:2008eq}
\nbbstauthor{C.~Klim\v{c}\'{i}k},
\nbbsttitle{``{On integrability of the Yang-Baxter $\sigma$-model}''},
\nbbstjournal{\doiref{10.1063/1.3116242}{J.~Math.~Phys.~50,~043508~(2009)}},
\nbbsteprint{\arxivref{0802.3518}{arxiv:0802.3518}}.

\bibitem{Novikov:1982ei}
\nbbstauthor{S.~Novikov},
\nbbsttitle{``{The Hamiltonian formalism and a many-valued analogue of Morse
  theory}''},
\nbbstjournal{\doiref{10.1070/RM1982v037n05ABEH004020}{Usp.~Mat.~Nauk~37N5,~3~(1982)}}.

\bibitem{Witten:1983tw}
\nbbstauthor{E.~Witten},
\nbbsttitle{``{Global aspects of current algebra}''},
\nbbstjournal{\doiref{10.1016/0550-3213(83)90063-9}{Nucl.~Phys.~B~223,~422~(1983)}}.

\bibitem{Witten:1983ar}
\nbbstauthor{E.~Witten},
\nbbsttitle{``{Non-abelian bosonization in two dimensions}''},
\nbbstjournal{\doiref{10.1007/BF01215276}{Commun.~Math.~Phys.~92,~455~(1984)}}.

\bibitem{Cherednik:1981df}
\nbbstauthor{I.~Cherednik},
\nbbsttitle{``{Relativistically invariant quasiclassical limits of integrable
  two-dimensional quantum models}''},
\nbbstjournal{\doiref{10.1007/BF01086395}{Theor.~Math.~Phys.~47,~422~(1981)}}.

\bibitem{Kawaguchi:2011mz}
\nbbstauthor{I.~Kawaguchi, D.~Orlando and K.~Yoshida},
\nbbsttitle{``{Yangian symmetry in deformed WZNW models on squashed
  spheres}''},
\nbbstjournal{\doiref{10.1016/j.physletb.2011.06.007}{Phys.~Lett.~B~701,~475~(2011)}},
\nbbsteprint{\arxivref{1104.0738}{arxiv:1104.0738}}.

\bibitem{Kawaguchi:2013gma}
\nbbstauthor{I.~Kawaguchi and K.~Yoshida},
\nbbsttitle{``{A deformation of quantum affine algebra in squashed
  Wess-Zumino-Novikov-Witten models}''},
\nbbstjournal{\doiref{10.1063/1.4880341}{J.~Math.~Phys.~55,~062302~(2014)}},
\nbbsteprint{\arxivref{1311.4696}{arxiv:1311.4696}}.

\bibitem{Delduc:2014uaa}
\nbbstauthor{F.~Delduc, M.~Magro and B.~Vicedo},
\nbbsttitle{``{Integrable double deformation of the principal chiral model}''},
\nbbstjournal{\doiref{10.1016/j.nuclphysb.2014.12.018}{Nucl.~Phys.~B~891,~312~(2015)}},
\nbbsteprint{\arxivref{1410.8066}{arxiv:1410.8066}}.

\bibitem{Borsato:2018idb}
\nbbstauthor{R.~Borsato and L.~Wulff},
\nbbsttitle{``{Non-abelian T-duality and Yang-Baxter deformations of
  Green-Schwarz strings}''},
\nbbstjournal{\doiref{10.1007/JHEP08(2018)027}{JHEP~1808,~027~(2018)}},
\nbbsteprint{\arxivref{1806.04083}{arxiv:1806.04083}}.

\bibitem{Klimcik:2019kkf}
\nbbstauthor{C.~Klim\v{c}\'{i}k},
\nbbsttitle{``{Dressing cosets and multi-parametric integrable
  deformations}''},
\nbbstjournal{\doiref{10.1007/JHEP07(2019)176}{JHEP~1907,~176~(2019)}},
\nbbsteprint{\arxivref{1903.00439}{arxiv:1903.00439}}.

\bibitem{Klimcik:2020fhs}
\nbbstauthor{C.~Klim\v{c}\'{i}k},
\nbbsttitle{``{Strong integrability of the bi-YB-WZ model}''},
\nbbsteprint{\arxivref{2001.05466}{arxiv:2001.05466}}.

\bibitem{Delduc:2013fga}
\nbbstauthor{F.~Delduc, M.~Magro and B.~Vicedo},
\nbbsttitle{``{On classical $q$-deformations of integrable $\sigma$-models}''},
\nbbstjournal{\doiref{10.1007/JHEP11(2013)192}{JHEP~1311,~192~(2013)}},
\nbbsteprint{\arxivref{1308.3581}{arxiv:1308.3581}}.

\bibitem{Maillet:1985fn}
\nbbstauthor{J.~M.~Maillet},
\nbbsttitle{``{Kac-Moody algebra and extended Yang-Baxter relations in the
  $O(N)$ non-linear $\sigma$-model}''},
\nbbstjournal{\doiref{10.1016/0370-2693(85)91075-5}{Phys.~Lett.~B~162,~137~(1985)}}.

\bibitem{Maillet:1985ek}
\nbbstauthor{J.~M.~Maillet},
\nbbsttitle{``{New integrable canonical structures in two-dimensional
  models}''},
\nbbstjournal{\doiref{10.1016/0550-3213(86)90365-2}{Nucl.~Phys.~B~269,~54~(1986)}}.

\bibitem{Maillet:1985ec}
\nbbstauthor{J.~M.~Maillet},
\nbbsttitle{``{Hamiltonian structures for integrable classical theories from
  graded Kac-Moody algebras}''},
\nbbstjournal{\doiref{10.1016/0370-2693(86)91289-X}{Phys.~Lett.~B~167,~401~(1986)}}.

\bibitem{Reyman:1988sf}
\nbbstauthor{A.~Reyman and M.~Semenov-Tian-Shansky},
\nbbsttitle{``{Compatible Poisson structures for Lax equations: an r-matrix
  approach}''},
\nbbstjournal{\doiref{10.1016/0375-9601(88)90707-4}{Phys.~Lett.~A~130,~456~(1988)}}.

\bibitem{Sevostyanov:1995hd}
\nbbstauthor{A.~Sevostyanov},
\nbbsttitle{``{The classical r matrix method for the nonlinear sigma model}''},
\nbbstjournal{\doiref{10.1142/S0217751X96001978}{Int.~J.~Mod.~Phys.~A~11,~4241~(1996)}},
\nbbsteprint{\arxivref{hep-th/9509030}{hep-th/9509030}}.

\bibitem{Vicedo:2010qd}
\nbbstauthor{B.~Vicedo},
\nbbsttitle{``{The Classical R-Matrix of AdS/CFT and its Lie Dialgebra
  Structure}''},
\nbbstjournal{\doiref{10.1007/s11005-010-0446-9}{Lett.~Math.~Phys.~95,~249~(2011)}},
\nbbsteprint{\arxivref{1003.1192}{arxiv:1003.1192}}.

\bibitem{Lacroix:2018njs}
\nbbstauthor{S.~Lacroix},
\nbbsttitle{``{Integrable models with twist function and affine Gaudin
  models}''},
\nbbsteprint{\arxivref{1809.06811}{arxiv:1809.06811}}.

\bibitem{Vicedo:2017cge}
\nbbstauthor{B.~Vicedo},
\nbbsttitle{``{On Integrable Field Theories as Dihedral Affine Gaudin
  Models}''},
\nbbstjournal{\doiref{10.1093/imrn/rny128}{International~Mathematics~Research~Notices~15,~4513~(2018)}},
\nbbsteprint{\arxivref{1701.04856}{arxiv:1701.04856}}.

\bibitem{Belavin:1982}
\nbbstauthor{A.~A.~Belavin and V.~G.~Drinfel'd},
\nbbsttitle{``{Solutions of the classical Yang-Baxter equation for simple Lie
  algebras}''},
\nbbstjournal{\doiref{10.1007/BF01081585}{Funct.~Anal.~Appl.~16,~159~(1982)}}.

\bibitem{Belavin:1984}
\nbbstauthor{A.~A.~Belavin and V.~G.~Drinfel'd},
\nbbsttitle{``{Triangle equations and simple Lie algebras}''},
\nbbstjournal{Sov.~Sci.~Rev.~C4,~93~(1984)}.

\bibitem{Stolin:1991a}
\nbbstauthor{A.~Stolin},
\nbbsttitle{``{Constant solutions of Yang-Baxter equation for sl(2) and
  sl(3)}''},
\nbbstjournal{\doiref{10.7146/math.scand.a-12370}{Math.~Scand.~69,~81~(1991)}}.

\bibitem{Ogievetsky:1992ph}
\nbbstauthor{O.~Ogievetsky},
\nbbsttitle{``{Hopf structures on the Borel subalgebra of sl(2)}''},
\nbbstjournal{Suppl.~Rend.~Circ.~Mat.~Palermo,~II.~Ser.~37,~185~(1994)},
in: \nbbsttitle{``{Proceedings of the 13th Winter School `Geometry and
  Physics', Zd\'{i}kov, Czech Republic, 1993}''},
pp.~185,
ed.: J.~Bure\v{s} and J.~Sou\v{c}ek,
Circolo Matematico di Palermo (1994),
Palermo.

\bibitem{Kulish:1998be}
\nbbstauthor{P.~Kulish, V.~Lyakhovsky and A.~Mudrov},
\nbbsttitle{``{Extended Jordanian twists for Lie algebras}''},
\nbbstjournal{\doiref{10.1063/1.532987}{J.~Math.~Phys.~40,~4569~(1999)}},
\nbbsteprint{\arxivref{math/9806014}{math/9806014}}.

\bibitem{Kulish:1999ua}
\nbbstauthor{P.~P.~Kulish, V.~D.~Lyakhovsky and M.~A.~del~Olmo},
\nbbsttitle{``{Chains of twists for classical Lie algebras}''},
\nbbstjournal{\doiref{10.1088/0305-4470/32/49/308}{J.~Phys.~A~32,~8671~(1999)}},
\nbbsteprint{\arxivref{math/9908061}{math/9908061}}.

\bibitem{Tolstoy}
\nbbstauthor{V.~N.~Tolstoy},
\nbbsttitle{``{Chains of extended Jordanian twists for Lie superalgebras}''},
\nbbsteprint{\arxivref{math/0402433}{math/0402433}},
in: \nbbsttitle{``{Proceedings of the 5th International Workshop
  `Supersymmetries and Quantum Symmetries', Dubna, Russia, 2003}''},
pp.~242,
ed.: E.~Ivanov and A.~Pashnev,
Dubna.

\bibitem{Drinfeld:1985rx}
\nbbstauthor{V.~Drinfeld},
\nbbsttitle{``{Hopf algebras and the quantum Yang-Baxter equation}''},
\nbbstjournal{Sov.~Math.~Dokl.~32,~254~(1985)}.

\bibitem{Jimbo:1985zk}
\nbbstauthor{M.~Jimbo},
\nbbsttitle{``{A q-Difference Analogue of U(g) and the Yang-Baxter
  Equation}''},
\nbbstjournal{\doiref{10.1007/BF00704588}{Lett.~Math.~Phys.~10,~63~(1985)}}.

\bibitem{Reshetikhin:1990ep}
\nbbstauthor{N.~Reshetikhin},
\nbbsttitle{``{Multiparameter Quantum Groups and Twisted Quasitriangular Hopf
  Algebras}''},
\nbbstjournal{\doiref{10.1007/BF00626530}{Lett.~Math.~Phys.~20,~331~(1990)}}.

\bibitem{Matsumoto:2014nra}
\nbbstauthor{T.~Matsumoto and K.~Yoshida},
\nbbsttitle{``{Lunin-Maldacena backgrounds from the classical Yang-Baxter
  equation -- towards the gravity/CYBE correspondence}''},
\nbbstjournal{\doiref{10.1007/JHEP06(2014)135}{JHEP~1406,~135~(2014)}},
\nbbsteprint{\arxivref{1404.1838}{arxiv:1404.1838}}.

\bibitem{Osten:2016dvf}
\nbbstauthor{D.~Osten and S.~J.~van~Tongeren},
\nbbsttitle{``{Abelian Yang-Baxter deformations and TsT transformations}''},
\nbbstjournal{\doiref{10.1016/j.nuclphysb.2016.12.007}{Nucl.~Phys.~B~915,~184~(2017)}},
\nbbsteprint{\arxivref{1608.08504}{arxiv:1608.08504}}.

\bibitem{Borsato:2016ose}
\nbbstauthor{R.~Borsato and L.~Wulff},
\nbbsttitle{``{Target space supergeometry of $\eta$ and $\lambda$-deformed
  strings}''},
\nbbstjournal{\doiref{10.1007/JHEP10(2016)045}{JHEP~1610,~045~(2016)}},
\nbbsteprint{\arxivref{1608.03570}{arxiv:1608.03570}}.

\bibitem{vanTongeren:2016eeb}
\nbbstauthor{S.~J.~van~Tongeren},
\nbbsttitle{``{Almost abelian twists and AdS/CFT}''},
\nbbstjournal{\doiref{10.1016/j.physletb.2016.12.002}{Phys.~Lett.~B~765,~344~(2017)}},
\nbbsteprint{\arxivref{1610.05677}{arxiv:1610.05677}}.

\bibitem{Chu}
\nbbstauthor{B.-Y.~Chu},
\nbbsttitle{``{Symplectic Homogeneous Spaces}''},
\nbbstjournal{\doiref{10.1090/S0002-9947-1974-0342642-7}{Trans.~Amer.~Math.~Soc.~197,~145~(1974)}}.

\bibitem{Lichnerowicz}
\nbbstauthor{A.~Lichnerowicz and A.~Medina},
\nbbsttitle{``{On Lie Groups with Left-Invariant Symplectic or K\"{a}hlerian
  Structures}''},
\nbbstjournal{\doiref{10.1007/BF00398959}{Lett.~Math.~Phys.~16,~225~(1988)}}.

\bibitem{Delduc:2019whp}
\nbbstauthor{F.~Delduc, S.~Lacroix, M.~Magro and B.~Vicedo},
\nbbsttitle{``{A unifying 2d action for integrable $\sigma$-models from 4d
  Chern-Simons theory}''},
\nbbstjournal{\doiref{10.1007/s11005-020-01268-y}{Lett.~Math.~Phys.~110,~1645~(2019)}},
\nbbsteprint{\arxivref{1909.13824}{arxiv:1909.13824}}.

\bibitem{Delduc:2019bcl}
\nbbstauthor{F.~Delduc, S.~Lacroix, M.~Magro and B.~Vicedo},
\nbbsttitle{``{Assembling integrable $\sigma$-models as affine Gaudin
  models}''},
\nbbstjournal{\doiref{10.1007/JHEP06(2019)017}{JHEP~1906,~017~(2019)}},
\nbbsteprint{\arxivref{1903.00368}{arxiv:1903.00368}}.

\bibitem{Vicedo:2015pna}
\nbbstauthor{B.~Vicedo},
\nbbsttitle{``{Deformed integrable $\sigma$-models, classical R-matrices and
  classical exchange algebra on Drinfel'd doubles}''},
\nbbstjournal{\doiref{10.1088/1751-8113/48/35/355203}{J.~Phys.~A~48,~355203~(2015)}},
\nbbsteprint{\arxivref{1504.06303}{arxiv:1504.06303}}.

\bibitem{Lacroix:2017isl}
\nbbstauthor{S.~Lacroix, M.~Magro and B.~Vicedo},
\nbbsttitle{``{Local charges in involution and hierarchies in integrable
  sigma-models}''},
\nbbstjournal{\doiref{10.1007/JHEP09(2017)117}{JHEP~1709,~117~(2017)}},
\nbbsteprint{\arxivref{1703.01951}{arxiv:1703.01951}}.

\bibitem{Klimcik:2017ken}
\nbbstauthor{C.~Klim\v{c}\'{i}k},
\nbbsttitle{``{Yang-Baxter $\sigma$-model with WZNW term as
  $\mathcal{E}$-model}''},
\nbbstjournal{\doiref{10.1016/j.physletb.2017.07.051}{Phys.~Lett.~B~772,~725~(2017)}},
\nbbsteprint{\arxivref{1706.08912}{arxiv:1706.08912}}.

\bibitem{Severa:2017kcs}
\nbbstauthor{P.~\v{S}evera},
\nbbsttitle{``{On integrability of 2-dimensional $\sigma$-models of Poisson-Lie
  type}''},
\nbbstjournal{\doiref{10.1007/JHEP11(2017)015}{JHEP~1711,~015~(2017)}},
\nbbsteprint{\arxivref{1709.02213}{arxiv:1709.02213}}.

\bibitem{Klimcik:1995dy}
\nbbstauthor{C.~Klim\v{c}\'{i}k and P.~\v{S}evera},
\nbbsttitle{``{Poisson-Lie T-duality and loop groups of Drinfeld doubles}''},
\nbbstjournal{\doiref{10.1016/0370-2693(96)00025-1}{Phys.~Lett.~B~372,~65~(1996)}},
\nbbsteprint{\arxivref{hep-th/9512040}{hep-th/9512040}}.

\bibitem{Klimcik:1996nq}
\nbbstauthor{C.~Klim\v{c}\'{i}k and P.~\v{S}evera},
\nbbsttitle{``{Non-abelian momentum winding exchange}''},
\nbbstjournal{\doiref{10.1016/0370-2693(96)00755-1}{Phys.~Lett.~B~383,~281~(1996)}},
\nbbsteprint{\arxivref{hep-th/9605212}{hep-th/9605212}}.

\bibitem{Costello:2013zra}
\nbbstauthor{K.~Costello},
\nbbsttitle{``{Supersymmetric gauge theory and the Yangian}''},
\nbbsteprint{\arxivref{1303.2632}{arxiv:1303.2632}}.

\bibitem{Costello:2019tri}
\nbbstauthor{K.~Costello and M.~Yamazaki},
\nbbsttitle{``{Gauge Theory And Integrability, III}''},
\nbbsteprint{\arxivref{1908.02289}{arxiv:1908.02289}}.

\bibitem{Vicedo:2019dej}
\nbbstauthor{B.~Vicedo},
\nbbsttitle{``{Holomorphic Chern-Simons theory and affine Gaudin models}''},
\nbbsteprint{\arxivref{1908.07511}{arxiv:1908.07511}}.

\bibitem{Costello:2013sla}
\nbbstauthor{K.~Costello},
\nbbsttitle{``{Integrable lattice models from four-dimensional field
  theories}''},
\nbbstjournal{\doiref{10.1090/pspum/088/01483}{Proc.~Symp.~Pure~Math.~88,~3~(2014)}},
\nbbsteprint{\arxivref{1308.0370}{arxiv:1308.0370}}.

\bibitem{Witten:2016spx}
\nbbstauthor{E.~Witten},
\nbbsttitle{``{Integrable lattice models from gauge theory}''},
\nbbstjournal{\doiref{10.4310/ATMP.2017.v21.n7.a10}{Adv.~Theor.~Math.~Phys.~21,~1819~(2017)}},
\nbbsteprint{\arxivref{1611.00592}{arxiv:1611.00592}}.

\bibitem{Costello:2017dso}
\nbbstauthor{K.~Costello, E.~Witten and M.~Yamazaki},
\nbbsttitle{``{Gauge Theory and Integrability, I}''},
\nbbsteprint{\arxivref{1709.09993}{arxiv:1709.09993}}.

\bibitem{Costello:2018gyb}
\nbbstauthor{K.~Costello, E.~Witten and M.~Yamazaki},
\nbbsttitle{``{Gauge Theory and Integrability, II}''},
\nbbsteprint{\arxivref{1802.01579}{arxiv:1802.01579}}.

\bibitem{Benini:2020skc}
\nbbstauthor{M.~Benini, A.~Schenkel and B.~Vicedo},
\nbbsttitle{``{Homotopical analysis of 4d Chern-Simons theory and integrable
  field theories}''},
\nbbsteprint{\arxivref{2008.01829}{arxiv:2008.01829}}.

\bibitem{Hoare:2016wsk}
\nbbstauthor{B.~Hoare and A.~Tseytlin},
\nbbsttitle{``{Homogeneous Yang-Baxter deformations as non-abelian duals of the
  $AdS_5$ $\sigma$-model}''},
\nbbstjournal{\doiref{10.1088/1751-8113/49/49/494001}{J.~Phys.~A~49,~494001~(2016)}},
\nbbsteprint{\arxivref{1609.02550}{arxiv:1609.02550}}.

\bibitem{Borsato:2016pas}
\nbbstauthor{R.~Borsato and L.~Wulff},
\nbbsttitle{``{Integrable Deformations of $T$-Dual $\sigma$ Models}''},
\nbbstjournal{\doiref{10.1103/PhysRevLett.117.251602}{Phys.~Rev.~Lett.~117,~251602~(2016)}},
\nbbsteprint{\arxivref{1609.09834}{arxiv:1609.09834}}.

\bibitem{Borsato:2017qsx}
\nbbstauthor{R.~Borsato and L.~Wulff},
\nbbsttitle{``{On non-abelian T-duality and deformations of supercoset string
  sigma-models}''},
\nbbstjournal{\doiref{10.1007/JHEP10(2017)024}{JHEP~1710,~024~(2017)}},
\nbbsteprint{\arxivref{1706.10169}{arxiv:1706.10169}}.

\bibitem{Borsato:2018spz}
\nbbstauthor{R.~Borsato and L.~Wulff},
\nbbsttitle{``{Marginal deformations of WZW models and the classical
  Yang-Baxter equation}''},
\nbbstjournal{\doiref{10.1088/1751-8121/ab1b9c}{J.~Phys.~A~52,~225401~(2019)}},
\nbbsteprint{\arxivref{1812.07287}{arxiv:1812.07287}}.

\bibitem{Hull:1989jk}
\nbbstauthor{C.~Hull and B.~J.~Spence},
\nbbsttitle{``{The Gauged Nonlinear $\sigma$-Model With Wess-Zumino Term}''},
\nbbstjournal{\doiref{10.1016/0370-2693(89)91688-2}{Phys.~Lett.~B~232,~204~(1989)}}.

\bibitem{Jack:1989ne}
\nbbstauthor{I.~Jack, D.~Jones, N.~Mohammedi and H.~Osborn},
\nbbsttitle{``{Gauging the General $\sigma$-Model With a Wess-Zumino Term}''},
\nbbstjournal{\doiref{10.1016/0550-3213(90)90099-Y}{Nucl.~Phys.~B~332,~359~(1990)}}.

\bibitem{Stolin:1991n}
\nbbstauthor{A.~Stolin},
\nbbsttitle{``{On rational solutions of Yang-Baxter equation for sl(n)}''},
\nbbstjournal{\doiref{10.7146/math.scand.a-12369}{Math.~Scand.~69,~57~(1991)}}.

\bibitem{Demulder:2017zhz}
\nbbstauthor{S.~Demulder, S.~Driezen, A.~Sevrin and D.~C.~Thompson},
\nbbsttitle{``{Classical and quantum aspects of Yang-Baxter Wess-Zumino
  models}''},
\nbbstjournal{\doiref{10.1007/JHEP03(2018)041}{JHEP~1803,~041~(2018)}},
\nbbsteprint{\arxivref{1711.00084}{arxiv:1711.00084}}.

\bibitem{Fateev:1996ea}
\nbbstauthor{V.~Fateev},
\nbbsttitle{``{The sigma model (dual) representation for a two-parameter family
  of integrable quantum field theories}''},
\nbbstjournal{\doiref{10.1016/0550-3213(96)00256-8}{Nucl.~Phys.~B~473,~509~(1996)}}.

\bibitem{Lukyanov:2012zt}
\nbbstauthor{S.~L.~Lukyanov},
\nbbsttitle{``{The integrable harmonic map problem versus Ricci flow}''},
\nbbstjournal{\doiref{10.1016/j.nuclphysb.2012.08.002}{Nucl.~Phys.~B~865,~308~(2012)}},
\nbbsteprint{\arxivref{1205.3201}{arxiv:1205.3201}}.

\bibitem{Hoare:2014pna}
\nbbstauthor{B.~Hoare, R.~Roiban and A.~Tseytlin},
\nbbsttitle{``{On deformations of $AdS_n \times S^n$ supercosets}''},
\nbbstjournal{\doiref{10.1007/JHEP06(2014)002}{JHEP~1406,~002~(2014)}},
\nbbsteprint{\arxivref{1403.5517}{arxiv:1403.5517}}.

\bibitem{Klimcik:2014bta}
\nbbstauthor{C.~Klimcik},
\nbbsttitle{``{Integrability of the Bi-Yang-Baxter $\sigma$-Model}''},
\nbbstjournal{\doiref{10.1007/s11005-014-0709-y}{Lett.~Math.~Phys.~104,~1095~(2014)}},
\nbbsteprint{\arxivref{1402.2105}{arxiv:1402.2105}}.

\bibitem{Delduc:2015xdm}
\nbbstauthor{F.~Delduc, S.~Lacroix, M.~Magro and B.~Vicedo},
\nbbsttitle{``{On the Hamiltonian integrability of the bi-Yang-Baxter
  $\sigma$-model}''},
\nbbstjournal{\doiref{10.1007/JHEP03(2016)104}{JHEP~1603,~104~(2016)}},
\nbbsteprint{\arxivref{1512.02462}{arxiv:1512.02462}}.

\bibitem{Delduc:2017fib}
\nbbstauthor{F.~Delduc, B.~Hoare, T.~Kameyama and M.~Magro},
\nbbsttitle{``{Combining the bi-Yang-Baxter deformation, the Wess-Zumino term
  and TsT transformations in one integrable $\sigma$-model}''},
\nbbstjournal{\doiref{10.1007/JHEP10(2017)212}{JHEP~1710,~212~(2017)}},
\nbbsteprint{\arxivref{1707.08371}{arxiv:1707.08371}}.

\bibitem{Cagnazzo:2012se}
\nbbstauthor{A.~Cagnazzo and K.~Zarembo},
\nbbsttitle{``{B-field in AdS$_3$/CFT$_2$ correspondence and integrability}''},
\nbbstjournal{\doiref{10.1007/JHEP11(2012)133}{JHEP~1211,~133~(2012)}},
\nbbsteprint{\arxivref{1209.4049}{arxiv:1209.4049}},
[Erratum: \href{http://dx.doi.org/10.1007/JHEP04(2013)003}{\textsf{JHEP 04, 003
  (2013)}}].

\bibitem{Delduc:2018xug}
\nbbstauthor{F.~Delduc, B.~Hoare, T.~Kameyama, S.~Lacroix and M.~Magro},
\nbbsttitle{``{Three-parameter integrable deformation of $\Integer_4$
  permutation supercosets}''},
\nbbstjournal{\doiref{10.1007/JHEP01(2019)109}{JHEP~1901,~109~(2019)}},
\nbbsteprint{\arxivref{1811.00453}{arxiv:1811.00453}}.

\bibitem{Babichenko:2009dk}
\nbbstauthor{A.~Babichenko, B.~Stefa\'{n}ski~jr. and K.~Zarembo},
\nbbsttitle{``{Integrability and the AdS$_3$/CFT$_2$ correspondence}''},
\nbbstjournal{\doiref{10.1007/JHEP03(2010)058}{JHEP~1003,~058~(2010)}},
\nbbsteprint{\arxivref{0912.1723}{arxiv:0912.1723}}.

\bibitem{Hoare:2014oua}
\nbbstauthor{B.~Hoare},
\nbbsttitle{``{Towards a two-parameter $q$-deformation of $AdS_3 \times S^3
  \times M^4$ superstrings}''},
\nbbstjournal{\doiref{10.1016/j.nuclphysb.2014.12.012}{Nucl.~Phys.~B~891,~259~(2015)}},
\nbbsteprint{\arxivref{1411.1266}{arxiv:1411.1266}}.

\bibitem{Fukushima:2020kta}
\nbbstauthor{O.~Fukushima, J.-i.~Sakamoto and K.~Yoshida},
\nbbsttitle{``{Comments on $\eta$-deformed principal chiral model from 4D
  Chern-Simons theory}''},
\nbbstjournal{\doiref{10.1016/j.nuclphysb.2020.115080}{Nucl.~Phys.~B~957,~115080~(2020)}},
\nbbsteprint{\arxivref{2003.07309}{arxiv:2003.07309}}.

\bibitem{Hoare:2020fye}
\nbbstauthor{B.~Hoare, N.~Levine and A.~A.~Tseytlin},
\nbbsttitle{``{Sigma models with local couplings: a new integrability--RG flow
  connection}''},
\nbbsteprint{\arxivref{2008.01112}{arxiv:2008.01112}}.

\bibitem{Bassi:2019aaf}
\nbbstauthor{C.~Bassi and S.~Lacroix},
\nbbsttitle{``{Integrable deformations of coupled $\sigma$-models}''},
\nbbstjournal{\doiref{10.1007/JHEP05(2020)059}{JHEP~2005,~059~(2020)}},
\nbbsteprint{\arxivref{1912.06157}{arxiv:1912.06157}}.

\bibitem{Delduc:2018hty}
\nbbstauthor{F.~Delduc, S.~Lacroix, M.~Magro and B.~Vicedo},
\nbbsttitle{``{Integrable Coupled $\sigma$ Models}''},
\nbbstjournal{\doiref{10.1103/PhysRevLett.122.041601}{Phys.~Rev.~Lett.~122,~041601~(2019)}},
\nbbsteprint{\arxivref{1811.12316}{arxiv:1811.12316}}.

\end{thebibliography}

\end{document}